\documentclass[prb, aps, superscriptaddress, twocolumn, showpacs, reprint, nobibnotes]{revtex4-1}
\usepackage{amsmath,amssymb,bm,wrapfig,mathrsfs}
\usepackage[hyperfootnotes=false]{hyperref}
\usepackage{graphicx,color}

\begin{document}
\title{
Brillouin-Wigner theory for high-frequency expansion in periodically driven systems:
Application to Floquet topological insulators
}
\author{Takahiro Mikami}
\affiliation{Department of Physics,  University of Tokyo, Hongo, Tokyo, 113-0033, Japan}
\author{Sota Kitamura}
\affiliation{Department of Physics,  University of Tokyo, Hongo, Tokyo, 113-0033, Japan}
\author{Kenji Yasuda}
\altaffiliation[Present address: ]{Department of Applied Physics, University of Tokyo, Hongo, Tokyo 113-8656, Japan.}
\affiliation{Department of Physics,  University of Tokyo, Hongo, Tokyo, 113-0033, Japan}
\author{Naoto Tsuji}
\altaffiliation[Present address: ]{RIKEN Center for Emergent Matter Science (CEMS), Wako, Saitama 351-0198, Japan.}
\affiliation{Department of Physics,  University of Tokyo, Hongo, Tokyo, 113-0033, Japan}
\author{Takashi Oka}
\altaffiliation[Present address: ]{Max-Planck-Institut f\"{u}r Physik komplexer Systeme and Max Planck
Institut f\"{u}r Chemische Physik fester Stoffe, D-01187 Dresden, Germany.}
\affiliation{Department of Applied Physics, University of Tokyo, Hongo, Tokyo 113-8656, Japan}
\author{Hideo Aoki}  
\affiliation{Department of Physics,  University of Tokyo, Hongo, Tokyo, 113-0033, Japan}
\date{\today}

\begin{abstract}
We construct a systematic high-frequency expansion for periodically driven 
quantum systems based on the Brillouin-Wigner (BW) perturbation theory, which generates 
an effective Hamiltonian on the projected zero-photon subspace in the Floquet theory,
reproducing the quasienergies and eigenstates of the original Floquet Hamiltonian
up to desired order in $1/\omega$, with $\omega$ being the frequency of the drive.
The advantage of the BW method is that it is not only efficient in deriving higher-order terms, but even enables us 
to write down the whole infinite series expansion, as compared to the van Vleck degenerate
perturbation theory. The expansion is also free from a spurious dependence on the driving phase, which has been an obstacle
in the Floquet-Magnus expansion.
We apply the BW expansion to various models of noninteracting electrons driven by circularly polarized light.
As the amplitude of the light is increased, the system undergoes a series of Floquet
topological-to-topological phase transitions, whose phase boundary in the high-frequency regime is well explained
by the BW expansion. As the frequency is lowered,
the high-frequency expansion breaks down at some point due to band touching with nonzero-photon sectors, 
where we find numerically even more intricate and richer Floquet topological phases spring out.  
We have then analyzed, with the Floquet dynamical mean-field theory, 
the effects of electron-electron interaction and energy dissipation.  
We have specifically revealed that phase transitions from Floquet-topological to Mott insulators emerge, 
where the phase boundaries can again be captured with the high-frequency expansion.
\end{abstract}

\draft

\maketitle

\section{Introduction} \label{sec:intro}

Periodically driven quantum systems are attracting intensive interests
from a viewpoint of dynamically controlling quantum phases of matter by external drives.
They indeed generate various phenomena, among which are
modulation of the tunneling amplitude of particles~\cite{Dunlap1986,Grossmann1991,Holthaus1992,Kierig2008},
superfluid-to-Mott insulator transitions~\cite{Eckardt2005,Creffield2006,Lignier2007},
magnetic frustrations~\cite{Struck2011},
artificial gauge fields~\cite{Struck2012} in Bose-Einstein condensates,
dynamical band flipping~\cite{Tsuji2009},
and Floquet topological insulators~
\cite{Oka2009,Oka2009d,Karch2010,Karch2011a,Oka2011b,Kitagawa2010,Kitagawa2011,Gu2011,Lindner2011,Rechtsman2013,Wang2013,Katan2013,Grushin14,Rudner13,Kundu2014,Dehghani14,Sentef2014,Jotzu14,Dehghani142,Klinovaja2015}.
The effect of a periodic drive is usually interpreted in terms of a change of the Hamiltonian to an effective static one
derived from the Floquet theory~\cite{Shirley1965,Sambe1973,DittrichBook,Grifoni1998}
and high-frequency expansions~\cite{Rahav2003,Casas2001,Blanes2009,Mananga2011,Goldman2014,Eckardt2015,Bukov2015}.
Specifically, topological phases sensitively depend on the parameters of the mapped Hamiltonian,
as we see in this paper (see also Ref.~\onlinecite{Kundu2014}).
Thus, it is imperative to understand the detailed behavior of the effective Hamiltonian for varied frequency
and amplitude of the periodic drive.

According to the Floquet theory~\cite{Shirley1965,Sambe1973,DittrichBook,Grifoni1998},
a problem of solving a time-dependent Schr\"odinger equation in the Hilbert space $\mathbb H$
is reduced to a \textit{time-independent} eigenvalue problem for the Floquet Hamiltonian
in an extended Hilbert space $\mathbb H\otimes \mathbb T$,
where $\mathbb T$ represents ``multi-photon-dressed'' states. This is a great advantage since 
the effective static Hamiltonian is usually much easier to treat than the original. 
A price to pay for that, however, is that now the dimension of $\mathbb T$ is infinite
by construction. If one wants to interpret periodically driven systems in terms of
a static Hamiltonian defined in the original Hilbert space $\mathbb H$, one has to further reduce 
the degrees of freedom from $\mathbb H\otimes\mathbb T$ to $\mathbb H$.

Such a mapping is indeed possible in the high-frequency regime.
Intuitively, when the frequency is high enough, the system cannot follow the rapid oscillation
of the external drive.
In this case, the system can be effectively regarded as a time-averaged one,
\begin{align}
H_\text{eff}
&\approx
H_0,
\end{align}
where $H_n={\mathscr T}^{-1}\int_0^{\mathscr T} dt\, e^{in\omega t} H(t)$ is the Fourier component of the original Hamiltonian,
with $\mathscr T$ and $\omega=2\pi/\mathscr T$ being the period and frequency of the drive, respectively.
There, the zero-photon-dressed states are decoupled from other dressed states.
One can then include virtual processes for electrons absorbing and emitting one photon
with a second-order perturbation theory to have
\begin{align}
H_\text{eff}
&\approx
H_0+\frac{[H_{-1},H_1]}{\omega}.
\label{eq:1/omega}
\end{align}
The eigenvalues of the effective Hamiltonian determine the non-time-periodic
part of the wave function, while the time-periodic part is obtained by applying the micromotion operator~\cite{Eckardt2005} to the eigenstates (see Sec.~\ref{subsec:otherexpansions}).
When it is applied to the tight-binding model on the honeycomb lattice with a circularly polarized light~\cite{Kitagawa2011},
one can show that the second term in Eq.~(\ref{eq:1/omega}) gives a complex phase to the second-nearest-neighbor hopping.
This enables one to map the problem to Haldane's model of a topological Chern insulator~\cite{Haldane1988},
providing a clear view for the Floquet topological transition.

The relation (\ref{eq:1/omega}) is a prelude to a high-frequency expansion: One can subsequently include
simultaneous $n$-photon absorption and emission processes, which give a correction of order $1/\omega$.
Higher-order perturbations should provide more complicated subleading terms of order $(1/\omega)^n$ ($n\ge 2$), 
but the question then is how one can systematically derive a consistent high-frequency expansion for the effective Hamiltonian.  
A conventional approach is based on the Floquet-Magnus expansion~\cite{Casas2001,Blanes2009,Mananga2011,Feldman1984}.
However, it is known that this method gives an effective Hamiltonian that explicitly depends on the phase of the drive.
Since the effective Hamiltonian should not have such a spurious dependence for time-periodic systems, 
it is desirable to have an expansion that does not suffer from this flaw.  
This point has been overcome by a van Vleck degenerate perturbation theory~\cite{Eckardt2015},
which is free from the phase dependence. While the van Vleck perturbation theory
is a systematic and consistent high-frequency expansion
that correctly reproduces the previous results~\cite{Rahav2003,Goldman2014},
it is not easy to derive higher-order terms in $1/\omega$ within this formalism.

This has motivated us in the present paper to develop an alternative approach to the consistent high-frequency expansion,
based on the Brillouin-Wigner (BW) perturbation theory~\cite{HubacBook}.
The idea is to project the whole Hilbert space $\mathbb H\otimes\mathbb T$ onto the zero-photon subspace
$\mathbb H\otimes\mathbb T_0\sim\mathbb H$. We define the ``wave operator" in the BW formalism that restores the eigenstates
$\in\mathbb H\otimes\mathbb T$ from the projected states $\in\mathbb H\otimes\mathbb T_0$.
There the wave operator can be determined recursively in powers of $1/\omega$. The effective Hamiltonian can then be 
expressed with the wave operator, which enables us to obtain the high-frequency expansion of the effective Hamiltonian in a transparent manner.
While each term in the expansion is generally different from the one in Floquet-Magnus and van Vleck approaches,
the BW formalism also correctly reproduces the quasienergies of the Floquet Hamiltonian up to a desired order in $1/\omega$.  
We reveal the relations among the BW, Floquet-Magnus, and van Vleck perturbative high-frequency expansions.
The advantage of the BW method is that (i) it does not suffer from a problem of the driving-phase dependence,
unlike the Floquet-Magnus expansion, and (ii) it is relatively easy to go to higher-order terms, as 
compared to the van Vleck perturbation theory. We can even formally write 
the whole infinite series. Another asset is that degenerate energy levels in unperturbed systems do not have to be treated separately in the BW perturbation theory. On the other hand, the drawback of the BW method is that the effective Hamiltonian derived in this way
either depends on the quasienergy self-consistently or does not necessarily remain Hermitian if the quasienergy dependence is removed.
While the latter may at first seem to be unphysical, we actually show that 
in the $n$th-order expansion [which is correct up to $(1/\omega)^n$]
the non-Hermitian part only has a contribution to the eigenvalues and right eigenstates of order $(1/\omega)^{n+1}$,
which can be neglected. Thus, our expansion is consistent and physical.

In order to demonstrate how useful the present method is,
we have applied it to various models of noninteracting electrons driven by circularly polarized lights, 
which turn out to generate Floquet topological insulators.
For the most simplified case of a Dirac field in a continuum space, 
an exact solution for the quasienergy spectrum can be 
obtained around the zero momentum, where we have checked that the BW approach correctly reproduces the 
high-frequency expansion of the exact results. We then examine tight-binding models on 
not only the honeycomb lattice, but also Lieb and kagom\'e lattices that accommodate flat bands raising an 
interesting problem about how the flat bands can possibly become topological. 
In the large frequency regime the BW expansion gives a good description of 
the Floquet topological-to-topological transitions among phases having various Chern numbers.  
As one decreases the frequency, the BW expansion breaks down at the point where
the photon-undressed band touches a photon-dressed one.
In this regime we have obtained Chern numbers numerically.
This has led us to a finding that the topological transitions start to take place in a tantalizingly intricate manner 
as we approach zero frequency, with a nearly fractal phase diagram.
For kagom\'e lattice we unravel that 
even the flat band becomes Floquet topological (with band bending).
The final motivation of the present work has to do 
with 
an important question: Do we really have the quantized 
Hall conductivity in the dynamically driven Floquet 
topological insulators?  
In equilibrium at zero temperature, where all the bands below (above) the Fermi energy are fully occupied (empty),
nonzero Chern numbers lead to quantization of the Hall conductivity.  
Since Floquet topological insulators arise in nonequilibrium, the distribution of electrons
should differ from the equilibrium one due to the external drive, which may give rise to derivation
from the quantization of the Hall conductivity.
Thus, we have to consider the effects of energy dissipation~\cite{Dehghani14,Dehghani142,Dehghani2015} which should affect the Hall conductivity.  
Moreover, in real situations many-body 
electron-electron interactions are present, 
and the competition between the topological 
phases and correlated phase such as Mott's insulator 
becomes an intriguing problem.  
To explore these, 
we have studied a dissipative Hubbard model defined on the honeycomb and Lieb lattices by attaching a heat bath 
and switching on the Hubbard interaction, 
with the Floquet dynamical mean-field theory 
(Floquet DMFT)~\cite{Tsuji2008,Aoki2014}. We have identified the condition for the Hall conductivity to approach the quantized value.  
We have also found that there exists a Floquet topological-to-Mott insulator transition (or crossover), as detected from
the double occupancy along with the Hall conductivity.
The crossover line, again, can be well explained by the high-frequency expansions.

The paper is organized as follows. In Sec.~\ref{sec:Heff}, we introduce the Brillouin-Wigner theory
for the high-frequency expansion in the Floquet formalism. We unveil the relations among the BW, Floquet-Magnus,
and van Vleck high-frequency expansions. By applying the BW method to the Dirac field in a circularly polarized light,
we confirm that the BW expansion correctly reproduces the high-frequency expansion of the exact solution for the Dirac field.
In Sec.~\ref{sec:honeycomb}, we study the honeycomb tight-binding model driven by a circularly polarized light.
The Floquet topological-to-topological transitions are analyzed with the BW high-frequency expansion. 
In Sec. \ref{sec:honeycomb}C we investigate the effects of many-body interaction and energy dissipation on the quantization of the Hall conductivity.
We further explore Floquet topological insulators on the Lieb lattice in Sec.~\ref{sec:Lieb} and on kagom\'e lattice in Sec.~\ref{sec:kagome}.
Finally, the conclusion is given in Sec.~\ref{sec:concl}.

\section{High-frequency expansion in the Floquet theory}
\label{sec:Heff}

\subsection{Floquet theory and the effective Hamiltonian}
Let us first overview the Floquet theory for a time-periodic Hamiltonian, $H(t+\mathscr T)=H(t)$, with $\mathscr T=2\pi/\omega$. 
According to the Floquet theory~\cite{Shirley1965,Sambe1973}, solving the time-dependent Schr\"{o}dinger equation, 
\begin{align}
i\partial_{t}|\Psi(t)\rangle=H(t)|\Psi(t)\rangle,
\label{Schroedinger}
\end{align}
in a Hilbert space $\mathbb{H}$ amounts to solving an eigenvalue problem, 
\begin{equation}
\sum_{n\in\mathbb{Z}}(H_{m,n}-m\omega\delta_{m,n})|u_{\alpha}^{n}\rangle=\epsilon_{\alpha}|u_{\alpha}^{m}\rangle,\label{eq:Heff-Floquet}
\end{equation}
in an expanded Hilbert space $\mathbb{H}\otimes\mathbb{T}$,
where the solutions for the original equation are given by 
\begin{equation}
|\Psi_{\alpha}(t)\rangle=\sum_{m\in\mathbb{Z}}e^{-i(\epsilon_{\alpha}+m\omega)t}|u_{\alpha}^{m}\rangle.
\end{equation}
Here $H_{m,n}={\mathscr T}^{-1}\int_0^{\mathscr T} dt \, H(t)e^{i(m-n)\omega t}$ is 
a Fourier transform of the Hamiltonian,
and $ \mathbb{T}=\oplus_m \mathbb T_m$ is the Hilbert space for $\mathscr{T}$-periodic functions spanned by 
$\mathbb T_m=\{e^{-im\omega t}\}$ ($m\in\mathbb{Z}$).

Note that an eigenvalue $\epsilon_\alpha$, called the quasienergy, and the corresponding eigenvector, $|\bm{u}_{\alpha}\rangle = \{ |u_\alpha^m\rangle\}_{m\in\mathbb{Z}}$, have a redundancy: For a solution with a quasienergy $\epsilon_\alpha$ and an eigenvector $|\bm{u}_\alpha \rangle$ for Eq.~(\ref{eq:Heff-Floquet}), 
one finds another solution with a quasienergy $\epsilon_\alpha + n\omega$ 
and an eigenvector $|\bm{u}_{\alpha}^{(n)}\rangle=\{\bm{u}_{\alpha}^{n+m}\}_{m\in\mathbb{Z}}$ for $n\in\mathbb{Z}$.
These solutions are linearly independent of each other in ${\mathbb H}\otimes{\mathbb T}$, but they all give the same solution for the original 
Eq.~(\ref{Schroedinger})
in the original Hilbert space ${\mathbb H}$. 
Due to this, we introduce an identification $|\bm{u}_{\alpha}^{(n)}\rangle\sim|\bm{u}_{\alpha}\rangle$ for $\forall n\in\mathbb{Z}$, with which we naturally have \\
\begin{center}
\renewcommand\arraystretch{1.5}
\begin{tabular}{ccc}
$\mathbb H\otimes\mathbb T/\sim$ & & $\mathbb H$ \\
\rotatebox{90}{$\cong$} & & \rotatebox{90}{$\cong$} \\
$\{|\bm{u}^\alpha\rangle \}_\alpha / \sim$ & $\cong$ & $\{ |\bm{u}^\alpha\rangle \}_{\epsilon_\alpha\in \Lambda}$,
\end{tabular}
\end{center}
where the energy interval $\Lambda=[-\omega/2,\omega/2)$ is the Floquet Brillouin zone (FBZ; a temporal analog 
of the BZ in the Bloch theory).
Since the shift $\epsilon_{\alpha}\to\epsilon_{\alpha}+n\omega$ and $|\bm{u}_{\alpha}\rangle\to|\bm{u}_{\alpha}^{(n)}\rangle$
generate ($\text{dim}\mathbb{T}$) linearly independent solutions in $\mathbb{H}\otimes\mathbb{T}$,
$\{|\bm{u}_{\alpha}\rangle\}_{\epsilon_{\alpha}\in\Lambda}$ spans a $(\dim\mathbb{H})$-dimensional subspace
of $\mathbb{H}\otimes\mathbb{T}$.

Then we can define an effective Hamiltonian $H_{\text{eff}}$ as an
operator acting on $\mathbb{H}$ rather than  on $\mathbb{H}\otimes\mathbb{T}$,
which reproduces all the inequivalent quasienergies, $\{\epsilon_{\alpha}\}_{\alpha}/\sim$
under $\epsilon_{\alpha}\sim\epsilon_{\alpha}+n\omega$. For this
purpose the Brillouin-Wigner theory is suitable: It provides
a general way for constructing the effective Hamiltonian projected
on a model space, which is a smaller Hilbert space one can choose
arbitrarily.

\subsection{Brillouin-Wigner theory for Floquet formalism}

So let us introduce the Brillouin-Wigner theory for the Floquet
eigenvalue problem, Eq.~(\ref{eq:Heff-Floquet}). A brief review of the Brillouin-Wigner theory in general for ordinary quantum mechanics is given in Appendix~\ref{app:BW}. In the following,
we suppress indices for $\mathbb{T}$ and write Eq.~(\ref{eq:Heff-Floquet}) in a matrix form as 
\begin{equation}
(\mathcal{H}-\mathcal{M}\omega)|\bm{u}_{\alpha}\rangle=\epsilon_{\alpha}|\bm{u}_{\alpha}\rangle,\label{eq:floquet-supr}
\end{equation}
where we define $[\mathcal M]_{mn}=m\delta_{mn}$.

In the Brillouin-Wigner theory, 
a key role is played by the wave operator $\mathit{\Omega}$,
which is a mapping from the model space to the original Hilbert space.
It relates the eigenvector $|\bm{u}_{\alpha}\rangle$
with its projection on the model space as 
\begin{equation}
|\bm{u}_{\alpha}\rangle=\mathit{\Omega}\mathcal{P}|\bm{u}_{\alpha}\rangle,
\label{eq:waveop-def}
\end{equation}
where $\mathcal{P}$ is the projection operator to the model space with $\mathcal P^2=\mathcal P$.
Note that $\text{rank}\,\mathit{\Omega}\le\dim\mathcal{P}$, so that we cannot satisfy
the relation Eq.~(\ref{eq:waveop-def}) for $\forall\alpha$,
but we can do so for at most $\dim\mathcal{P}$ eigenvectors. Once we obtain
such an operator, 
\begin{align}
H_{\rm eff}
&=
\mathcal{P}(\mathcal{H}-\mathcal{M}\omega)\mathit{\Omega}\mathcal{P},
\label{eq:Heff general}
\end{align}
gives the effective Hamiltonian, whose eigenstates and eigenenergies
are given by $\mathcal{P}|\bm{u}_{\alpha}\rangle$ and $\epsilon_{\alpha}$, respectively.
Thus, the central problem for constructing the effective Hamiltonian
is how to construct the wave operator systematically. 

In order to obtain the effective Hamiltonian in a static form, we
choose the model space as a diagonal one with respect to $\mathcal{M}$,
i.e., the zero-photon subspace $\mathbb H\otimes\mathbb T_0\sim\mathbb H$.
Due to the redundancy, we can choose it as a photon vacuum, $[\mathcal{P}]_{m,n}=\delta_{mn}\delta_{m0},$
without loss of generality.  With this choice, $\mathcal{P}$ averages out the micromotion (time-periodic oscillation of state vectors) of quasienergy eigenstates, and the wave operator $\mathit{\Omega}$ recovers it.

First we derive a general form of the wave operators for a single
eigenvector $|\bm{u}_{\alpha}\rangle$ in a self-consistent manner.
We define $\mathcal{Q}\equiv 1-\mathcal P$, with $[\mathcal{Q}]_{m,n}=\delta_{mn}(1-\delta_{m0})$,
and operate it to Eq.~(\ref{eq:floquet-supr}) to have
\begin{equation}
\mathcal{Q}|\bm{u}_{\alpha}\rangle=\frac{\mathcal{Q}}{\epsilon_{\alpha}+\mathcal{M}\omega}\mathcal{H}|\bm{u}_{\alpha}\rangle.
\end{equation}
Note that $\mathcal Q$ commutes with $\mathcal M$, 
which implies
\begin{align}
|\bm{u}_{\alpha}\rangle & =\mathcal{P}|\bm{u}_{\alpha}\rangle+\frac{\mathcal{Q}}{\epsilon_{\alpha}+\mathcal{M}\omega}\mathcal{H}|\bm{u}_{\alpha}\rangle\nonumber \\
& =\left(1-\frac{\mathcal{Q}}{\epsilon_{\alpha}+\mathcal{M}\omega}\mathcal{H}\right)^{-1}\mathcal{P}|\bm{u}_{\alpha}\rangle.
\label{eq:projection}
\end{align}
This indicates that the wave operator for a single eigenvector, $\mathit{\Omega}_{\alpha}$,
can be expressed with a common function as $\mathit{\Omega}_{\alpha}=\mathit{\Omega}(\epsilon_{\alpha})$
for $\forall\alpha$, where 
\begin{equation}
\mathit{\Omega}(\epsilon)=\left(1-\frac{\mathcal{Q}}{\epsilon+\mathcal{M}\omega}\mathcal{H}\right)^{-1}\mathcal{P}.
\label{eq:waveop-eps}
\end{equation}
From Eq.~(\ref{eq:Heff general}) and $\mathcal P\mathcal M=0$, we can express the effective Hamiltonian as 
\begin{align}
H_{\text{eff}}(\epsilon)
&=
\mathcal{P}\mathcal{H}\left(1-\frac{\mathcal Q}{\epsilon+\mathcal{M}\omega}\mathcal{H}\right)^{-1}\mathcal{P}.
\end{align}

Since the effective Hamiltonian obtained here contains, unusually but as generally the case in the BW theory, the eigenenergy itself,
we can treat all the eigenvectors simultaneously in a self-consistent manner:
One regards $\epsilon$ as an unknown and diagonalizes $H_{\text{eff}}(\epsilon)$ to obtain eigenenergies, $E_{i}(\epsilon)$ ($i=1,\dots,\dim\mathbb{H})$, and solves a self-consistent set of equations $\epsilon=E_{i}(\epsilon)$. 
In general, $\epsilon=E_{i}(\epsilon)$ is a nonlinear equation of order $\dim\mathbb T$,
so that the total number of solutions becomes $\dim\mathbb H\otimes\mathbb T$,
reproducing all the quasienergies.
An exception is the case of $\mathcal{P}|\bm{u}_{\alpha}\rangle=0$, where one cannot restore the eigenvectors of the Floquet Hamiltonian $\mathcal H-\mathcal M\omega$ 
from those of $H_{\rm eff}$ by using Eq.~(\ref{eq:waveop-def}).

By iterating the first line in Eq.~(\ref{eq:projection}), we can explicitly represent
$\mathit{\Omega}(\epsilon)$ and $H_{\text{eff}}(\epsilon)$
as an infinite series, 
\begin{widetext}
\begin{gather}
[\mathit{\Omega}(\epsilon)]_{m,0}=\delta_{m,0}+(1-\delta_{m,0})\left[\frac{H_{m,0}}{\epsilon+m\omega}+\sum_{N=1}^{\infty}\sum_{n_{1},\dots,n_{N}\neq0}\frac{H_{m,n_{1}}(\prod_{i=1}^{N-1}H_{n_{i},n_{i+1}})H_{n_{N},0}}{(\epsilon+m\omega)\prod_{i=1}^{N}(\epsilon+n_{i}\omega)}\right],
\label{eq:Omega-eps-series}\\
H_{\text{eff}}(\epsilon)=H_{0,0}+\sum_{N=1}^{\infty}\sum_{n_{1},\dots,n_{N}\neq0}\frac{H_{0,n_{1}}(\prod_{i=1}^{N-1}H_{n_{i},n_{i+1}})H_{n_{N},0}}{\prod_{i=1}^{N}(\epsilon+n_{i}\omega)}.
\label{eq:heff-eps-series}
\end{gather}
\end{widetext}
In the above, we can further expand 
\begin{align}
\frac{1}{\epsilon+m\omega}
&=
\frac{1}{m\omega}\sum_{r=0}^\infty \left(-\frac{\epsilon}{m\omega}\right)^r
\quad (m\neq 0),
\\
\frac{1}{\epsilon+n_i\omega}
&=
\frac{1}{n_i\omega}\sum_{r_i=0}^\infty \left(-\frac{\epsilon}{n_i\omega}\right)^{r_i}
\quad (n_i\neq 0).
\end{align}
Hence, Eqs.~(\ref{eq:Omega-eps-series}) and (\ref{eq:heff-eps-series}) can be regarded
as a formal expression for the high-frequency expansion.
This is the first key result in the present paper.

\subsection{High-frequency expansion of\\ 
the effective Hamiltonian}

In the above the wave operator $\mathit\Omega(\epsilon_\alpha)$ (\ref{eq:Omega-eps-series})
depends on $\alpha$, so that it satisfies Eq.~(\ref{eq:waveop-def}) only for a single eigenvector $|\bm u_\alpha\rangle$.
However, the wave operator can be made independent of $\alpha$ as follows.
Using Eq.~(\ref{eq:waveop-eps}), we have
\begin{equation}
\mathit{\Omega}(\epsilon)=\mathcal{P}+\frac{\mathcal{Q}}{\mathcal{M}\omega}\mathcal{H}\mathit{\Omega}(\epsilon)-\frac{\mathcal{Q}}{\mathcal{M}\omega}\mathit{\Omega}(\epsilon)\epsilon,\label{eq:waveop-eq-eps}
\end{equation}
where $(\mathcal Q/\mathcal M\omega)_{mn}=\delta_{mn}/(m\omega)$ for $m,n\neq 0$
and $=0$ otherwise.
When $\mathit\Omega(\epsilon)$ (\ref{eq:waveop-eq-eps}) is operated on $\mathcal P|\bm u_\alpha\rangle$
in Eq.~(\ref{eq:waveop-def}), we can replace 
$\mathit{\Omega}(\epsilon)\epsilon$ on the right-hand side of Eq.~(\ref{eq:waveop-eq-eps})
with $\mathit{\Omega}(\epsilon)\mathcal{P}\mathcal{H}\mathit{\Omega}(\epsilon)$, 
since $\mathcal P\mathcal H\mathit\Omega(\epsilon)\mathcal P=H_{\rm eff}$ (\ref{eq:Heff general}).
This yields an $\epsilon$-independent equation for the wave operator,
\begin{equation}
\mathit{\Omega}=\mathcal{P}+\frac{\mathcal{Q}}{\mathcal{M}\omega}\mathcal{H}\mathit{\Omega}-\frac{\mathcal{Q}}{\mathcal{M}\omega}\mathit{\Omega}\mathcal{P}\mathcal{H}\mathit{\Omega}.\label{eq:waveop-eq}
\end{equation}
This in turn defines an $\epsilon$-independent effective Hamiltonian, which we call  $H_\text{BW}$,
constructed from the solution of Eq.~(\ref{eq:waveop-eq}), $\mathit{\Omega}_\text{BW}$.
One can readily show that for any eigenvector $|v_{\alpha}^{0}\rangle$ of $H_\text{BW}$, $|\bm{v}_\alpha\rangle\equiv\{[\mathit{\Omega}_\text{BW}]_{m,0}|v_{\alpha}^{0}\rangle\}_{m\in\mathbb{Z}}$
must be a solution of Eq.~(\ref{eq:floquet-supr}); i.e., all the
eigenvectors of $H_\text{BW}$ are the projection of the original
eigenvectors onto the model space with the true quasienergies.

However, while $\mathit{\Omega}(\epsilon_{\alpha})$ and $H_{\text{eff}}(\epsilon_{\alpha})$
can be constructed to satisfy Eq.~(\ref{eq:waveop-eq-eps}) for each
($\dim\mathbb{H}\otimes\mathbb{T}$) eigenvector, $H_\text{BW}$
contains only ($\dim\mathbb{H}$) eigenvectors. This implies that
the solution of Eq.~(\ref{eq:waveop-eq}) is not unique. Moreover,
whether $H_\text{BW}$ contains redundant solutions ($\epsilon+n\omega$
for $\epsilon$) is unclear. We can circumvent
this ambiguity by expanding $\mathit{\Omega}_\text{BW}$ in
a $1/\omega$ series.
Then we obtain a recursion relation, 
\begin{align}
\mathit{\Omega}_\text{BW}&=\sum_{N=0}^{\infty}\mathit{\Omega}_\text{BW}^{(N)},
\\
\mathit{\Omega}_\text{BW}^{(0)}&=\mathcal{P},
\\
\mathit{\Omega}_\text{BW}^{(1)}&=\frac{\mathcal{Q}}{\mathcal{M}\omega}\mathcal{H}\mathcal{P},
\\
\mathit{\Omega}_\text{BW}^{(N+1)}&=\frac{\mathcal{Q}}{\mathcal{M}\omega}\mathcal{H}\mathit{\Omega}_\text{BW}^{(N)}
-\sum_{M=1}^{N}\frac{\mathcal{Q}}{\mathcal{M}\omega}\mathit{\Omega}_\text{BW}^{(M)}\mathcal{P}\mathcal{H}\mathit{\Omega}_\text{BW}^{(N-M)}
\notag
\\
&\quad
(N\ge 1),
\label{eq:recursion}
\end{align}
which provides a perturbative solution for Eq.~(\ref{eq:waveop-eq})
without any indefiniteness. For $\omega\rightarrow\infty$, $\mathit{\Omega}_\text{BW}=\mathcal{P}$
and $H_\text{BW}=H_{0,0}$, so that all the solutions belong to
the FBZ and redundant ones are excluded. Therefore, the effective
Hamiltonian constructed perturbatively reproduces all the inequivalent quasienergies.
The effective Hamiltonian is explicitly given as
\begin{widetext}
\begin{subequations}
\begin{align}
H_\text{BW}
&=
\sum_{n=0}^\infty H_\text{BW}^{(n)},
\label{eq:Heff-formula}
\\
H_\text{BW}^{(0)}
&=
H_{0,0},
\label{eq:Heff-formula0}
\\
H_\text{BW}^{(1)}
&=
\sum_{\{n_{i}\}\neq0}
\frac{H_{0,n_{1}}H_{n_{1},0}}{n_{1}\omega},
\label{eq:Heff-formula1}
\\
H_\text{BW}^{(2)}
&=
\sum_{\{n_{i}\}\neq0}
\left(
\frac{H_{0,n_{1}}H_{n_{1},n_{2}}H_{n_{2},0}}{n_{1}n_{2}\omega^{2}}-\frac{H_{0,n_{1}}H_{n_{1},0}H_{0,0}}{n_{1}^{2}\omega^{2}}
\right),
\label{eq:Heff-formula2}
\\
H_\text{BW}^{(3)}
&=
\sum_{\{n_{i}\}\neq0}\bigg[
\frac{H_{0,n_{1}}H_{n_{1},n_{2}}H_{n_{2},n_{3}}H_{n_{3},0}}{n_{1}n_{2}n_{3}\omega^{3}}
+\frac{H_{0,n_{1}}H_{n_{1},0}H_{0,0}H_{0,0}}{n_{1}^{3}\omega^{3}}
-\frac{H_{0,n_{1}}H_{n_{1},0}H_{0,n_{2}}H_{n_{2},0}}{n_{1}^{2}n_{2}\omega^{3}}
\notag
\\
&\quad
-\frac{H_{0,n_{1}}H_{n_{1},n_{2}}H_{n_{2},0}H_{0,0}}{n_{1}n_{2}\omega^{3}}\left(\frac{1}{n_{1}}+\frac{1}{n_{2}}\right)
\bigg],
\label{eq:Heff-formula3}
\\
H_\text{BW}^{(4)}
&=
\sum_{\{n_{i}\}\neq0}\bigg[
\frac{H_{0,n_{1}}H_{n_{1},n_{2}}H_{n_{2},n_{3}}H_{n_{3},n_{4}}H_{n_{4},0}}{n_{1}n_{2}n_{3}n_{4}\omega^{4}}
-\frac{H_{0,n_{1}}H_{n_{1},0}H_{0,0}H_{0,0}H_{0,0}}{n_{1}^{4}\omega^{4}}
-\frac{H_{0,n_{1}}H_{n_{1},0}H_{0,n_{2}}H_{n_{2},n_{3}}H_{n_{3},0}}{n_{1}^{2}n_{2}n_{3}\omega^{4}}
\notag
\\
&\quad
+\frac{H_{0,n_{1}}H_{n_{1},0}H_{0,0}H_{0,n_{2}}H_{n_{2},0}}{n_{1}^{3}n_{2}\omega^{4}}
+\frac{H_{0,n_{1}}H_{n_{1},0}H_{0,n_{2}}H_{n_{2},0}H_{0,0}}{n_{1}^{2}n_{2}\omega^{4}}\left(\frac{1}{n_{1}}+\frac{1}{n_{2}}\right)
\notag
\\
&\quad
-\frac{H_{0,n_{1}}H_{n_{1},n_{2}}H_{n_{2},0}H_{0,n_{3}}H_{n_{3},0}}{n_{1}n_{2}n_{3}\omega^{4}}\left(\frac{1}{n_{1}}+\frac{1}{n_{2}}\right)
-\frac{H_{0,n_{1}}H_{n_{1},n_{2}}H_{n_{2},n_{3}}H_{n_{3},0}H_{0,0}}{n_{1}n_{2}n_{3}\omega^{4}}\left(\frac{1}{n_{1}}+\frac{1}{n_{2}}+\frac{1}{n_{3}}\right)
\notag
\\
&\quad
+\frac{H_{0,n_{1}}H_{n_{1},n_{2}}H_{n_{2},0}H_{0,0}H_{0,0}}{n_{1}n_{2}\omega^{4}}\left(\frac{1}{n_{1}^{2}}+\frac{1}{n_{1}n_{2}}+\frac{1}{n_{2}^{2}}\right)\bigg].
\label{eq:Heff-formula4}
\end{align}\label{eq:Heff-formula0-4}
\end{subequations}
\end{widetext}
We note that the same series can be obtained from Eq.~(\ref{eq:heff-eps-series})
by expanding the denominator with respect to $1/\omega$ and replacing 
$H_{n_{N},0}\epsilon_{\alpha}$ with $H_{n_{N},0}H_{\text{eff}}$ iteratively.

We note here a property of this formalism:
$H_\text{BW}$ reproduces all the quasienergies $\{\epsilon_\alpha\}$ (in FBZ)
and eigenvectors projected on the zero-photon subspace $\{|u^0_{\alpha}\rangle\}$, with $|u_\alpha^0\rangle=\mathcal P|\bm u_\alpha\rangle$ for Eq.~(\ref{eq:floquet-supr}).
By operating $\mathit{\Omega}_\text{BW}$ we can also construct the original
solution for Eq.~(\ref{Schroedinger}) as
\begin{equation}
|\Psi_{\alpha}(t)\rangle=e^{-i\epsilon_{\alpha}t}\Xi(t)|u_{\alpha}^{0}\rangle,\label{eq:Xi}
\end{equation}
where 
\begin{align}
\Xi(t)\equiv\sum_{m\in\mathbb{Z}}e^{-im\omega t}[\mathit{\Omega}_\text{BW}]_{m,0},
\end{align}
whose explicit form in the $1/\omega$ expansion is obtained by replacing $H_\text{BW}^{(0)}$ with 1 and $H_{0,n_1}$ with $e^{-in_1\omega t}$ in Eqs.~(\ref{eq:Heff-formula0-4}):
\begin{equation}
	\Xi(t)=	1+	\sum_{\{n_{i}\}\neq0}\frac{e^{-in_1 \omega t}H_{n_{1},0}}{n_{1}\omega}+\dots.
\end{equation}
Namely, $\Xi(t)$ acts as a counterpart to the so-called micromotion operator (see next section).

On the other hand, $\{|u_{\alpha}^{0}\rangle\}_{\alpha}$
does not, in general, form an orthonormal set, namely $\langle u_{\alpha}^{0}|u_{\beta}^{0}\rangle$
can deviate from $\delta_{\alpha\beta}$.
This implies that the effective Hamiltonian is not necessarily Hermitian,
although all the eigenvalues are real. In terms of projected eigenvectors,
the orthonormal relation between eigenvectors should be represented
as 
\begin{equation}
\langle u_{\alpha}^{0}|[\mathit{\Omega}_\text{BW}^{\dagger}\mathit{\Omega}_\text{BW}]_{0,0}|u_{\beta}^{0}\rangle=\langle\bm{u}_{\alpha}|\bm{u}_{\beta}\rangle=\delta_{\alpha\beta};
\end{equation}
thus, the effective Hamiltonian is Hermitian if $[\mathit{\Omega}_\text{BW}^{\dagger}\mathit{\Omega}_\text{BW}]_{0,0}$
is constant.

\subsection{Relations with other series expansions}\label{subsec:otherexpansions}

Having constructed the Brillouin-Wigner theory for the effective Hamiltonian as a $1/\omega$ series, 
we are now in position to make a comparison with similar series-expansion techniques for obtaining 
effective static Hamiltonians for periodically driven systems, such as the Floquet-Magnus expansion\cite{Casas2001,Mananga2011} and the van Vleck degenerate perturbation theory\cite{Eckardt2015,Bukov2015}. 

First, we recapitulate these expansions. 
They are derived from an ansatz for the time-evolution operator,
\begin{equation}
U(t,t^{\prime})=e^{-i\Lambda(t)}e^{-iF(t-t^{\prime})}e^{i\Lambda(t^{\prime})},\label{eq:timeevol}
\end{equation}
with a time-independent operator $F$ and a time-periodic operator $\Lambda(t)=\Lambda(t+\mathscr T)$.
Let $|\alpha\rangle$ be an eigenstate of $F$ with an eigenvalue $\epsilon_{\alpha}$; then
$|\Psi_{\alpha}(t)\rangle=e^{-i\epsilon_{\alpha}t}e^{-i\Lambda(t)}|\alpha\rangle$ satisfies 
$U(t,t^{\prime})|\Psi_{\alpha}(t^\prime)\rangle=|\Psi_{\alpha}(t)\rangle$, so that $|\Psi_{\alpha}(t)\rangle$ is the eigenstate of the quasienergy.
Namely, the original time-dependent problem is reduced to diagonalization of $F$. Here $F$ behaves as an effective static Hamiltonian and $e^{-i\Lambda(t)}$ as a so-called micromotion operator. 

From an equation of motion, $i\partial_{t}U(t,t^{\prime})=H(t)U(t,t^{\prime})$, $\Lambda(t)$ and $F$ satisfy a differential equation, $e^{i\Lambda(t)}[H(t)-i\partial_{t}]e^{-i\Lambda(t)}=F$, which is rewritten as~\cite{Casas2001,Mananga2011}
\begin{equation}
\frac{d}{dt}\Lambda(t)=\sum_{k=0}^{\infty}\frac{B_{k}}{k!}(-i)^{k}(ad_{\Lambda})^{k}[H(t)+(-1)^{k+1}F],\label{eq:MagnusEoM}
\end{equation}
where $B_{k}$'s are the Bernoulli numbers, while $ad_{\Lambda}X=[\Lambda,X]$
is an adjoint operator. The equation can be solved recursively via
series expansions as
\begin{align}
\Lambda(t)=\sum_{k=0}^{\infty}\Lambda^{(k)}(t),\quad F=\sum_{k=0}^{\infty}F^{(k)},
\end{align}
where one assigns an order 1 for $H(t)$ and the order $k+1$ for $\Lambda^{(k)}(t)$
and $F^{(k)}$. Comparing both sides of Eq.~(\ref{eq:MagnusEoM}) with
each order, one obtains recursive relations for $\Lambda_{k}$ and
$F_{k}$, which can be solved explicitly from lower orders.

Let us note that the boundary condition for the time-evolution operator, $U(t,t)=1$, is automatically satisfied 
when we have Eq.~(\ref{eq:timeevol}), so that there is an arbitrariness in the boundary condition
for $\Lambda(t)$. If $\Lambda(t_{0})=0$ is chosen as the boundary condition, 
one arrives at the Floquet-Magnus expansion,
\begin{widetext}
\begin{subequations}
\begin{align}
F_{\text{FM}}
&=
\sum_{n=0}^\infty F_{\rm FM}^{(n)},
\\
F_{\text{FM}}^{(0)}&= H_{0},\\
F_{\text{FM}}^{(1)}&= \sum_{m\neq0}\frac{[H_{-m},H_{m}]}{2m\omega}+\sum_{m\neq0}\frac{[H_{m},H_{0}]}{m\omega}e^{-im\omega t_{0}},\\
F_{\text{FM}}^{(2)}&= \sum_{m\neq0}\frac{[[H_{-m},H_{0}],H_{m}]}{2m^{2}\omega^{2}}+\sum_{m\neq0}\sum_{n\neq0,m}\frac{[[H_{-m},H_{m-n}],H_{n}]}{3mn\omega^{2}}-\sum_{m\neq0}\frac{[[H_{m},H_{0}],H_{0}]}{m^{2}\omega^{2}}e^{-im\omega t_{0}}\notag\\
 &\quad -\sum_{m,n\neq0}\frac{[[H_{m},H_{n}],H_{-n}]}{3mn\omega^{2}}e^{-im\omega t_{0}}+\sum_{m,n\neq0}\frac{[[H_{n},H_{-n}],H_{m}]}{3mn\omega^{2}}e^{-im\omega t_{0}}-\sum_{m\neq0}\sum_{n\neq0,m}\frac{[[H_{n},H_{m-n}],H_{0}]}{2mn\omega^{2}}e^{-im\omega t_{0}}\notag\\
 &\quad +\sum_{m,n\neq0}\frac{[[H_{m},H_{n}],H_{0}]}{2mn\omega^{2}}e^{-i(m+n)\omega t_{0}}-\sum_{m,n\neq0}\frac{[[H_{m},H_{0}],H_{n}]}{2mn\omega^{2}}e^{-i(m+n)\omega t_{0}},
\end{align}
\end{subequations}
\begin{subequations}
	\begin{align}
	\Lambda_{\text{FM}}(t)= & \sum_{n=0}^{\infty}\Lambda_{\text{FM}}^{(n)}(t),\\
	i\Lambda_{\text{FM}}^{(1)}(t)= & \sum_{m\neq0}\frac{H_{m}}{m\omega}e^{-im\omega t_{0}}-\sum_{m\neq0}\frac{H_{m}}{m\omega}e^{-im\omega t},\\
	i\Lambda_{\text{FM}}^{(2)}(t)= & \sum_{m\neq0}\sum_{n\neq0}\frac{[H_{n},H_{m}]}{2mn\omega^{2}}e^{-i(m+n)\omega t_{0}}-\sum_{m\neq0}\sum_{n\neq0,m}\frac{[H_{n},H_{m-n}]}{2mn\omega^{2}}e^{-im\omega t_{0}}-\sum_{m\neq0}\frac{[H_{m},H_{0}]}{m^{2}\omega^{2}}e^{-im\omega t_{0}}\nonumber \\
	& -\sum_{m\neq0}\sum_{n\neq0}\frac{[H_{n},H_{m}]}{2mn\omega^{2}}e^{-in\omega t_{0}}e^{-im\omega t}+\sum_{m\neq0}\sum_{n\neq0,m}\frac{[H_{n},H_{m-n}]}{2mn\omega^{2}}e^{-im\omega t}+\sum_{m\neq0}\frac{[H_{m},H_{0}]}{m^{2}\omega^{2}}e^{-im\omega t}.
	\end{align}
\end{subequations}
One can see that $F_{\rm FM}^{(n)}$ generally depends on $t_0$ or, equivalently, the phase of the periodic drive.
If one chooses $\int_{0}^{\mathscr T}dt\Lambda(t)=0$ as the boundary condition,
one can remove the $t_0$ dependence. In fact,
it yields the high-frequency expansion~\cite{Mananga2011}
as derived from the van Vleck degenerate perturbation theory~\cite{Eckardt2015,Bukov2015},
\begin{subequations}
\begin{align}
F_\text{vV}=&
\sum_{n=0}^\infty F_\text{vV}^{(n)},
\label{eq:Heff-vV}
\\
F_\text{vV}^{(0)}
=&H_{0},
\label{eq:Heff-vV0}
\\
F_\text{vV}^{(1)}
=&
\sum_{m\neq0}\frac{[H_{-m},H_{m}]}{2m\omega},
\label{eq:Heff-vV1}
\\
F_\text{vV}^{(2)}
=&
\sum_{m\neq0}\frac{[[H_{-m},H_{0}],H_{m}]}{2m^{2}\omega^{2}}
+\sum_{m\neq0}\sum_{n\neq0,m}\frac{[[H_{-m},H_{m-n}],H_{n}]}{3mn\omega^{2}},
\label{eq:Heff-vV2}
\\
F_\text{vV}^{(3)}
=&
\sum_{m\neq0}\frac{[[[H_{-m},H_{0}],H_{0}],H_{m}]}{2m^{3}\omega^{3}}
+\sum_{m\neq0}\sum_{n\neq0,m}\frac{[[[H_{-m},H_{0}],H_{m-n}],H_{n}]}{3m^{2}n\omega^{3}}
\notag
\\
&+\sum_{m\neq0}\sum_{n\neq0,m}\frac{[[[H_{-m},H_{m-n}],H_{0}],H_{n}]}{4mn^{2}\omega^{3}}
-\sum_{m,n\neq0}\frac{[[[H_{-m},H_{m}],H_{-n}],H_{n}]}{12mn^{2}\omega^{3}}
\notag
\\
&+\sum_{m\neq0}\sum_{n\neq0,m}\frac{[[H_{-m},H_{0}],[H_{m-n},H_{n}]]}{12m^{2}n\omega^{3}}
+\sum_{m,n\neq0}\sum_{l\neq0,m,n}\frac{[[[H_{-m},H_{m-l}],H_{l-n}],H_{n}]}{6lmn\omega^{3}}
\notag
\\
&+\sum_{m,n\neq0}\sum_{l\neq0,m-n}\frac{[[[H_{-m},H_{m-n-l}],H_{l}],H_{n}]}{24lmn\omega^{3}}
+\sum_{m,n\neq0}\sum_{l\neq0,m,n}\frac{[[H_{-m},H_{m-l}],[H_{l-n},H_{n}]]}{24lmn\omega^{3}},
\label{eq:Heff-vV3}
\end{align}
\end{subequations}
\begin{subequations}
	\begin{align}
	\Lambda_{\text{vV}}(t)= & \sum_{n=0}^{\infty}\Lambda_{\text{vV}}^{(n)}(t),\\
	i\Lambda_{\text{vV}}^{(1)}(t)= & -\sum_{m\neq0}\frac{H_{m}}{2m\omega}e^{-im\omega t},\\
	i\Lambda_{\text{vV}}^{(2)}(t)= & \sum_{m\neq0}\sum_{n\neq0,m}\frac{[H_{n},H_{m-n}]}{2mn\omega^{2}}e^{-im\omega t}+\sum_{m\neq0}\frac{[H_{m},H_{0}]}{m^{2}\omega^{2}}e^{-im\omega t},\\
	i\Lambda_{\text{vV}}^{(3)}(t)= & -\sum_{m\neq0}\frac{[[H_{m},H_{0}],H_{0}]}{m^{3}\omega^{3}}e^{-im\omega t}+\sum_{m\neq0}\sum_{n\neq0}\frac{[H_{m},[H_{-n},H_{n}]]}{4m^{2}n\omega^{3}}e^{-im\omega t}\nonumber \\
	& -\sum_{m\neq0}\sum_{n\neq0,m}\frac{[[H_{n},H_{0}],H_{m-n}]}{2mn^{2}\omega^{3}}e^{-im\omega t}-\sum_{m\neq0}\sum_{n\neq0,m}\frac{[[H_{n},H_{m-n}],H_{0}]}{2m^{2}n\omega^{3}}e^{-im\omega t}\nonumber\\&-\sum_{m\neq0}\sum_{n\neq0}\sum_{l\neq0,n,m}\frac{[[H_{n},H_{l-n}],H_{m-l}]}{4mnl\omega^{3}}e^{-im\omega t}-\sum_{m\neq0}\sum_{n\neq0}\sum_{l\neq0,m-n}\frac{[H_{n},[H_{l},H_{m-n-l}]]}{12mnl\omega^{3}}e^{-im\omega t}.
	\end{align}
\end{subequations}
\end{widetext}
One can see that there appear many more terms (especially if the commutators are expanded) 
at higher order in the Floquet-Magnus and van Vleck expansions
than in the BW expansion. In addition, deriving higher-order terms in the Floquet-Magnus
and van Vleck expansions requires tedious calculations, whereas higher-order terms 
can be efficiently computed in the BW expansion by using the simple recursion relation (\ref{eq:recursion}).
These are the advantages of the BW method over the others.

The Floquet-Magnus and van Vleck expansions are related with each other by a unitary transformation: By comparing the form of $U(t_{0}+\mathscr T,t_{0})$ in two expressions, one can confirm that 
\begin{equation}
F_{\text{FM}}=e^{-i\Lambda_{\text{vV}}(t_{0})}F_{\text{vV}}e^{i\Lambda_{\text{vV}}(t_{0})},
\end{equation}
which also leads to a relation between $\Lambda(t)$'s, $e^{-i\Lambda_{\text{FM}}(t)}=e^{-i\Lambda_{\text{vV}}(t)}e^{i\Lambda_{\text{vV}}(t_{0})}$.

Now let us compare these expansions with the present BW formalism. Here we start with discussing how the time-evolution operator is expressed in terms of the Brillouin-Wigner theory with $H_\text{BW}$ and $\Xi(t)$. 
To this end, we first introduce $\Xi(t)^{-1}$: This operator has an explicit expression,
\begin{equation}
\Xi(t)^{-1}=e^{-iH_\text{BW}t}\sum_{\alpha}|u_{\alpha}^{0}\rangle\langle\Psi_{\alpha}(t)|,
\end{equation}
which is indeed the inverse of $\Xi(t)$, as one can confirm from Eq.~(\ref{eq:Xi}).
We note that $\mathit{\Omega}_\text{BW}$, which operates on the model space with $\text{rank}\,\mathit{\Omega}_\text{BW}=\dim\mathbb{H}<\dim\mathbb{H}\otimes\mathbb{T}$, does not have an inverse, while  $\Xi(t)$, operating on the original space with $\text{rank}\,\Xi(t)=\dim\mathbb{H}$, has one. 

\begin{figure}[t]
\begin{center}
\includegraphics[width=8cm]{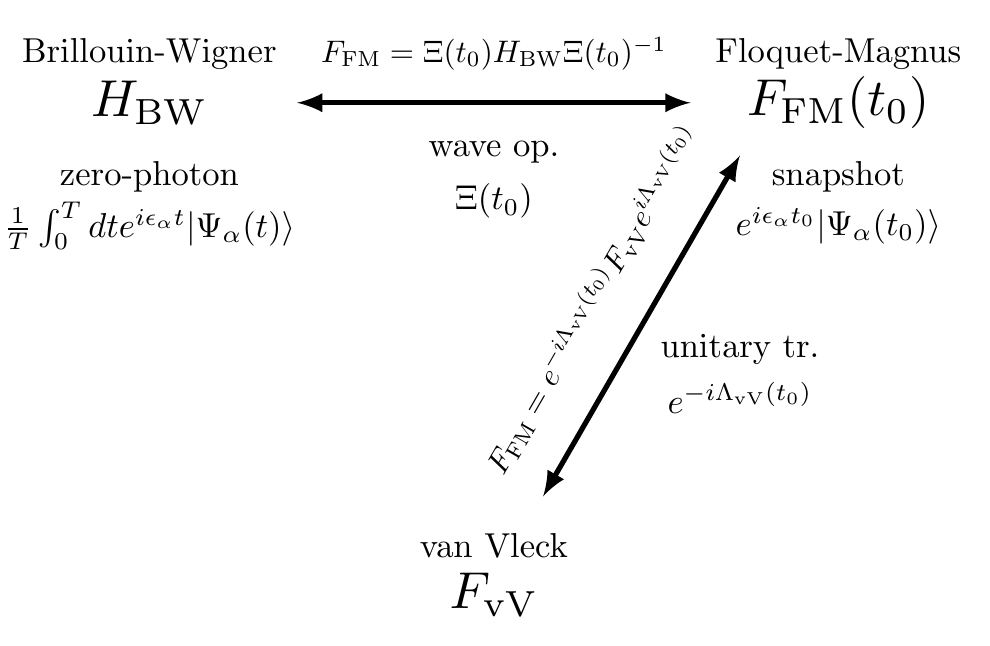}
\caption{
Relations among the Brillouin-Wigner, Floquet-Magnus, and van Vleck perturbative high-frequency expansions.
}
\label{relation}
\end{center}
\end{figure}

The time-evolution operator is then expressed as
\begin{equation}
U(t,t^{\prime})=\sum_{\alpha}|\Psi_{\alpha}(t)\rangle\langle\Psi_{\alpha}(t^{\prime})|=\Xi(t)e^{-iH_\text{BW}(t-t^{\prime})}\Xi(t^{\prime})^{-1}.
\end{equation}
Then we can again compare the form of $U(t_{0}+\mathscr T,t_{0})$ to yield a relation of the effective Hamiltonian $H_\text{BW}$ with that of the Floquet-Magnus expansion,
\begin{equation}
F_{\text{FM}}=\Xi(t_{0})H_\text{BW}\Xi(t_{0})^{-1}.
\end{equation}
In general $\Xi(t_{0})$ is not unitary [$\Xi^{\dagger}(t_{0})\neq\Xi(t_{0})^{-1}$], which corresponds to 
$H_\text{BW}$ being not necessarily Hermitian. $\Xi(t)^{-1}$ can be explicitly computed as a $1/\omega$ series from its equation of motion, 
\begin{equation}
-i\partial_{t}\Xi(t)^{-1}=\Xi(t)^{-1}H(t)-H_\text{BW}\Xi(t)^{-1},
\end{equation}
and related with $\Lambda(t)$ via
\begin{equation}
\Xi(t)^{-1}=\frac{1}{\mathscr T}\int_{0}^{\mathscr T}dt^{\prime}e^{-i\Lambda(t^{\prime})}e^{i\Lambda(t)}.
\end{equation}
We summarize the relations among the Brillouin-Wigner, Floquet-Magnus, and van Vleck expansions in Fig.~\ref{relation}.

As we have seen, while several versions of systematic expansions for the effective static Hamiltonian can be considered, these expression are related by certain transformations, and their eigenvalue spectra are equivalent to each other. However, note that the infinite series is, in practice, truncated at some finite order, which makes the equivalence up to the truncation order.

\subsection{Remarks on application to noninteracting many-particle systems}
While the formalism we have presented is applicable to general periodic systems, we are in position to make some remarks on the application of the method to noninteracting but many-particle systems.  
Let us consider a Hamiltonian in a second-quantized form,
\begin{equation}
\mathcal{H}(t)=\sum_{i,j}J_{i,j}(t)c_{i}^{\dagger}c_{j},
\label{eq:Heff-TightBinding}
\end{equation}
where $J_{i,j}(t)=J_{j,i}^{\ast}(t)$ is a one-body element with a temporal periodicity, $J_{i,j}(t+\mathscr T)=J_{i,j}(t)$. 
Floquet components of the Hamiltonian (\ref{eq:Heff-TightBinding}) read
\begin{align}
	\mathcal{H}_{m,n} & =\sum_{i,j}J_{i,j}^{m-n}c_{i}^{\dagger}c_{j},\notag\\
	J_{i,j}^{m-n} & =\int_{0}^{\mathscr T}\frac{dt}{\mathscr T}J_{i,j}(t)e^{i(m-n)\omega t}=(J_{j,i}^{n-m})^{\ast}.
	\label{eq:Heff-Jmn}
\end{align}
By substituting these into Eqs.~(\ref{eq:Heff-formula})--(\ref{eq:Heff-formula2}) and reorganizing the terms into the normal-ordered form, one obtains the effective Hamiltonian for Eq.~(\ref{eq:Heff-TightBinding}) as
\begin{multline}
\mathcal{H}_\text{BW}=\sum_{i,j} \left( J^{(0)}_{i,j} + J^{(1)}_{i,j} + J^{(2)}_{i,j} \right) c^\dagger_i c_j\\
+\sum_{i,j,k,l}V^{(2)}_{i,j,k,l}c_{i}^{\dagger}c_{j}^{\dagger}c_{k}c_{l} + J\mathcal{O}\left( \frac{J^3}{\omega^3}\right),
\label{eq:Heff-EffectiveTightBinding-full}
\end{multline}
where
\begin{subequations}
	\begin{align}
	J_{i,j}^{(0)} &= J^{0}_{i,j}, \\
	J_{i,j}^{(1)} &= \sum_{n\neq0} \sum_k \frac{ J_{i,k}^{-n} J_{k,j}^{n} }{n\omega}, \label{eq:Heff-J1}\\
	J_{i,j}^{(2)} &= \sum_{n\neq 0} \sum_{k,l} \left(\sum_{m\neq0} \frac{J_{i,k}^{-m} J_{k,l}^{m-n} J_{l,j}^{n}}{mn\omega^2}
	- \frac{J_{i,k}^{-n} J_{k,l}^{n}J_{l,j}^{0}}{n^2\omega^2}\right), \label{eq:Heff-J2}\\
	V^{(2)}_{i,j,k,l}&=\sum_{n\neq0}\sum_{m}\frac{(J_{i,m}^{0}J_{m,l}^{-n}-J_{i,m}^{-n}J_{m,l}^{0})J_{j,k}^{n}}{n^{2}\omega^{2}}.  \label{eq:Heff-V2}
	\end{align}\label{eq:Heff-EffectiveTightBinding}
\end{subequations}
We can notice that a two-particle term, Eq.~(\ref{eq:Heff-V2}), appears in the effective Hamiltonian (\ref{eq:Heff-EffectiveTightBinding-full}) even though the original Hamiltonian (\ref{eq:Heff-TightBinding}) is noninteracting.  
Terms involving more than two particles also appear in higher orders.

Such a property is absent in other (Floquet-Magnus and van Vleck) expansions: 
These expansions, for noninteracting systems, are composed of commutators of one-body operators, which inductively guarantees the effective Hamiltonian to be noninteracting. On the other hand, if we go over to many-body interacting systems, these expansions can also generate many-body terms, e.g., three-body interactions from $[[[H_{-m},H_{0}],H_{0}],H_{m}]/(2m^{3}\omega^{3})$ in $F_{\text{vV}}^{(3)}$ [Eq.~(\ref{eq:Heff-vV3})], due to two-body terms in $H_0$.

The emergent many-particle terms such as Eq.~(\ref{eq:Heff-V2}) in noninteracting problems comes from the fact that the projection on the zero-photon space is imposed on the entire many-particle states rather than on each orbital. Let us see this in detail.  
For the original time-dependent problem, once we obtain one-particle eigenstates of quasienergy, $|\Psi_{\alpha}(t)\rangle$, we can easily construct many-particle eigenstates as single Slater determinants for fermions (permanents for bosons). For instance, two-particle eigenstates are given as
\begin{equation}
|\Psi_{\alpha}(t)\rangle|\Psi_{\beta}(t)\rangle\mp|\Psi_{\beta}(t)\rangle|\Psi_{\alpha}(t)\rangle \label{eq:two-particle}
\end{equation} 
with $-(+)$ for fermions (bosons).

Such a relation between one-particle and many-particle eigenstates does \textit{not} hold for the zero-photon projections: 
The one-particle eigenstates of Eq.~(\ref{eq:Heff-EffectiveTightBinding-full}) are given as $|u_\alpha^0\rangle=\sum_i u_{\alpha,i}^0c_i^\dagger|0\rangle$, which satisfy a Schr\"odinger equation, 
\begin{equation}
\sum_{j} \left[ J^{(0)}_{i,j} + J^{(1)}_{i,j} + J^{(2)}_{i,j} +J\mathcal{O}\left( \frac{J^3}{\omega^3}\right) \right] u_{\alpha,j}^0=\epsilon_\alpha u_{\alpha,i}^0. \label{eq:onebody-sch-eq-zp}
\end{equation}
We can construct the Slater determinants (permanents) as $|u_\alpha^0\rangle|u_\beta^0\rangle\mp|u_\beta^0\rangle|u_\alpha^0\rangle$, but these are not the eigenstates of Eq.~(\ref{eq:Heff-EffectiveTightBinding-full}) in general: The eigenstates should be the zero-photon projections of Eq.~(\ref{eq:two-particle}), namely 
\begin{equation}
\sum_m\left(|u_\alpha^m\rangle|u_\beta^{-m}\rangle\mp|u_\beta^m\rangle|u_\alpha^{-m}\rangle\right)\label{eq:two-particle-zp},
\end{equation} with $|u_\alpha^m\rangle$ being the $m$-photon projections of $|\Psi_{\alpha}(t)\rangle$.
This discrepancy is compensated by the emergent many-particle terms.
In other words, the Hamiltonian with eigenstates that cannot be represented as single Slater determinants should have many-particle terms, which can indeed hold for Eq.~(\ref{eq:two-particle-zp}) in general.

However, for noninteracting systems, once we obtain the zero-photon projection of one-particle eigenstates $|u_{\alpha}^{0}\rangle$, we can restore original eigenstates $|\Psi_{\alpha}(t)\rangle$ by the wave operator $\Xi(t)$ [see Eq.~(\ref{eq:Xi})], so that we can readily construct Eq.~(\ref{eq:two-particle}). 
Namely, without loss of accuracy, one can restrict state vectors to one-particle ones, where the many-particle terms have no influence [see Eq.~(\ref{eq:onebody-sch-eq-zp})].
Hence, although we use the second-quantized form in the following for convenience, we do not show many-particle terms explicitly for noninteracting problems.
We note that for interacting systems the emergent many-particle terms should be included explicitly, but such terms also emerge in other expansions in these cases.

\subsection{An example: Dirac field in a circularly polarized light}\label{subsec:dirac}

To see how the expansions work, let us first take a simplest example, the Dirac field driven by a circularly polarized light~\cite{Oka2009},
whose quasienergy spectrum around $\Gamma$ point ($\bm{k}=0$) can be derived analytically.
We take the Dirac Hamiltonian,
\begin{equation}
H_{\bm{k}}^{\text{Dirac}}(t)=\begin{pmatrix}0 & k_{x}-ik_{y}+Ae^{-i\omega t}\\ k_{x}+ik_{y}+Ae^{i\omega t} & 0 \end{pmatrix},\label{eq:Heff-Dirac}
\end{equation}
where $\bm{k}=(k_{x},k_{y})$ is the momentum and $\bm{A}(t)=(A\cos\omega t,A\sin\omega t)$ is the vector potential, for which
we take a circularly polarized light (CPL).  
This makes the system topological, with a 
topological gap $\Delta$, as shown in Ref.~\onlinecite{Oka2009}. 
The quasienergy spectrum is expanded as
\begin{align}
\epsilon^2=\left(\frac{\Delta}{2}\right)^2+\frac{\Delta\kappa}{2} k^2+O(k^4),\label{eq:Dirac-esquare} 
\end{align}
where $\kappa$ determines the curvature of the band dispersion at $\bm{k}=0$.
These quantities are analytically given as
\begin{subequations}
\begin{align}
\Delta
& =\sqrt{\omega^2+4A^2} - \omega, \label{eq:Heff-Dirac_gap} 
\\
\kappa
&
=\frac{(\omega^{2}+A^{2})}{A^{2}\sqrt{\omega^{2}+4A^{2}}}.
\label{eq:Heff-Dirac_gamma}
\end{align} \label{eq:Heff-Dirac_gap_kappa}
\end{subequations}
The expression for $\kappa$, Eq.~(\ref{eq:Heff-Dirac_gamma}), is obtained here with the $\bm{k}\cdot\bm{p}$-perturbation theory around $\bm{k}=0$
[see Appendix \ref{app:Dirac}, Eqs.~(\ref{eq:Dirac_LargeW_gap}) and (\ref{eq:Dirac_LargeW_kappa})]. 
Equation (\ref{eq:Dirac-esquare}) can be expanded in $A/\omega$ as
\begin{align}
\epsilon^{2}
&=
\left(\frac{A^{4}}{\omega^{2}}-\frac{2A^{6}}{\omega^{4}}+\frac{5A^{8}}{\omega^{6}}+\dots\right)
\notag
\\
&\quad
+\left(1-\frac{2A^{2}}{\omega^{2}}+\frac{7A^{4}}{\omega^{4}}-\frac{25A^{6}}{\omega^{6}}+\dots\right)k^{2}+\mathcal{O}(k^{4}).\label{eq:Dirac-esquare-expanded}
\end{align}

In the Floquet matrix form for the Hamiltonian (\ref{eq:Heff-Dirac}), 
the elements are
\begin{align}
H_{0}=\begin{pmatrix}0 & k_{-}\\ k_{+} & 0 \end{pmatrix},\,
H_{1}=\begin{pmatrix}0 & A\\ 0 & 0 \end{pmatrix},\,
H_{-1}=\begin{pmatrix}0 & 0\\ A & 0 \end{pmatrix},
\end{align}
with $k_{\pm}\equiv k_{x}\pm ik_{y}$, and zero otherwise.
In this example, the effective Hamiltonian derived by the high-frequency
expansion with Brillouin-Wigner method up to 
the fourth order in $1/\omega$, $H_{\text{BW}}^{(\le4)}=\sum_{i=0}^4H_{\text{BW}}^{(i)}$, has a simple form, 
\begin{subequations}
\begin{align}
H_{\text{BW}}^{(0)} & =\begin{pmatrix}0 & k_{-}\\ k_{+} & 0 \end{pmatrix},\\
H_{\text{BW}}^{(1)} & =-\frac{A^{2}}{\omega}\begin{pmatrix}1 & 0\\ 0 & -1 \end{pmatrix},\\
H_{\text{BW}}^{(2)} & =-\frac{A^{2}}{\omega^{2}}\begin{pmatrix}0 & k_{-}\\ k_{+} & 0 \end{pmatrix},\\
H_{\text{BW}}^{(3)} & =\frac{A^{2}(A^{2}-2k^{2})}{\omega^{3}}\begin{pmatrix}1 & 0\\ 0 & -1 \end{pmatrix},\\
H_{\text{BW}}^{(4)} & =\frac{A^{2}(A^{2}-4k^{2})}{\omega^{4}}\begin{pmatrix}0 & k_{-}\\ k_{+} & 0 \end{pmatrix}.
\end{align}\label{eq:Heff-Dirac-expanded}
\end{subequations}
The quasienergies can be obtained by calculating the eigenvalue as 
\begin{align}
(\epsilon_{\text{BW}}^{(\le4)})^{2}
&=
\left(\frac{A^{4}}{\omega^{2}}-\frac{2A^{6}}{\omega^{4}}+\frac{A^{8}}{\omega^{6}}\right)
\notag
\\
&\quad
+\left(1-\frac{2A^{2}}{\omega^{2}}+\frac{7A^{4}}{\omega^{4}}-\frac{6A^{6}}{\omega^{6}}+\dots\right)k^{2}+\mathcal{O}(k^{4}),
\end{align}
which agree with the exact expression, Eqs.~(\ref{eq:Dirac-esquare-expanded}), up to the truncation order.

We note that the effective Hamiltonian derived by the van Vleck method [Eqs.~(\ref{eq:Heff-vV})--(\ref{eq:Heff-vV3})] has the same form as Eqs.~(\ref{eq:Heff-Dirac-expanded}), while that in the Floquet-Magnus expansion has $t_0$ dependence and results in
\begin{align}
(\epsilon_{\text{FM}}^{(\le4)})^{2}
&=\left(\frac{A^{4}}{\omega^{2}}-\frac{2A^{6}}{\omega^{4}}+\frac{15A^{8}}{\omega^{6}}+\dots\right)
\notag
\\
&\quad
+\text{Re}\left(\frac{40A^{7}}{\omega^{6}}\frac{k_{+}}{k}e^{-it_{0}}+\dots\right)k
\notag
\\
&\quad
+\left(1-\frac{2A^{2}}{\omega^{2}}+\frac{7A^{4}}{\omega^{4}}-\frac{40A^{6}}{\omega^{6}}+\dots\right)k^{2}
\notag
\\
&\quad
+\text{Re}\left(-\frac{90A^{5}}{\omega^{6}}\frac{k_{+}}{k}e^{-it_{0}}+\dots\right)k^{3}+\mathcal{O}(k^{4}),
\end{align}
which also agree with the exact expression up to the truncation order, although $t_0$-dependent terms appear in a higher order.

As we have confirmed, the expansions are consistent with the exact results. 
Compared to the Floquet-Magnus and van Vleck methods, the terms appearing in higher orders
of the Brillouin-Wigner expansion have simpler form and their number is smaller, 
so that one can obtain them easily. 
Later in the present paper, 
it will turn out that the high-frequency expansion
plays an essential role in understanding the mechanism of Floquet topological-to-topological transitions that takes place in lattice systems.  
In a broader context, we emphasize that our formalism is general enough to be applicable to any quantum systems
obeying the time-periodic Schr\"odinger equation. 
For further examples of the appearance of non-Hermitian and many-particle terms, see also Appendix~\ref{app:drawback}.

\section{Honeycomb lattice in a circularly polarized light}
\label{sec:honeycomb}

Now let us explore the honeycomb tight-binding model in circularly polarized light (CPL) in
a wide range of amplitude and frequency, which we show
exhibits a variety of topological transitions.
The Hamiltonian is
\begin{equation}
\mathcal{H}^\text{honeycomb}(t) = \sum_{i,j}^\text{NN} J_{i,j}(t) c^\dagger_i c_j,
\label{eq:honeycomb-Ht}
\end{equation}
where the sum is taken over the nearest-neighbor (NN) honeycomb sites. 
We introduce a time-dependent spatially uniform electric field ${\bm{E}}(t)=-\partial_{t}{\bm{A}}(t)$
with $\bm{A}(t) = {}^t(A\cos\omega t, A\sin\omega t)$, which is plugged into the hopping 
as the Peierls phase,
\begin{align}
J_{i,j}(t)
&=
J_{i,j}\exp\left(-i\int_{{\bm{R}}_{j}}^{{\bm{R}}_{i}}{\bm{A}}(t)\cdot d{\bm{r}}\right),
\label{Peierls}
\end{align}
where $\bm R_i$ is the position of site $i$.
Explicitly, we have 
\begin{equation}
J_{i,j}(t) = J e^{-i \bm{A}(t) \cdot \bm{e}_l } = J e^{-iA\sin(\omega t - 2\pi l/3)},
\label{eq:honeycomb-Jt}
\end{equation}
where the NN unit vector $\bm{e}_l = {}^t (\cos\phi_l, \sin\phi_l)$ with $\phi_l= \pi/2 + 2\pi l/3$
corresponds to the hopping from site $j$ to $i$.
We use $J$ as the unit of energy and set $J=1$ hereafter. We also note that $A$ is a dimensionless parameter in this notation.
Throughout this section we consider the half-filled case (i.e., one particle per site).

Performing the Fourier transform with respect to site indices, one can express the Hamiltonian in the momentum space as
\begin{subequations}
\begin{align}
\mathcal{H} (t) &= \sum_{\bm{k}\sigma} \psi^\dagger_{\bm{k}\sigma} H(\bm{k}+\bm{A}(t)) \psi_{\bm{k}\sigma},\\
H(\bm{k}) &=
\sum_{l=0}^2 \begin{pmatrix}
0 & J e^{-i \bm{k} \cdot \bm{e}_l} \\
J e^{i \bm{k} \cdot \bm{e}_l} & 0
\end{pmatrix},
\end{align}
\end{subequations}
where $\psi_{\bm{k}\sigma}^\dagger = (c^\dagger_{A\bm{k}\sigma}, c^\dagger_{B\bm{k}\sigma})$ with $A$ and $B$ the sublattice indices.

\subsection{Effective Hamiltonian in\\ 
the Brillouin-Wigner expansion}

\begin{figure}
\begin{center}
\includegraphics[width=0.6\hsize]{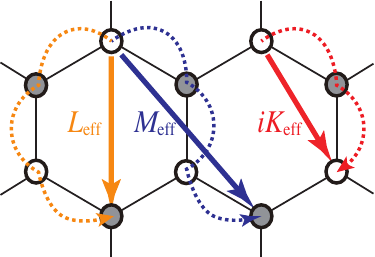}
\caption{
Photo-assisted two-step and three-step hopping processes (dotted lines) 
in the effective Hamiltonian (\ref{eq:honeycomb-Heff}) derived from the BW expansion
for the honeycomb lattice driven by a circularly polarized light.}
\label{fig:hopping_diagram}
\end{center}
\end{figure}

We apply the Brillouin-Wigner expansion to the honeycomb 
tight-binding model (\ref{eq:honeycomb-Ht}) 
by plugging Eq.~(\ref{eq:honeycomb-Jt}) to Eq.~(\ref{eq:Heff-Jmn}) to have Floquet matrices, 
\begin{align}
J^{m,n}_{i,j}
&= J \mathcal{J}_{m-n}(A) e^{i2\pi (m-n) l / 3} ,
\end{align}
where $\mathcal{J}_n(A)$ is the $n$th Bessel function of the first kind.
We can then use the formula (\ref{eq:Heff-EffectiveTightBinding}) 
to obtain an effective Hamiltonian up to $J \mathcal{O}\left({J^3}/{\omega^3}\right)$ as
\begin{multline}
\mathcal{H}_\text{BW}^{\rm honeycomb} = \sum_{i,j}^\text{NN} J_\text{eff} c^\dagger_i c_j
+\sum_{i, j}^\text{NNN}i \tau_{i,j} K_\text{eff} c^\dagger_i c_j
\\
+\sum_{i,j}^\text{L-path}L_\text{eff} \; c^\dagger_i c_j
+\sum_{i,j}^\text{M-path}M_\text{eff} \; c^\dagger_i c_j 
+J \mathcal{O}\left(\frac{J^3}{\omega^3}\right).
\label{eq:honeycomb-Heff}
\end{multline}
Here $\tau_{i,j}=\pm 1$ in the next-nearest-neighbor (NNN) hopping term represents the chirality of the hopping path
from $j$ to $i$, where $+$~($-$) is assigned to the clockwise (counterclockwise) path on each hexagon.
The third-neighbor hoppings are represented by the third and fourth terms 
on the right-hand side of Eq.~(\ref{eq:honeycomb-Heff}), 
where 
``L-path'' and ``M-path'' stand for those respectively labeled with 
$L_{\rm eff}$ and $M_{\rm eff}$ in Fig.~\ref{fig:hopping_diagram}.

The effective 
hopping amplitudes have explicit forms,
\begin{widetext}
\begin{subequations}
\begin{align}
J_\text{eff}
&= 
J\mathcal{J}_0(A)-\frac{J^3}{\omega^2}
\left[
\sum_{n\neq0}\frac{\mathcal{J}_{n}^2(A)\mathcal{J}_{0}(A)}{n^2} \left(2\cos \frac{2\pi n}{3}+3 \right)
+\sum_{m,n\neq0}\frac{\mathcal{J}_{m}(A)\mathcal{J}_{n}(A)\mathcal{J}_{m+n}(A)}{mn} \left(4\cos \frac{2\pi n}{3} +1 \right)
\right],
\label{eq:honeycomb-Jeff}
\\
iK_\text{eff} &= -i\frac{J^2}{\omega}\sum_{n\neq 0} \frac{\mathcal{J}_n^2(A)}{n}\sin\frac{2\pi n}{3},
\label{eq:honeycomb-Keff}
\\
L_\text{eff}&=-\frac{2J^3}{\omega^2}
\left[
\sum_{n\neq0}\frac{\mathcal{J}_{n}^2(A)\mathcal{J}_{0}(A)}{n^2}  \cos \frac{2\pi n}{3}
+\sum_{m,n\neq0}\frac{\mathcal{J}_{m}(A)\mathcal{J}_{n}(A)\mathcal{J}_{m+n}(A)}{mn} \cos \frac{2\pi(m-n)}{3}
\right] ,
\label{eq:honeycomb-Leff}
\\
M_\text{eff}&=-\frac{J^3}{\omega^2}
\left[
\sum_{n\neq0}\frac{\mathcal{J}_{n}^2(A)\mathcal{J}_{0}(A)}{n^2} \cos\frac{2\pi n}{3}
+\sum_{m,n\neq0}\frac{\mathcal{J}_{m}(A)\mathcal{J}_{n}(A)\mathcal{J}_{m+n}(A)}{mn} \cos\frac{2\pi (m+n)}{3}
\right].
\label{eq:honeycomb-Meff}
\end{align}
\label{eq:honeycomb-effs}
\end{subequations}
\end{widetext}
Note that the NN and third-neighbor hoppings take real values ($J_\text{eff}, L_\text{eff}$, and $M_\text{eff}$), while the NNN hopping is purely imaginary
($iK_\text{eff}$). 
The van Vleck expansion gives equivalent results, but requires tedious calculations 
to arrive at the compact form for the coefficients in Eqs.~(\ref{eq:honeycomb-Jeff})--(\ref{eq:honeycomb-Meff}).

The zeroth-order term in $J_\text{eff}$ comes from the time average of the original Hamiltonian (\ref{eq:honeycomb-Ht}), where the bare hopping $J$ is renormalized into $J\mathcal{J}_0(A)$, as is known in the context of the dynamical localization and Floquet theory\cite{Dunlap1986,Tsuji2008}. 

The effective NNN hopping $iK_\text{eff}$ is derived from the two-step photo-assisted hopping processes represented in the first-order $J_{i,j}^{(1)}$ (\ref{eq:Heff-J1}). The derivation of the leading-order NNN hopping by Kitagawa \textit{et al.}~\cite{Kitagawa2011} in the weak-field regime just corresponds to 
the single-photon absorption ($n=1$), here extended to 
multiphoton processes in Eq.~(\ref{eq:Heff-J1}).
Thus, Eq.~(\ref{eq:honeycomb-Keff}) is the full expression for arbitrary strength of the field.

\subsection{Numerical results for the Chern number}
\label{sec:honeycomb-chern}
While the first two terms in the effective Hamiltonian (\ref{eq:honeycomb-Heff}) correspond to the Haldane's model, we have also $L_\text{eff}$, $M_\text{eff}$ along with higher-order terms, so the question is whether and 
how the terms beyond the leading order can induce further topological-to-topological phase transitions in strong ac fields. Hence, we directly obtain such transitions by numerically 
calculating the Chern number for the Floquet bands.

According to the previous Floquet topological formalism~\cite{Oka2009}, the Hall conductivity in ac fields is expressed 
as a nonequilibrium extension of 
the Thouless-Kohmoto-Nightingale-den Nijs (TKNN) formula,
\begin{equation}
\sigma_{xy}=\frac{2e^2}{\hbar}\int\frac{d\bm{k}}{(2\pi)^2}\sum_\alpha  \\
f_\alpha(\bm{k})[\nabla_{\bm{k}}\times \bm{A}_\alpha(\bm{k})]_z ,
\label{eq:TKNN_formula}
\end{equation}
where $\bm{A}_\alpha(\bm{k})=-i\frac{1}{\mathscr{T}}\int^{\mathscr{T}}_0 dt \langle\bm{u}_{\alpha}| \nabla_{\bm{k}}|\bm{u}_{\alpha}\rangle $ is a gauge potential defined in terms of the Floquet states, while 
$f_\alpha(\bm{k})$ is the nonequilibrium distribution function.  
The factor of 2 
comes from the spin degeneracy.  The expression for $\sigma_{xy}$ 
involves the Chern number for the Floquet band,
\begin{equation}
C_\alpha=\int \frac{d\bm{k}}{2\pi} [\nabla_{\bm{k}}\times \bm{A}_\alpha(\bm{k})]_z .
\label{eq:Chern_number}
\end{equation}
Note that in an ac field $f_\alpha(\bm{k})$ should differ from the 
equilibrium Fermi-Dirac distribution, so that the 
Hall conductivity is not necessarily quantized, although the Chern numbers always take integer values. 
The problem of the nonequilibrium distribution is considered in Sec.~\ref{dissipation}.
Here let us concentrate on 
the topological phases as defined by the Chern numbers.

\begin{figure}[t]
\begin{center}
\includegraphics[width=\hsize]{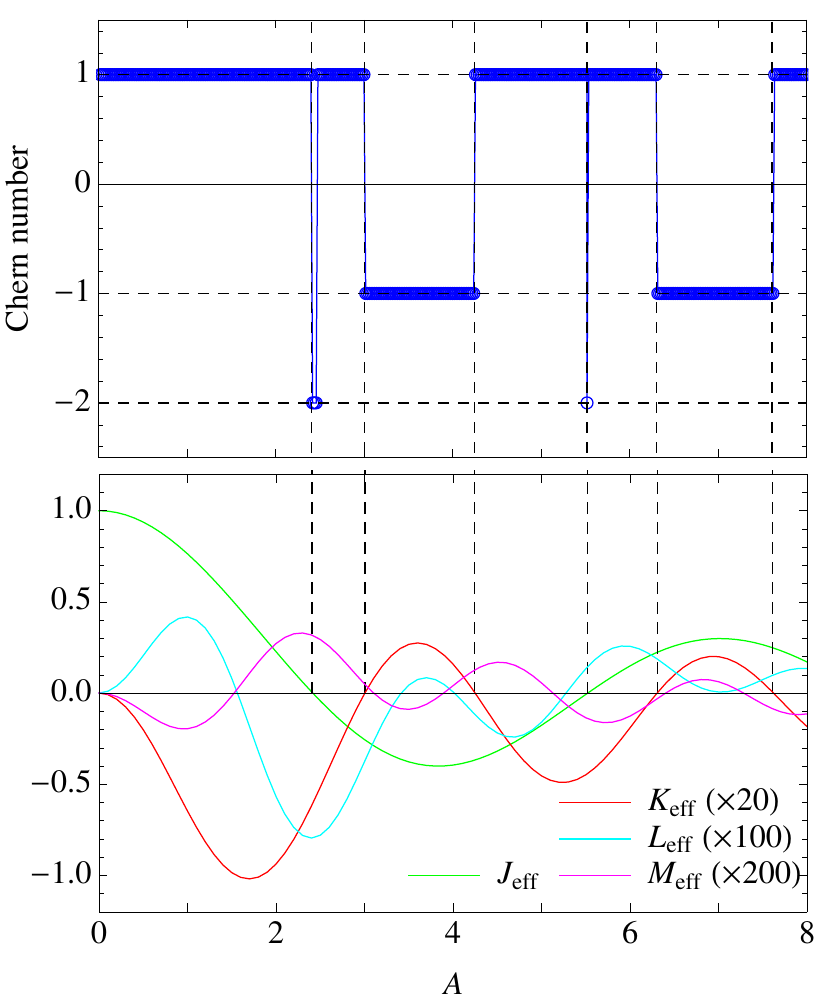}
\caption{
(Top) The Chern number for the lower band in the zero-photon sector against $A$ in the honeycomb tight-binding model 
driven by circularly polarized light with frequency $\omega=10$.
(Bottom) The hopping parameters in the effective Hamiltonian (\ref{eq:honeycomb-Heff})
against $A$. Dashed lines mark some zeros of these.
The values of $K_\mathrm{eff}$, $L_\mathrm{eff}$, and $M_\mathrm{eff}$ are scaled for visibility. }
\label{fig:Chern_numbers}
\end{center}
\end{figure}

We obtain numerically exact results for Chern numbers of the Floquet bands
by applying the numerical method due to Fukui, Hatsugai, and Suzuki \cite{Fukui2005}
to the full Floquet Hamiltonian, $\mathcal H_{m,n}-m\omega\delta_{m,n}$ [Eq.~(\ref{eq:Heff-Floquet})],
with a truncation for the matrix size. As the field $A$ becomes stronger or the frequency $\omega$ becomes smaller, one needs larger matrix sizes.
Figure~\ref{fig:Chern_numbers} displays the Chern number for the lower band in the zero-photon sector
as a function of the amplitude $A$ of CPL
for a frequency $\omega=10$.
We can immediately see that the system undergoes a series of topological-to-topological phase transitions as $A$ is increased.  

If we look more closely at Fig.~\ref{fig:Chern_numbers}, we can notice two features: (i) Unless the renormalized real NN hopping $J_\mathrm{eff}$ is too close to zero, the Chern number 
takes $1$ when the effective imaginary NNN hopping $iK_\mathrm{eff}$ has $K_\mathrm{eff}<0$ or takes 
$-1$ when $K_\mathrm{eff}>0$~\cite{JotzuSuppl},
while (ii) in the vicinity of the zeros of $J_\mathrm{eff}$, the Chern number 
jumps to $-2$. In the following, we show that these strange features can 
be clearly understood from the effective Hamiltonian (\ref{eq:honeycomb-Heff}).

When $J_\mathrm{eff}$ is not close to zero, the third-neighbor hoppings are so small that we can ignore the third and fourth terms in (\ref{eq:honeycomb-Heff}). In other words, the Brillouin-Wigner expansion for the effective Hamiltonian up to $J\mathcal O(J/\omega)$ is sufficient in this case.
Then Hamiltonian (\ref{eq:honeycomb-Heff}) is nothing but Haldane's  model \cite{Haldane1988,Kitagawa2011} 
with a pure imaginary NNN hopping. In Haldane's  model, the Chern number for the lower band is given by $-\text{sign}\, K_\text{eff}$, which is consistent with our observation (i) above.

\begin{figure}
\begin{center}
\includegraphics[width=\columnwidth]{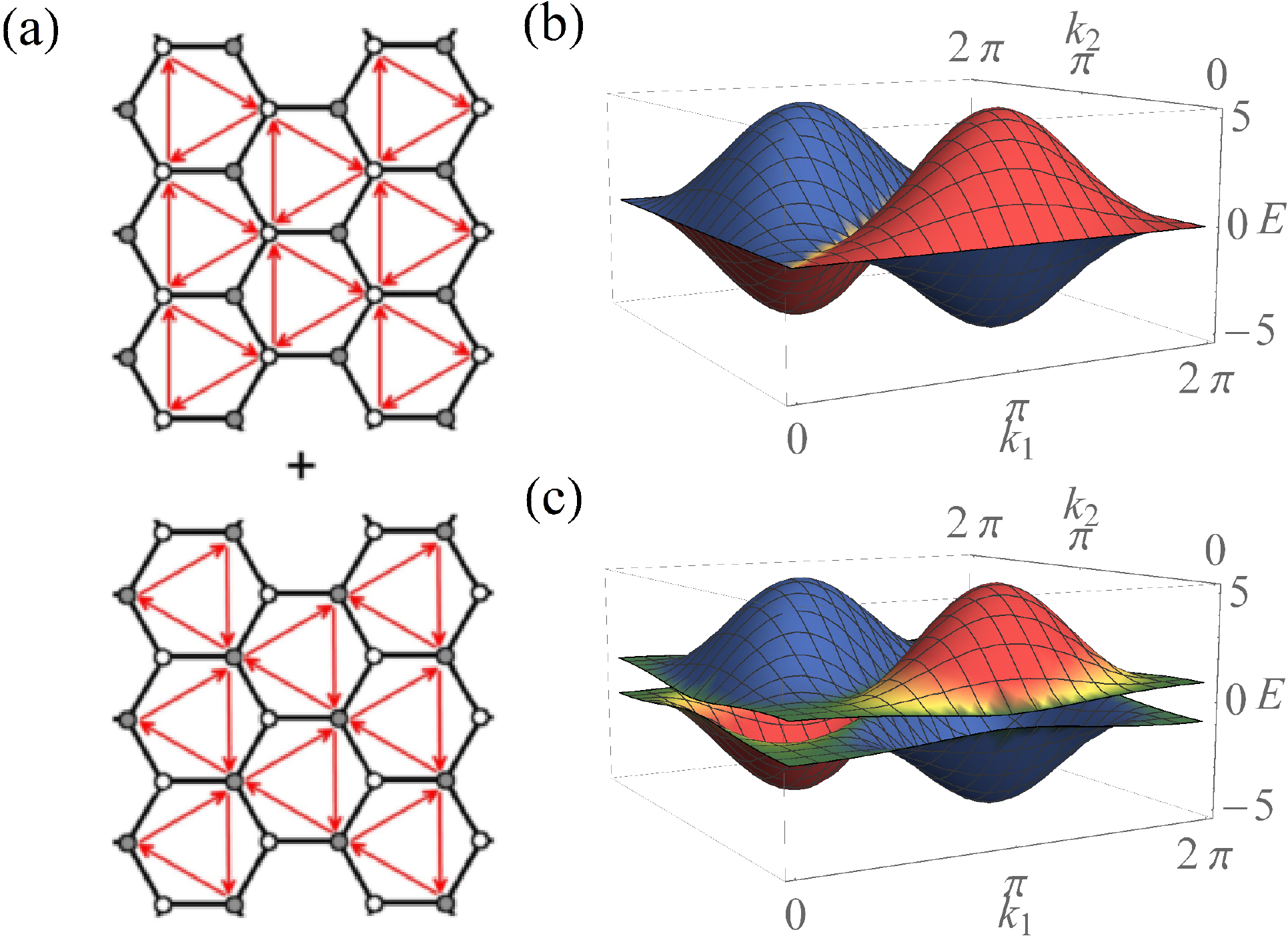}
\caption{(a) 
Honeycomb lattices with only NNN hoppings (arrows) 
as a superposition of a triangular lattice on 
A sublattice (open circles) and that on B sublattice (solid circles). 
(b) Corresponding band dispersion. 
(c) The band dispersion when we have small NN and third-neighbor hoppings in addition.}
\label{fig:NNN_plus_others}
\end{center}
\end{figure}

Conversely, in the region where the NN hopping is vanishingly small ($J_\text{eff}\simeq0$), the NNN hopping should dominate the physics. 
As shown in Fig.~\ref{fig:NNN_plus_others}~(a), the honeycomb lattice with only NNN hoppings may be regarded 
as a superposition of two separate triangular tight-binding lattices. This 
system has a dispersion relation as shown in Fig.~\ref{fig:NNN_plus_others}~(b), 
where the two bands are degenerate along nodal lines with $k_1=0$,  $k_2=0$, and $k_1=k_2$. 
Here $k_1 = \bm{k}\cdot(\bm{e}_1-\bm{e}_0)$ and $k_2 = \bm{k}\cdot(\bm{e}_2-\bm{e}_0)$.
In addition, third-neighbor terms are present along with NN terms when we are close to,
but not exactly at, a zero of $J_{\rm eff}$, and
they connect A and B sublattices to open a gap, where the topological quantities become well-defined.

\begin{figure}
\begin{center}
\includegraphics[width=\columnwidth]{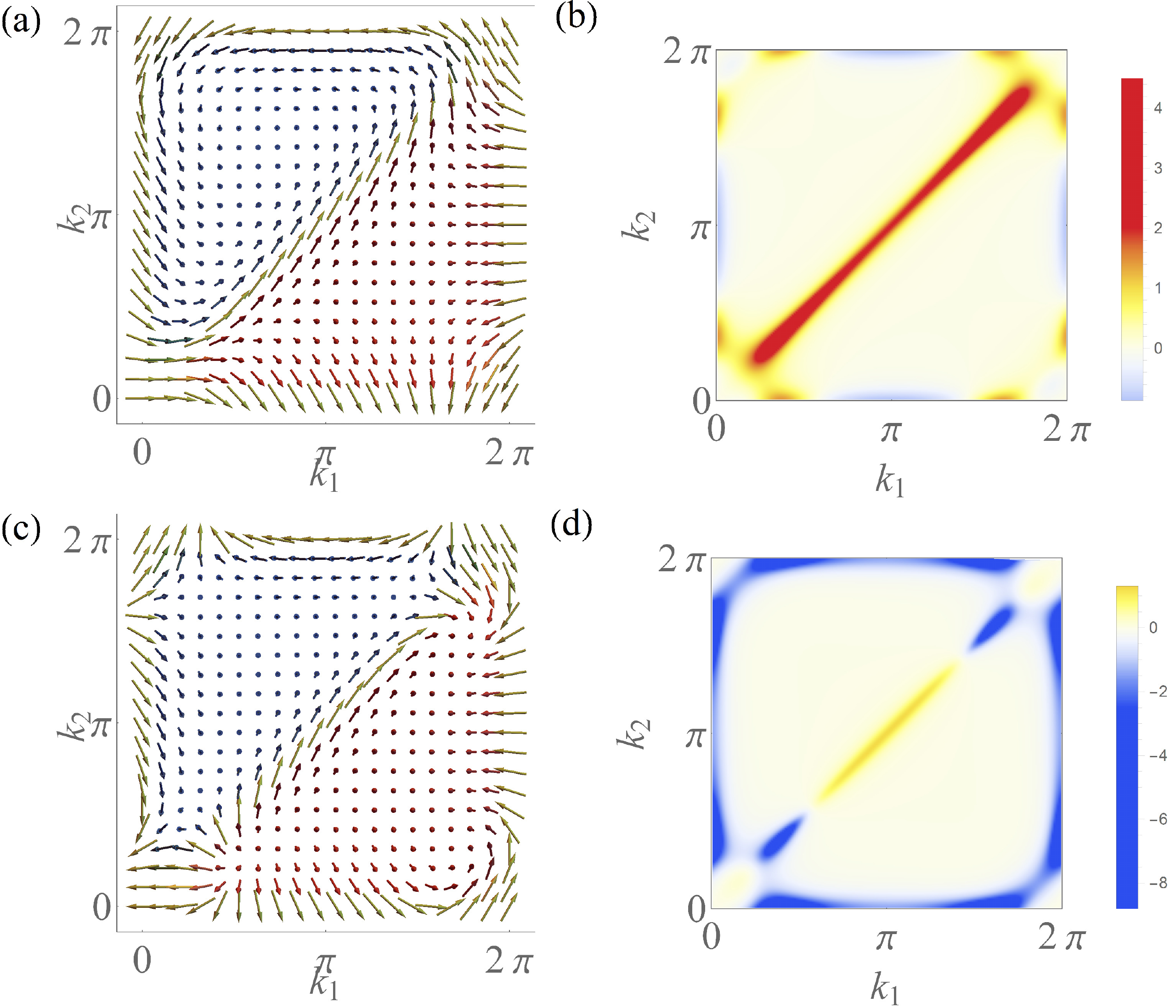}
\caption{Texture of $\hat{\bm{R}}=(R_x,R_y,R_z)$ (a),(c) and the associated Berry's curvature $-[\nabla_{\bm{k}}\times\bm{A}_\alpha(\bm{k})]_z$ (b),(d) plotted against $(k_1,k_2)$ at $A=2.40$ with the Chern number $C=+1$ (a),(b), and at $A=2.41$ with $C=-2$ (c),(d).
The color of each arrow represents the value of $R_z$, where red (blue) is assigned for $R_z=+1$ ($-1$).
}
\label{fig:berry}
\end{center}
\end{figure}

For the honeycomb lattice, $\mathcal{H}_\text{eff}$ is a 2$\times$2 matrix, so that it can be expressed as $\mathcal{H}_\text{eff}=\bm{R}\cdot \bm{\sigma}$, where $\bm R$ is a three-dimensional vector, and $\bm{\sigma}$ is the Pauli matrix.
The Chern number (\ref{eq:Chern_number}) is then expressed as
\begin{equation}
C_\pm=\pm\int \frac{d\bm{k}}{4\pi}\hat{\bm{R}} \cdot \left( \frac{d\hat{\bm{R}}}{dk_x} \times \frac{d\hat{\bm{R}}}{dk_y}\right) ,
\label{eq:winding_number}
\end{equation}
where $\hat{\bm{R}}=\bm{R}/|\bm{R}|$, and the sign $+(-)$ is taken for the upper (lower) band~\cite{HatsugaiAoki}. 
If we regard $\hat{\bm{R}}(\bm{k})$ as a mapping from $\bm{k}$ in 
the Brillouin zone (BZ) to a unit sphere, Eq.~(\ref{eq:winding_number}) represents the number of times $\hat{\bm{R}}$ wraps the unit sphere, which should be reflected in 
the texture of $\hat{\bm{R}}$ in BZ. 
Since $K_\text{eff}$ is negative around the zeros of $J_\text{eff}$ in Fig.~\ref{fig:Chern_numbers}, 
$\hat{\bm{R}}$ points mostly upwards (downwards) in the lower (upper) triangular areas enclosed by the zero-gap lines, $k_1=0$,  $k_2=0$ and $k_1=k_2$. Along the zero-gap lines, $\hat{\bm{R}}$ lies in the $xy$ plane [as depicted in Figs.~\ref{fig:berry}~(a) and \ref{fig:berry}(c) below]. Therefore, from geometric considerations, the Chern number for the lower band is equal to the winding number of $(\hat{R}_x,\hat{R}_y)$ when $(k_1,k_2)$ moves around the lower triangle along $(0,0)\rightarrow(2\pi,2\pi)\rightarrow(0,2\pi)\rightarrow(0,0)$.

Now we are ready to discuss the Chern number at $J_\text{eff}\simeq0$. 
We consider a case where $K_\text{eff}<0$ with only one of $L_\text{eff}$ or 
$M_\text{eff}=0$ is nonzero: 
When $L_\text{eff}$ is finite, we have $C=-2$, while we have 
$C=+1$ when $M_\text{eff}$ is finite (or $L_\text{eff}=M_\text{eff}=0$). 
The result indicates that Chern number changes accordingly as the relative strengths of $K_\text{eff}, L_\text{eff}$, and $M_\text{eff}$ are varied.

The numerical result in Fig.~\ref{fig:Chern_numbers} shows that $C=+1$ at $A=2.40$, while $C$ jumps to $C=-2$ at $A=2.41$. This agrees with the 
above argument, which can be reinforced from the texture of $\hat{\bm R}$. At $A=2.40$, for which 
$J_\text{eff}=0.0073, K_\text{eff}=-0.0310, L_\text{eff}=-0.0080$ and $M_\text{eff}=0.0016$, the texture of $\hat{\bm{R}}$ as defined in Eq.~(\ref{eq:winding_number}) is displayed in Fig.~\ref{fig:berry}(a). This shows that 
the texture of $\hat{\bm{R}}$ is indeed related to Berry's curvature [Fig.~\ref{fig:berry}(b)] and 
finally to the Chern number (which is $C=+1$ here).  When we vary $A$ by a tiny 
amount, $A=2.40\rightarrow2.41$, at which $J_\text{eff}=0.0021, K_\text{eff}=-0.0304, L_\text{eff}=-0.0080$ and $M_\text{eff}=0.0016$, the $\hat{\bm{R}}$ texture and Berry curvature suddenly
becomes as in Figs.~\ref{fig:berry}(c) and \ref{fig:berry}(d), giving a jump to $C=-2$, which shows that 
the $\hat{\bm{R}}$ texture can drastically change for a slight change in $A$. 
Thus, we can understand the observation (ii) above.

\begin{figure}[t]
\begin{center}
\includegraphics[width=8.5cm]{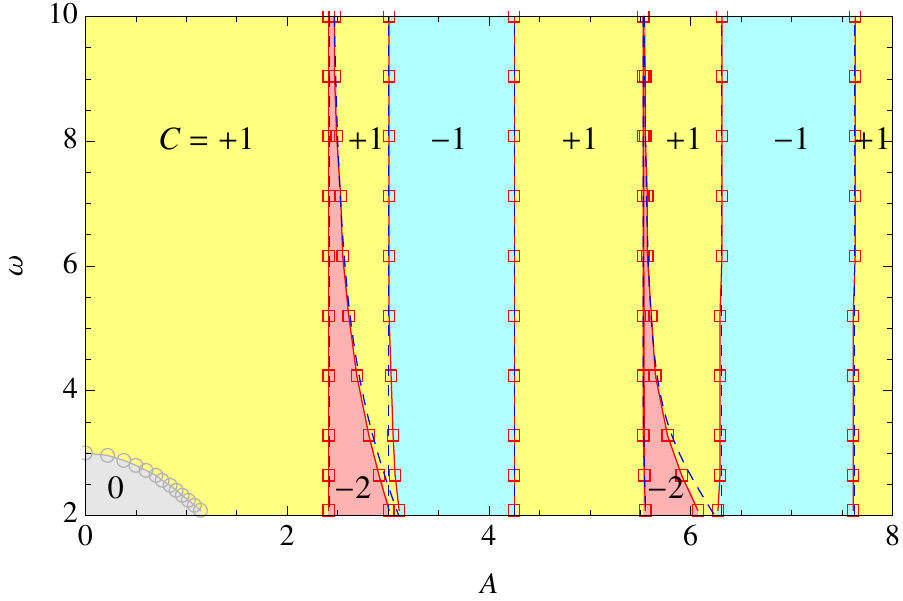}
\caption{
The phase diagram for the tight-binding model on the honeycomb lattice
driven by circularly polarized light with the amplitude $A$ and frequency $\omega\ge 2$.
Each phase is characterized by the Chern number $C$,
obtained by summation over all the bands 
below $E=0$. 
Solid lines marked by {\tiny$\square$}'s represent
the phase boundaries obtained by numerically exact calculations, 
while the dashed lines are obtained from the Brillouin-Wigner expansion up to $J\mathcal O(J^2/\omega^2)$.
Touching between the $n=+1$- and $n=-1$-photon Floquet sidebands occurs at the boundary marked by $\circ$'s (see text).}
\label{fig:phase_diagram}
\end{center}
\end{figure}

\begin{figure*}[t]
\begin{center}
\hspace{-1.5cm}
\includegraphics[width=19cm]{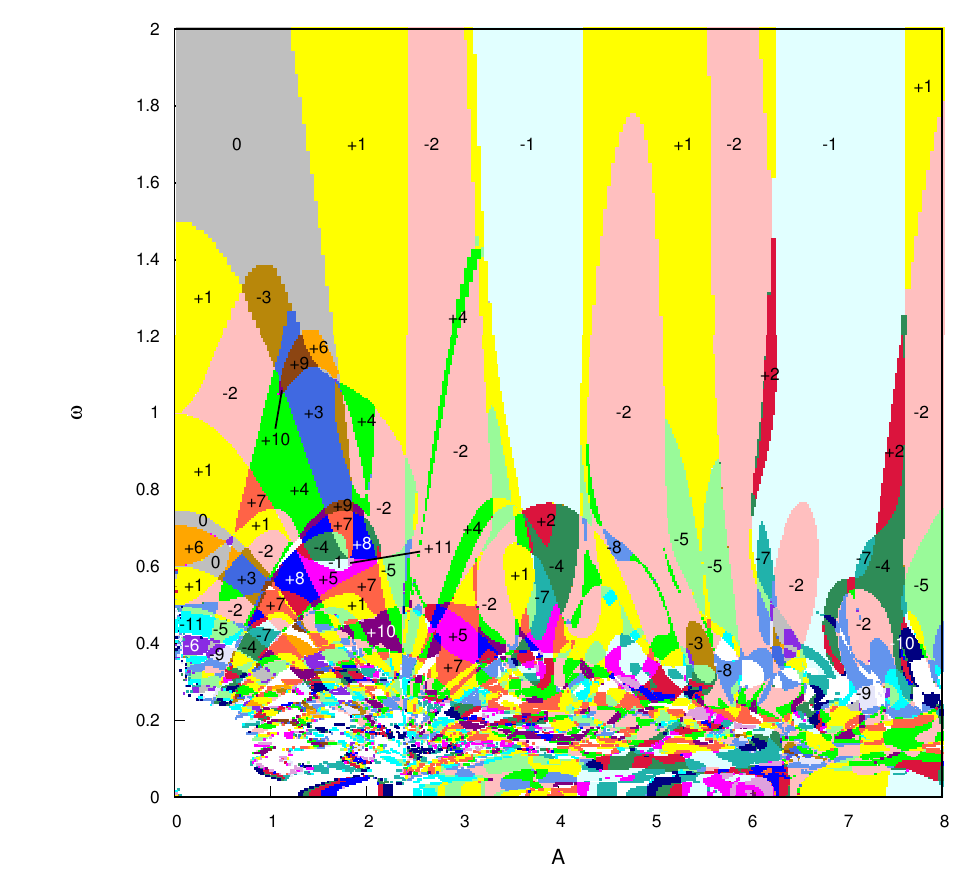}
\caption{
For lower frequencies $\omega\le 2$
the phase diagram for the honeycomb tight-binding model driven by circularly polarized light
is plotted against the amplitude $A$ and frequency $\omega$,
where the numbers represent the Chern number summed over all the bands below $E=0$.
Chern number is calculated with the method of
Fukui, Hatsugai, and Suzuki~\cite{Fukui2005}, where we take $200\times 200$ $k$ points and 50 Floquet bands in total. 
Different colors correspond to different Chern numbers $C$ as indicated,
while white regions are those having even higher values of $|C|$.
}
\label{fig:phase-diagram-details}
\end{center}
\end{figure*}

In Fig.~\ref{fig:phase_diagram},
we display the phase diagram for the Chern number summed over all the bands below the energy $E=0$ on the space of $A$ and $\omega(\ge 2)$,
with solid lines marked with {\tiny$\square$}'s representing numerically exact results.
One can see that that the $C=-2$ regions expand into the $C=\pm1$ regions
as the frequency $\omega$ of CPL is reduced.
This is because the amplitude of the third-neighbor hopping is $J\mathcal{O}(J^2/\omega^2)$
in the Brillouin-Wigner expansion, so that the relative weight of third-neighbor hopping increases as $\omega$ is decreased.  
Indeed, the phase boundaries obtained from the Brillouin-Wigner expansion up to $J\mathcal O(J^2/\omega^2)$
(dashed lines in Fig.~\ref{fig:phase_diagram})
agree well with the numerically exact results.
As we decrease $\omega$, 
band touching begins to occur between different numbers of photon sectors.
The Chern number summed over all the bands below the energy $E=0$
deviates from that obtained by the BW expansion projecting to the zero-photon subspace 
when the Floquet sidebands in $n=\pm 1$-photon sectors touch each other,
as indicated with the boundary marked with $\circ$'s in Fig.~\ref{fig:phase_diagram}.
In this region, the method in terms of the effective Hamiltonian is no longer valid.
For $A\to 0$, in particular, the band touching between $n=\pm 1$ sectors 
takes place at $\omega=W/2=3$ ($W$, bandwidth).
There may be other factors that cause the breakdown of the BW expansion.
It may have a finite radius of convergence.
For small frequency, the nonzero-photon components projected out in the BW method can contribute to the Chern number,
which may invalidate the BW expansion.

The phase boundaries can be calculated analytically  within the BW expansion in the region where the Floquet band touching does not occur.  
To this end, we diagonalize the effective Hamiltonian (\ref{eq:honeycomb-Heff})
in the momentum space. The eigenvalues take an especially simple form 
at the symmetric points in the Brillouin zone:
\begin{align}
&\pm 3(J_{\rm eff}+L_{\rm eff}+2M_{\rm eff})
\quad
\mbox{for }\bm k=\bm\Gamma=(0,0),
\\
&\pm 3\sqrt{3}K_{\rm eff}
\quad
\mbox{for }\bm k=\bm K=\left(\frac{4\pi}{3\sqrt{3}},0\right),
\\
&\pm(J_{\rm eff}-3L_{\rm eff}+2M_{\rm eff})
\quad
\mbox{for }\bm k=\bm M=\left(\frac{\pi}{\sqrt{3}},\frac{\pi}{3}\right).
\end{align}
The topological transitions occur when the band gap closes.
If we assume that the gap closes at the symmetric points of the Brillouin zone,
each phase boundary of the topological transitions is determined by
one of the following three conditions:
\begin{align}
&J_{\rm eff}+L_{\rm eff}+2M_{\rm eff}
=0,
\label{eq:honeycomb-boundary1}
\\
&K_{\rm eff}
=0,
\label{eq:honeycomb-boundary2}
\\
&J_{\rm eff}-3L_{\rm eff}+2M_{\rm eff}
=0.
\label{eq:honeycomb-boundary3}
\end{align}
From Eq.~(\ref{eq:honeycomb-boundary2}), $A$ is determined irrespective of $\omega$.
This is why the phase boundaries of the BW expansion between $C=\pm 1$ phases in Fig.~\ref{fig:phase_diagram} are straight. From Eqs.~(\ref{eq:honeycomb-boundary1}) and (\ref{eq:honeycomb-boundary3}), $\omega$ is given, respectively, as a function of $A$,
\begin{align}
\omega&=J\sqrt{-\frac{J_{\rm eff}^{(2)}(A)+L_{\rm eff}^{(2)}(A)+2M_{\rm eff}^{(2)}(A)}{J_{\rm eff}^{(0)}(A)}},
\label{eq:honeycomb-omega1}
\\
\omega&=J\sqrt{-\frac{J_{\rm eff}^{(2)}(A)-3L_{\rm eff}^{(2)}(A)+2M_{\rm eff}^{(2)}(A)}{J_{\rm eff}^{(0)}(A)}},
\label{eq:honeycomb-omega2}
\end{align}
where we have defined $J_{\rm eff}(\omega,A)=J_{\rm eff}^{(0)}(A)+(J/\omega)^2 J_{\rm eff}^{(2)}(A)$, $L_\text{eff}(\omega,A)=(J/\omega)^2 L_\text{eff}^{(2)}(A)$, and $M_\text{eff}(\omega,A)=(J/\omega)^2 M_\text{eff}^{(2)}(A)$ [see Eqs.~(\ref{eq:honeycomb-effs})].
These coincide with the dashed lines in Fig.~\ref{fig:phase_diagram}. 
Since $J_{\rm eff}^{(0)}(A)=J\mathcal{J}_0(A)$, $\omega$ in Eqs.~(\ref{eq:honeycomb-omega1}) and (\ref{eq:honeycomb-omega2})
diverges at the zeros of $\mathcal{J}_0(A)$, indicating that the topological transitions to $C=-2$ phases take place at arbitrary high frequencies.  
This feature cannot be captured by the high-frequency expansions up to $J\mathcal O(J/\omega)$, which reduces the effective Hamiltonian
to Haldane's model, even in the high-frequency regime.
If we further assume that the band touching occurs between $n=\pm 1$-photon Floquet sidebands at $\bm\Gamma$, the boundary
between $C=0$ and $C=+1$ phases is determined by
\begin{align}
\omega=3(J_{\rm eff}+L_{\rm eff}+2M_{\rm eff}),
\end{align}
which agrees with the curve marked by $\circ$'s in Fig.~\ref{fig:phase_diagram}.

We remark that the jump of the Chern number at the Floquet topological transitions
can be estimated by the method of localization: The Berry curvature localizes
near the band touching point in the Brillouin zone when a topological transition takes place. Thus, we can Taylor expand the effective
Hamiltonian as a function of momentum around the band touching point,
and calculate the integral of the Berry curvature. Since the result should be an integer,
this suffices to know the change of the Chern number. We do not go into details of
the calculation in the present paper.

If we turn to the lower-frequency region ($\omega\le 2$), something spectacular occurs:
we display the phase diagram in this region in Fig.~\ref{fig:phase-diagram-details},
which is obtained by the method of Ref.~\onlinecite{Fukui2005} with $200\times 200$ $k$ points and 50 Floquet bands. 
Note that as one reduces the frequency of the driving field one needs a larger Floquet matrix size
(typically in the order of $W/\omega$). Here the result can be trusted for $\omega\gtrsim 0.2$,
which we have checked numerically.
A similar phase diagram has been obtained in Ref.~\onlinecite{Kundu2014},
while in Ref.~\onlinecite{Piskunow2015} the Chern number in the small $A$ region is investigated analytically.
The present result exhibits a 
tantalizingly intricate phase diagram: The phase boundaries become 
increasingly complex as we go into the lower-$\omega$ region, 
which is physically due to increasingly complex folding of the Floquet bands with many band touchings.  
We can see that the phase diagram contains the phases with the Chern number as large as $+9$ [around $(A,\omega)\sim(1.3,1.1)$] or 
$-11$ [around $(A,\omega)\sim(0.2,0.45)$].
This greatly extends the possibility of Floquet engineering to derive high Chern numbers.
It would be particularly relevant to a realistic situation when one considers a graphene irradiated by CPL,
where the hopping energy is $J\sim 2.8$ eV~\cite{CastroNeto2009}.
If visible light ($\omega\sim 1.5$ eV) is used, then $\omega/J\sim 0.5$ and $A\ll 1$, where
the phase diagram is already quite complicated (the Chern number is not necessarily $\pm 1$,
unlike in the Haldane model.  
Another observation is that the jump of the Chern number at each phase boundary cannot be freely taken but
seems to be limited to a certain set of integers ($\pm 1,\pm 2,\pm 3,\pm 6,\pm 12,\dots$).
The absolute value of the jump is conserved along the smoothly connected phase boundaries.
The sign of the jump is flipped when one meets a tricritical point (where one phase boundary is terminated)
on smoothly connected boundaries.

\subsection{Effects of electron-electron interaction and energy dissipation}
\label{dissipation}

We have clarified the topological phase transitions in terms of 
the Chern numbers for an isolated, noninteracting tight-binding model so far.  However, we have 
to note that the Hall conductivity for ac-field-driven systems are 
given by the nonequilibrium version of the 
TKNN formula~\cite{Oka2009}, Eq.~(\ref{eq:TKNN_formula}),
which involves $f_{\alpha}(\bm{k})$, the nonequilibrium distribution function for the Floquet band $\alpha$.
Unlike in equilibrium, the distribution function is nonuniversal and can 
significantly depend on the details of the system. Thus, we cannot assume that the Floquet bands are completely filled or empty.  
Another important ingredient is the electron-electron interaction, which should be present in real systems. 
In these situations we can pose important questions. 
(i) In what condition does the photo-induced Hall conductivity approach quantized values under the effects of interaction and dissipation?
(ii) Can the electron-electron interaction introduce new quantum phases in the Floquet topological phase diagram?  
To answer these questions,
we go over in this section to a model in which we attach a heat bath to the system, and switch on an electron-electron (Hubbard on-site) interaction, to determine the nonequilibrium distribution for interacting systems.
For this purpose, we employ the Floquet dynamical mean-field theory (DMFT)~\cite{Tsuji2008,Tsuji2009,Aoki2014}.

\subsubsection{Floquet DMFT for multiband systems}

\begin{figure}
\begin{center}
\includegraphics[width=\hsize]{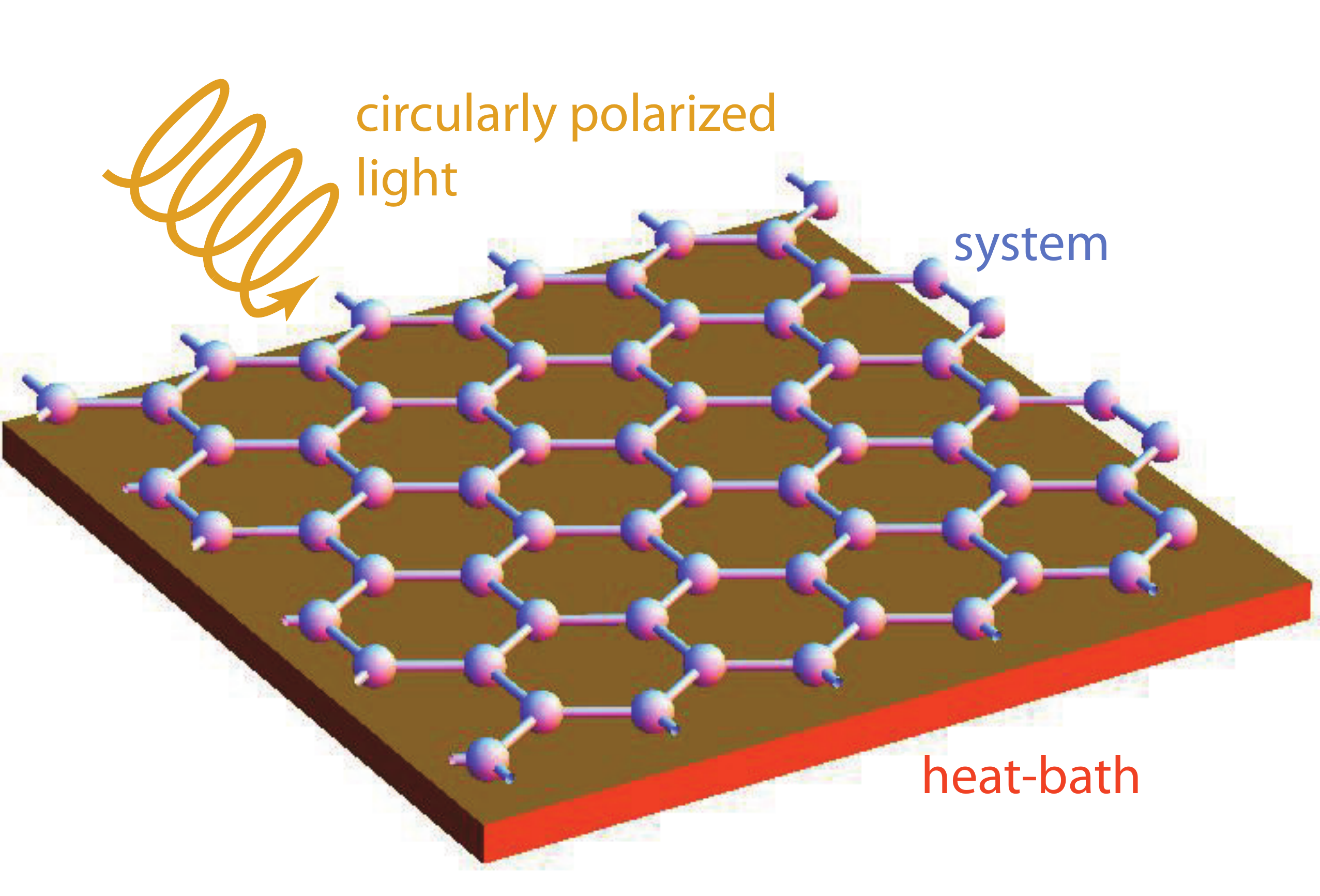}
\caption{A honeycomb lattice irradiated by a 
circularly polarized light and attached to a bath is schematically 
shown.}
\label{fig:honeycomb-CPL2}
\end{center}
\end{figure}

So let us start with a theoretical framework for treating nonequilibrium steady states in interacting multiband systems with dissipation.
First, we introduce a dissipative environment to 
which the system is coupled, and assume that 
the system will be in a nonequilibrium steady state where an energy balance between the driving field and the dissipation to the bath is achieved.  
To implement the dissipation, we introduce a fermionic heat reservoir \cite{Tsuji2009,Aoki2014} attached to each site of the system (``B\"{u}ttiker bath"~\cite{Buttiker1985,Buttiker1986}) 
that represents a dissipative environment such as substrates (see Fig.~\ref{fig:honeycomb-CPL2}). 
To implement the electron-electron interaction, we introduce the repulsive Hubbard interaction.
The total Hamiltonian then reads
\begin{subequations}
\begin{align}
\mathcal{H}_\text{tot}(t) &= \mathcal{H}_\text{sys}(t) + \mathcal{H}_\text{mix} + \mathcal{H}_\text{bath}, \\
\mathcal{H}_\text{sys}(t) &= \sum_{\langle i,j\rangle,\sigma} \left[ J_{i,j}(t) c^\dagger_{i,\sigma} c_{j,\sigma} + \text{h.c.} \right] + \sum_{i} U {\hat n}_{i,\uparrow} {\hat n}_{i,\downarrow}, \\
\mathcal{H}_\text{mix} &= \sum_{i,p,\sigma}\left( V_{p} c^\dagger_{i,\sigma}b_{i,p,\sigma} + \text{h.c.} \right), \\
\mathcal{H}_\text{bath} &= \sum_{i,p,\sigma} \nu_p b^\dagger_{i,p,\sigma} b_{i,p,\sigma},
\end{align}
\label{eq:honeycomb-Htot}
\end{subequations}
where $\sigma=\uparrow,\downarrow$ labels the electron's spin, $\hat n_{i,\sigma}=c_{i,\sigma}^\dagger c_{i,\sigma}$ is the electron number operator, $b^\dagger_{i,p,\sigma}$ is the creation operator of fermions in the B\"{u}ttiker bath 
with an energy dispersion $\nu_p$ ($p$ is a momentum), and $V_p$ is the hybridization between the system and the bath. 
The electron-electron interaction is introduced as
the on-site repulsion $U$ to investigate the effect of electron correlations.

Due to the dissipation, we can envisage that the system reaches a steady state on a time scale of $\Gamma^{-1}$, where $\Gamma$ is the energy scale of the dissipation as given in Eq.~(\ref{eq:honeycomb-Gamma}).
We can expect that the steady state is time-periodic with period $\mathscr T$ at long enough time
in the presence of dissipation and/or many-body interaction~\cite{Hausinger2010,Russomanno2012}, 
that is, the real-time Green's function $G_{\bm{k},a,b}(t,t') = -i\langle \mathcal{T} c_{\bm{k},a}(t) c^\dagger_{\bm{k},b}(t') \rangle$ (with $\mathcal T$ representing the time-ordering operator and $a, b$ the band indices)
satisfies a periodicity $G_{\bm k,a,b}(t+\mathscr T,t'+\mathscr T)=G_{\bm k,a,b}(t,t')$.
In this situation,
we can utilize the Floquet Green's function technique~\cite{Tsuji2008, Tsuji2009}.
The Floquet Green's function, $\hat{G}_{\bm{k}}(\nu)$,
is given by a double Fourier transform of $G_{\bm{k},a,b}(t,t')$ as
\begin{align}
[ {\hat G}_{\bm{k}}(\nu) &]_{a,b;m,n} = \frac{1}{\mathscr T}\int_0^{\mathscr T} dt \int_{-\infty}^\infty dt' \notag\\
&\times G_{\bm{k},a,b} \left( t_+, t_- \right)
e^{i (\nu + m \omega) t_+ - i (\nu + n \omega) t_- },
\end{align} 
with $t_\pm = t \pm t'/2$. Note that $[ {\hat G}_{\bm{k}}(\nu) ]_{m,n}$ is an $N_b\times N_b$ matrix 
with $N_b$ ($=2$ for the honeycomb lattice) being the number of the original bands.
 
Since the Hamiltonian is quadratic with respect to the heat-bath fermions, one can analytically integrate out the bath degrees of freedom, 
where the heat bath is incorporated in the self-energy as~\cite{Tsuji2009}
\begin{align}
\begin{pmatrix}
{\hat \Sigma}_\text{diss}^R(\nu) &
{\hat \Sigma}_\text{diss}^K(\nu) \\
0 &
{\hat \Sigma}_\text{diss}^A(\nu) \\
\end{pmatrix}
=
\begin{pmatrix}
-i\Gamma &
-2i\Gamma {\hat F}(\nu) \\
0 &
i\Gamma
\end{pmatrix}
.
\end{align}
Here we have defined the dissipation rate due to the coupling to the heat bath as
\begin{align}
\Gamma(\nu) = \sum_p \pi |V_p|^2 \delta(\nu - \nu_p).
\label{eq:honeycomb-Gamma}
\end{align}
If we assume that the bath is in equilibrium with a temperature $T=1/\beta$, we have 
${\hat F}(\nu) = \tanh[\frac \beta 2 (\nu+n\omega)] \delta_{mn}\delta_{ab}$ in the Keldysh component.  
The Dyson's equation for Floquet Green's functions reads~\cite{Tsuji2009}
\begin{align}
\begin{pmatrix}
\hat{G}^R_{\bm{k}}(\nu) & \hat{G}^K_{\bm{k}}(\nu)\\
0 & \hat{G}^A_{\bm{k}}(\nu)
\end{pmatrix}^{-1}
&=
\begin{pmatrix}
\hat{G}^{R0}_{\bm{k}}(\nu)^{-1} & [\hat{G}^{0}_{\bm{k}}(\nu)^{-1}]^K \\
0 & \hat{G}^{A0}_{\bm{k}}(\nu)^{-1}
\end{pmatrix} \notag\\
+
\begin{pmatrix}
i\Gamma \hat{1} & 2i\Gamma \hat{F}(\nu) \\
0 & -i\Gamma \hat{1}
\end{pmatrix}
&-
\begin{pmatrix}
\hat{\Sigma}^R(\nu) & \hat{\Sigma}^K(\nu) \\
0 & \hat{\Sigma}^A
\end{pmatrix}, 
\label{eq:honeycomb-Dyson}
\end{align}
where we have omitted the $\nu$ dependence of $\Gamma(\nu)$ 
for simplicity.
Each matrix element in the above $2\times2$ matrix equation has Floquet indices $m,n=0,\pm1,\pm2,\cdots$ and band indices $a,b$. The retarded and advanced Green's functions are explicitly given as 
\begin{align}
&[G^{R0}_{\bm{k}}(\nu)^{-1}]_{m,n;a,b} = (\nu+n\omega + i\eta)\delta_{m,n}\delta_{a,b} - H_{m,n;a,b}^0(\bm{k}), \\
&[G^{A0}_{\bm{k}}(\nu)]_{m,n;a,b} = [G^{R0}_{\bm{k}}(\nu)]_{n,m;b,a}^\ast,
\end{align}
where $\eta$ is a positive infinitesimal and $\hat H^0$ is the Floquet matrix representation for the noninteracting part of the system Hamiltonian. The Keldysh component $[\hat{G}^{0}_{\bm{k}}(\nu)^{-1}]^K$ in Eq.~(\ref{eq:honeycomb-Dyson}) is proportional to $i\eta$. Since there exists a nonzero contribution ${\hat \Sigma}^K_\text{diss}(\nu)$ to the Keldysh part, we can set $[\hat{G}^{0}_{\bm{k}}(\nu)^{-1}]^K\to 0$ to solve Eq.~(\ref{eq:honeycomb-Dyson}). 

The self-energy from the Hubbard interaction is implemented by DMFT \cite{Georges1996,Freericks2006,Tsuji2008,Tsuji2009} 
with the second-order iterated perturbation theory (IPT) used for the impurity solver,
\begin{align}
&[\Sigma(t,t')]_{a,b}^{\lessgtr} = \delta_{a,b} U^2 \mathcal{G}^\lessgtr_{a,a}(t,t') \mathcal{G}^\gtrless_{a,a}(t',t) \mathcal{G}^\lessgtr_{a,a}(t,t'),
\end{align}
where $\mathcal G_{a,b}(t,t')$ is the Weiss Green's function defined by
\begin{align}
\mathcal{G}_{a,b}(t,t')
&=
\sum_{m,n}\int_{-\omega/2}^{\omega/2} \frac{d\nu}{2\pi} e^{-i(\nu+m\omega)t+i(\nu+n\omega)t'}
\mathcal G_{m,n;a,b}(\nu),
\\
\hat{\mathcal G}(\nu)
&=
[\hat G_{\rm loc}^{-1}(\nu) + \hat\Sigma(\nu)]^{-1},
\end{align}
and $\lessgtr$ denote the lesser ($<$) or greater ($>$) component.
$\hat G_{\rm loc}(\nu)$ is the local Green's function,
\begin{align}
\hat G_{\rm loc}(\nu)
&=
\frac{1}{\mathcal N}\sum_{\bm k} \hat G_{\bm k}(\nu),
\end{align}
with $\mathcal N$ the number of $k$ points.
The lesser and greater components of the self-energy 
are connected to the retarded, advanced, and Keldysh components as
\begin{align}
&\Sigma^K (t,t') = \Sigma^< (t,t') + \Sigma^> (t,t'), \\
&\text{Im}\,\Sigma^R (t,t') = \frac i2 \left[ \Sigma^< (t,t') - \Sigma^> (t,t') \right].
\end{align}
IPT is known to be applicable to particle-hole symmetric cases~\cite{Tsuji2013},
and in this paper we apply it to the honeycomb and Lieb-Hubbard models at half filling.
Although IPT is based on the weak-coupling perturbation theory, it is also known to
correctly reproduce the strong-coupling limit and qualitatively describe the Mott transition~\cite{Zhang1993,Georges1996}.

The Hall conductivity $\sigma_{xy}$ is then expressed in terms of Floquet Green's functions in the nonequilibrium linear-response formula as
\begin{align}
\sigma_{xy} (\nu)
&= \text{Re}\, \frac{1}{\nu} \frac{1}{{\mathcal N}S_\text{cell}}\text{tr}\, \sum_{\bm{k}}
\int_{-\omega/2}^{\omega/2}
\frac{d\nu'}{2\pi} \notag\\
\times\big[
&\hat{v}_{\bm{k}}^{i} \hat{G}_{\bm{k}}^\text{R}(\nu'+\nu) \hat{v}_{\bm{k}}^{j} \hat{G}_{\bm{k}}^<(\nu') +
\hat{v}_{\bm{k}}^{i} \hat{G}_{\bm{k}}^<(\nu') \hat{v}_{\bm{k}}^{j} \hat{G}_{\bm{k}}^\text{A}(\nu'-\nu)
\big] \label{eq:honeycomb-cond}
\end{align}
up to bubble diagrams. Here $\hat{\bm v}_{\bm{k}}$ is the current operator in a Floquet matrix form, 
\begin{align}
[\hat{\bm{v}}_{\bm{k}}]_{m,n} \equiv \frac{1}{\mathscr T} \int_0^{\mathscr T} e^{i(m-n)\omega t} \frac{\partial H^0(\bm{k}+\bm{A}(t)) }{\partial \bm{k}} dt, 
\end{align}
and $S_\text{cell}$($=\sqrt 3 /2$ for honeycomb) the area of a unit cell.  Note that, 
while the vertex correction should, in general, be considered in the strongly correlated regime, the approximation (\ref{eq:honeycomb-cond}) turns out to be good unless the driving frequency $\omega$ is too close to the probe frequency $\nu$~\cite{Tsuji2009}.

\begin{figure}[t]
\begin{center}
\includegraphics[width=\hsize]{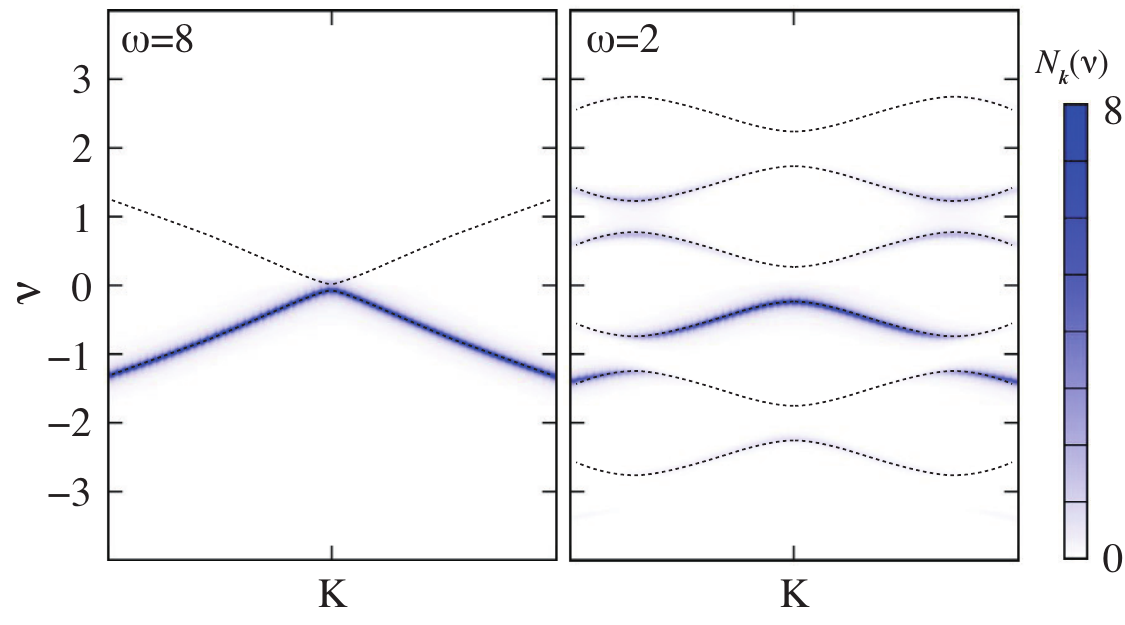}
\caption{
The electron occupation $N_{\bm{k}}(\nu)$ (blue shading) and quasienergy spectrum (dotted black lines) around the $K$-point for the honeycomb tight-binding model ($U=0$) driven by circularly polarized light with $(\omega, A)=(8, 0.5)$ (off-resonant regime; 
left panel) and $(2, 0.5)$ (resonant; right panel) and coupled to the heat bath with $\Gamma=T=0.04$.
The horizontal axis is a line in the Brillouin zone along $\bm k=(4\pi/3\sqrt{3},t)$ with $-1\le t\le 1$.}
\label{fig:honeycomb-band}
\end{center}
\end{figure}

\subsubsection{Results}
\label{sec:honeycomb-results}

We first examine how the electron distribution deviates from the equilibrium Fermi-Dirac distribution 
when the honeycomb lattice is continuously irradiated by CPL and reaches a nonequilibrium steady state in the noninteracting case.
To this end, we calculate the electron occupation, 
\begin{align}
N_{\bm{k}}(\nu) = \frac 1{2\pi} \text{tr}\,\text{Im}\, \int_0^{\mathscr T} \frac{dt}{\mathscr T} \int_{-\infty}^\infty dt' \,  G_{\bm{k}}^< \left( t + \frac {t'}{2}, t - \frac {t'}{2} \right) e^{i\nu t'},
\end{align}
where $G^<_{\bm{k},a,b}(t,t') = i\langle c^\dagger_{\bm{k},b}(t') c_{\bm{k},a} (t) \rangle$
and the trace is taken over the band indices.
In Fig.~\ref{fig:honeycomb-band}, we show the results for $\omega=8$ (left panel) and $\omega=2$ (right).
In the case of $\omega=8$, the frequency is much larger than the electronic bandwidth $W=6$
of the honeycomb tight-binding model. Hence,
hybridization between different Floquet bands becomes weak, and interband transitions of electrons 
are suppressed (``off-resonant"). As a result,
the lower Floquet band remains almost fully occupied, while the other bands are empty. 
This makes $\sigma_{xy}$ to be almost quantized into the Chern number of the occupied band
as shown below.
In the ``resonant" ($\omega<W$) case of $\omega=2$, by contrast, 
Floquet bands start to hybridize with each other. Then
photo-induced carriers are generated around the Floquet Brillouin zone boundaries $\nu =\pm \omega/2$, and 
the distribution significantly deviates from the equilibrium one.
In this situation, the quantization of  $\sigma_{xy}$ is no longer maintained.
Thus, one of the necessary conditions for the quantization of  $\sigma_{xy}$
turns out to be the frequency of the drive being greater than the bandwidth.

\begin{figure}[t]
\begin{center}
\includegraphics[width=\hsize]{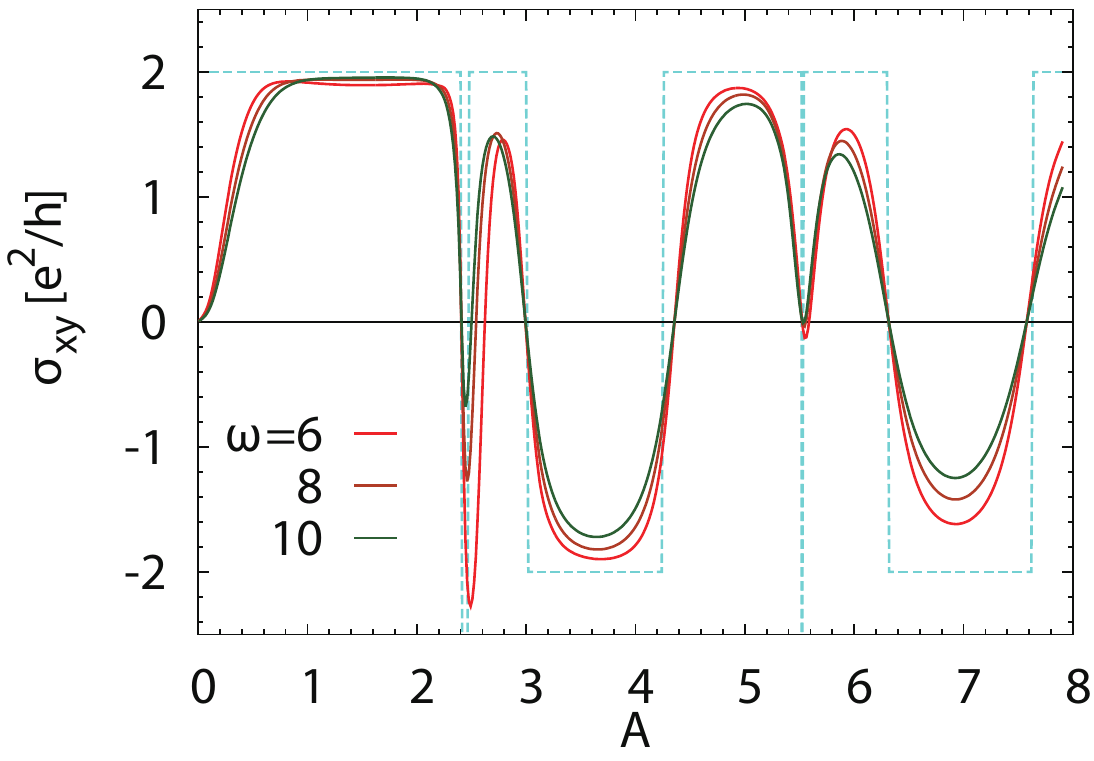}
\caption{
The Hall conductivity $\sigma_{xy}$ against the amplitude $A$ of circularly polarized light
irradiated to the honeycomb tight-binding model ($U=0$) at half filling
for several values of the frequency $\omega$.
The system is coupled to the heat bath with $\Gamma=T=0.04$.
The dashed line is the Chern number ($\times2$) summed over all the bands below $E=0$
for $\omega=10$ (Fig.~\ref{fig:Chern_numbers}).
}
\label{fig:honeycomb-Sxy-Keff}
\end{center}
\end{figure}

In the rest of this section (and also in the rest of the paper where we treat
the interaction and dissipation), we concentrate 
on the off-resonant case ($\omega\ge W$).
Let us first look at the noninteracting ($U=0$) case in Fig.~\ref{fig:honeycomb-Sxy-Keff}, which displays $\sigma_{xy}$ for the tight-binding model coupled to
the heat bath with $\Gamma=T=0.04$ as a function of the field amplitude $A$. 
The frequency of the ac field $\omega$ is varied from $6$ to $10$.
The result indicates the way in which 
$\sigma_{xy}$ approaches the Hall conductivity quantized in $e^2/h$ 
(times the spin degeneracy $2$).  We can see that 
the jump of $\sigma_{xy}$ at the topological-to-topological transitions
are rounded as compared to that of the Chern number shown in Fig.~\ref{fig:Chern_numbers},
where $C=\pm 1$ is determined by the sign of $K_\text{eff}$, and 
$C$ sharply jumps to $-2$ in the vicinity of the zeros of $\mathcal{J}_0(A)$, as predicted from the effective Hamiltonian [Eq.~(\ref{eq:honeycomb-Heff})].
Although the region with $C=-2$ is very narrow, the transition is quite visible as a prominent dip in $\sigma_{xy}$.

For the quantization to be approached, the photo-induced gap should be sufficiently larger than the temperature $T$ and the dissipation $\Gamma$ in order to suppress excitations over the gap induced by thermal fluctuations or coupling to the bath.
The band gap at the $K$ point is given by $\Delta_\text{G} = 6\sqrt{3} |K_\text{eff}|$, where 
in the low-amplitude regime $|K_\text{eff}|\sim (\sqrt{3}/8)(J^2A^2/\omega)$.
For $\Delta_\text{G}>T, \Gamma$ to be realized, 
we have to satisfy $\omega\lesssim J^2A^2/T, J^2A^2/\Gamma$.
If $\omega$ becomes smaller than 
the effective bandwidth $W |\mathcal{J}_0(A)|$ ($W=6J$), however, Floquet bands hybridize with each other
between different photon sectors,
and photo-induced electrons and holes appear at the Floquet Brillouin zone boundary. Therefore, we should choose a moderate value of $\omega$ larger than the effective bandwidth, while keeping the band gap larger than $T$ and $\Gamma$.
The $\omega$ dependence of $\sigma_{\it xy}$ in Fig.~\ref{fig:honeycomb-Sxy-Keff} is consistent with this argument: 
$\omega\ge 6 \ge W|\mathcal{J}_0(A)|$ always holds in Fig.~\ref{fig:honeycomb-Sxy-Keff}, so that Floquet band hybridization between different photon sectors
is small. On the other hand, for $A\lesssim \sqrt{\omega T/J^2}\sim 0.5-0.6$ and $A$ close to the zeros of $K_{\rm eff}$,
$\Delta_G$ becomes smaller than $T, \Gamma$, so that the quantization is degraded. 
The dependence of $\sigma_{xy}$ on $\omega$ in Fig.~\ref{fig:honeycomb-Sxy-Keff} 
mainly comes from a difference in the gap size.

\begin{figure}[t]
\begin{center}
\includegraphics[width=\hsize]{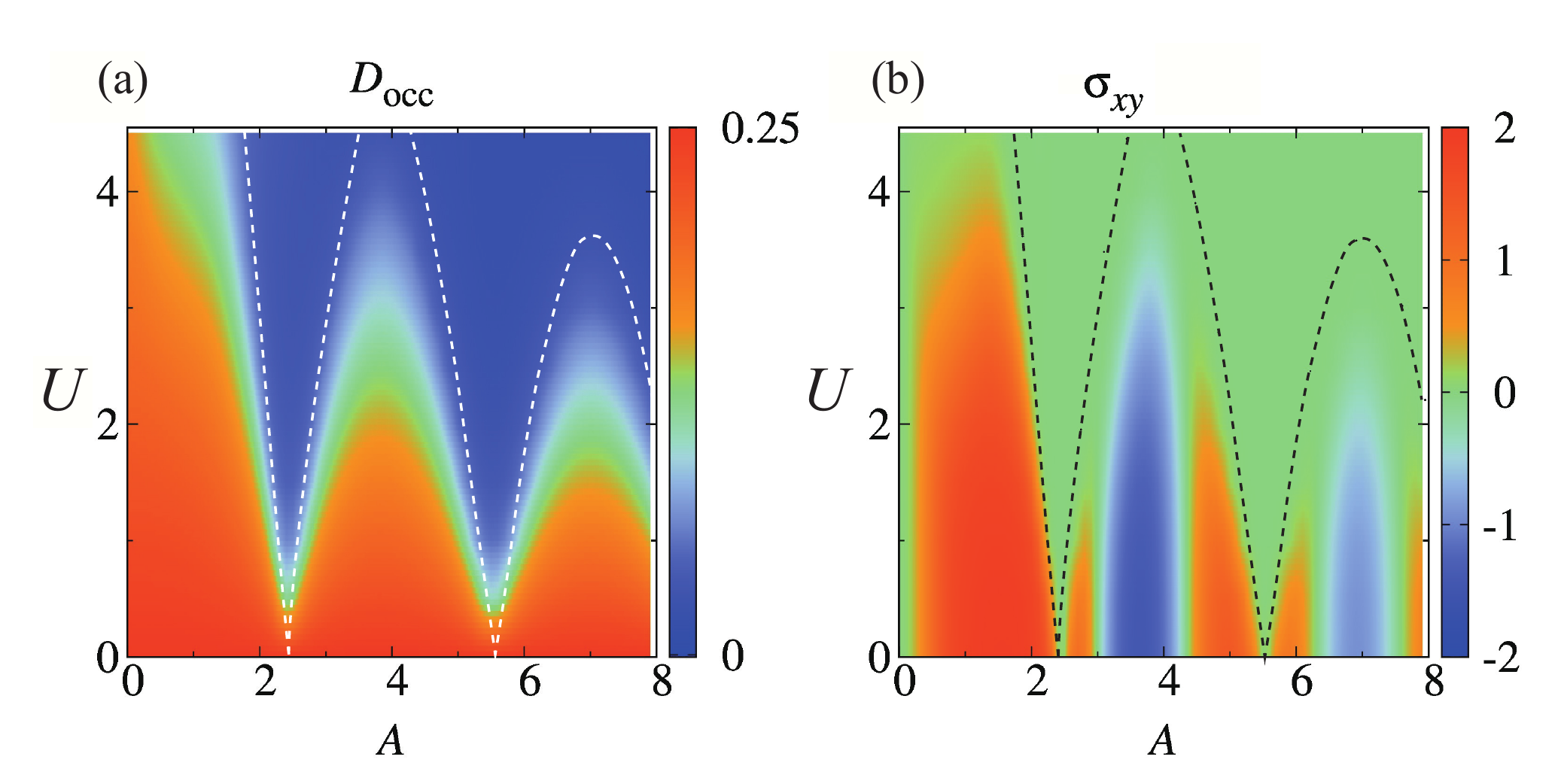}
\caption{Color-coded
(a) double occupancy $D_\text{occ}$ and (b) Hall conductivity $\sigma_{xy}$ 
against the amplitude $A$ of the circularly polarized light and the Hubbard repulsive interaction 
$U$ for the honeycomb Hubbard model driven by circularly polarized light 
(with a frequency $\omega=10$ here)
and coupled to the heat bath (with $\Gamma=T=0.1$ here).
The dashed lines indicate the effective critical value $U=U_c \mathcal{J}_0(A)$, where $U_c$ is the critical $U$ for the Mott transition in  equilibrium.}
\label{fig:honeycomb-Docc-Sxy}
\end{center}
\end{figure}

Finally in this section, we turn to study the effect of the electron repulsion $U$ by means of Floquet DMFT for the Hubbard model.
To identify the phases, we focus on two quantities to characterize the 
honeycomb Hubbard model: One is the double occupancy $D_\text{occ}$ defined by
\begin{align}
D_\text{occ}
&=
\frac{1}{\mathscr T}\int_0^{\mathscr{T}} dt \left\langle n_{i,\uparrow}(t) n_{i,\downarrow}(t) \right\rangle,
\end{align}
which is a measure of the Mott insulator.
The other is the Hall conductivity $\sigma_{\it xy}$, which is a measure of the topological property.

In Fig.~\ref{fig:honeycomb-Docc-Sxy}, we plot these quantities 
in the parameter space of $A$ and $U$ to determine the 
topological \textit{and} correlation phase diagram.  
The result, which is shown here for the off-resonant case with $\omega=10$,
 indeed exhibits intriguing features.  The double occupancy in 
Fig.~\ref{fig:honeycomb-Docc-Sxy}(a) shows that the 
Mott insulator phase, which can be identified as the low-$D_\text{occ}$ region 
(blue in the color coding), appears with a characteristic boundary.  
Since the other phase is topological, as we discuss below, the boundary 
actually signifies a Mott-insulator-to-topological-insulator transition (or crossover).
The oscillating shape of the boundary reminds us of the Bessel function, 
and, indeed, the transition is clearly understood in terms of the effective bandwidth of the system in ac fields, as is pointed out for the Bose-Hubbard model driven by ac fields~\cite{Eckardt2005}.
Namely, in the Floquet Hamiltonian (\ref{eq:honeycomb-Heff}), an external field gives the factor $\mathcal{J}_0(A)$ to the NN hopping and causes a shrinking of the bandwidth $W$~\cite{Tsuji2008}.
Thus, the effective electron correlation strength $U/W$ is renormalized 
into $U/(W |\mathcal{J}_0(A)|)$, and a field-induced Mott transition takes place 
at $U_c/W =U/(W |\mathcal{J}_0(A)|)$ (dashed line in Fig.~\ref{fig:honeycomb-Docc-Sxy}), where $U_c$ is the critical value $U_c\simeq 12$ 
for the Mott transition in the honeycomb lattice in equilibrium ($A=0$).
Hence, the Mott transition induced by off-resonant external fields found 
here may be interpreted as a nonequilibrium analog of the bandwidth-controlled Mott transition \cite{Imada1998}. Since $|\mathcal{J}_0(A)|$ vanishes toward its 
zeros, we can note that the field-induced phase transition should take place no matter how small $U$ may be, for example, in graphene with $U\simeq 1.6$~\cite{Schueler2013}. 
The magnitude of the Hall conductivity $\sigma_{xy}$ tends to become small as $U$ is increased [Fig.~\ref{fig:honeycomb-Docc-Sxy}(b)]. When the system goes into the Mott phase, $\sigma_{xy}$ approaches zero,
and the topological nature disappears.

\section{Lieb lattice in a circularly polarized light}
\label{sec:Lieb}

\begin{figure}[t]
\begin{center}
\includegraphics[width=\hsize]{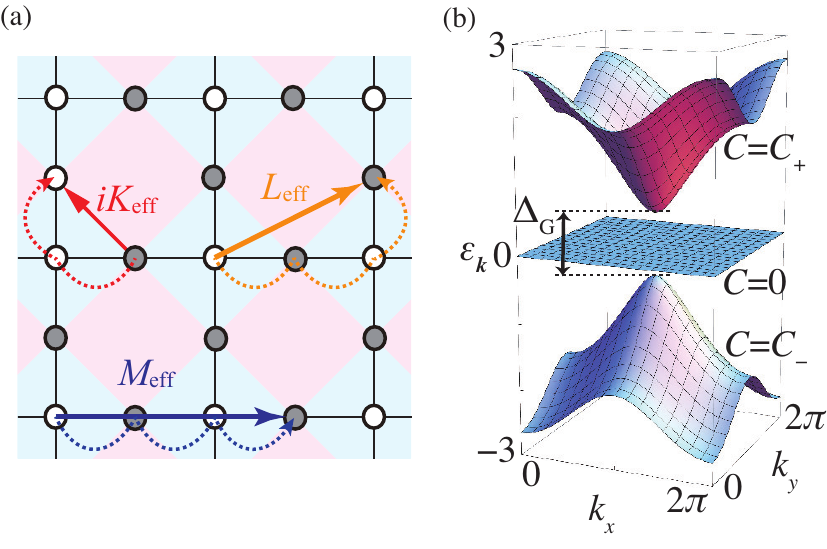}
\caption{
(a) The effective hoppings in the effective Hamiltonian (\ref{eq:Lieb-Heff}) derived from the BW expansion
for the Lieb lattice driven by a circularly polarized light.  See the text for blue and pink regions.  
(b) Schematic band structure for the effective Hamiltonian (\ref{eq:Lieb-Heff}) with the 
topological gap $\Delta_\text{G} = 8 |K_\text{eff} |$ and Chern number $C_\pm = \pm \text{sign} K_\text{eff}$.}
\label{fig:Lieb-lattice}
\end{center}
\end{figure}

\begin{figure}[t]
\begin{center}
\includegraphics[width=8.3cm]{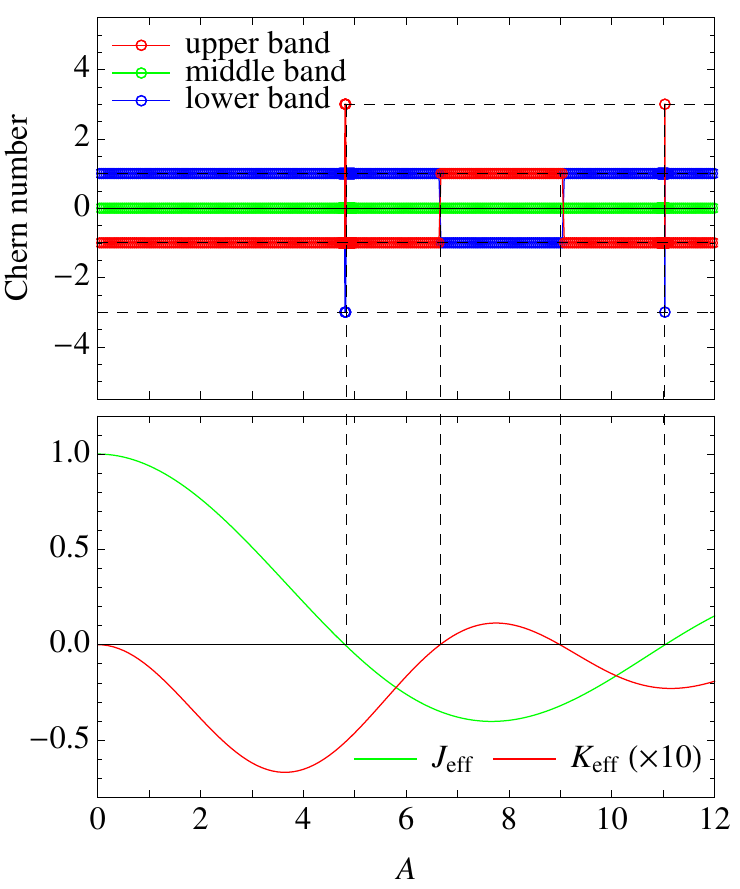}
\caption{
(Top) Chern number for each band in the zero-photon sector of the Lieb tight-binding model
driven by circularly polarized light with frequency $\omega=10$.
(Bottom) The hopping parameters for the corresponding effective Hamiltonian [Eq.~(\ref{eq:Lieb-Heff})].
The value of $K_\text{eff}$ is rescaled for visibility. 
Dashed lines mark some zeros of $J_{\rm eff}$ and $K_{\rm eff}$.}
\label{lieb-chern-number}
\end{center}
\end{figure}

As the second example of Floquet topological insulators generated by CPL,
we consider the tight-binding model on the Lieb lattice [Fig.~\ref{fig:Lieb-lattice}(a)].
The Hamiltonian is
\begin{align}
\mathcal H^{\rm Lieb}(t)
&=
\sum_{i,j}^{\rm NN} J_{i,j}(t) c_i^\dagger c_j,
\end{align}
where $J_{i,j}(t)$ is given by Eq.~(\ref{Peierls}). The length of NN bonds is set to be $1/2$
(i.e., the length of sides of squares is $1$).
The band structure consists of one flat band
and two dispersive bands, where the latter forms a Dirac cone right at
the energy of the flat band ($E=0$).
The Lieb lattice was originally conceived to study the flat-band ferromagnetism,\cite{Lieb1989} but in the present 
context poses an interesting problem of what would be the topological nature of the flat band in nonequilibrium.  

As we have shown above, CPL generally gives rise to NNN hopping with a phase factor. If we apply the Brillouin-Wigner expansion up to $J\mathcal O(J^2/\omega^2)$ [Eqs.~(\ref{eq:Heff-formula})-(\ref{eq:Heff-formula2})] to the Lieb model in CPL, 
we obtain a non-Hermitian effective Hamiltonian, in contrast to the van Vleck perturbation expansion.
If we divide the effective Hamiltonian into Hermitian and anti-Hermitian parts as
$H_{\rm BW}^{\rm h.}+H_{\rm BW}^{\rm a.h.}$, the latter part is $J\mathcal O(J^2/\omega^2)$.
We can show that the anti-Hermitian part does not contribute to the quasienergies 
up to $J\mathcal O(J^2/\omega^2)$: We treat $H_{\rm BW}^{\rm a.h.}$ as a perturbation
to $H_{\rm BW}^{\rm h.}$. Let us denote the eigenvalue and eigenstate of $H_{\rm BW}^{\rm h.}$
as $\varepsilon$ and $v$, respectively. The first-order correction to the quasienergy
is given by $v^\dagger H_{\rm BW}^{\rm a.h.}v$, which is purely imaginary since $H_{\rm BW}^{\rm a.h.}$ is anti-Hermitian. As we discussed in Sec.~\ref{sec:Heff}, the BW expansion is guaranteed to 
produce the correct real quasienergy up to a given order [$J\mathcal O(J^2/\omega^2)$
in the present case]. Thus, the purely imaginary contribution to the quasienergy
from $H_{\rm BW}^{\rm a.h.}$ should be of higher order than $J\mathcal O(J^2/\omega^2)$.
The second-order perturbation gives higher-order corrections as well.
In this way, as long as the quasienergy is concerned up to a given order,
it is sufficient to consider the Hermitian part of the effective Hamiltonian.

\begin{figure}[t]
\begin{center}
\includegraphics[width=\hsize]{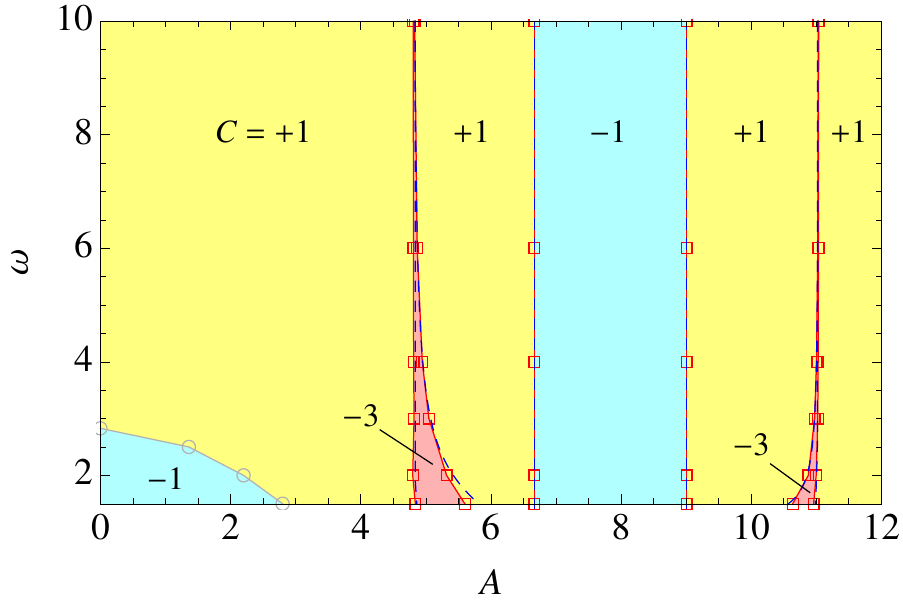}
\caption{
The phase diagram for the tight-binding model on the Lieb lattice
driven by a circularly polarized light with the amplitude $A$ and frequency $\omega\ge 1.5$.
Each phase is characterized by the Chern number $C$,
obtained by summation over all the bands below $E=0$ (the flat band lying at $E=0$ always has
zero Chern number).
Solid lines marked by {\tiny$\square$} represent
the phase boundaries obtained by numerically exact calculations, 
while the dashed lines are obtained from the Brillouin-Wigner expansion up to $J\mathcal O(J^2/\omega^2)$.
Touching between the $n=+1$- and $n=-1$-photon Floquet sidebands occurs at the boundary marked by $\circ$'s.}
\label{fig:Lieb-phase-diagram}
\end{center}
\end{figure}

For the case of the Lieb lattice, the Hermitian part of the effective Hamiltonian 
obtained from the BW expansion is
\begin{align}
\mathcal H_{\rm BW}^{\rm Lieb}
&= 
\sum_{i,j}^{\rm NN} J_{\rm eff} c_i^\dagger c_j
+\sum_{i,j}^{\mbox{\scriptsize K-path}} i\tau_{i,j}K_{\rm eff}c_i^\dagger c_j
+\sum_{i,j}^{\mbox{\scriptsize L-path}} L_{\rm eff} c_i^\dagger c_j
\notag
\\
&\quad
+\sum_{i,j}^{\mbox{\scriptsize M-path}} M_{\rm eff} c_i^\dagger c_j
+J\mathcal O\left(\frac{J^3}{\omega^3}\right), 
\label{eq:Lieb-Heff}
\end{align}
where $\tau_{i,j}=+(-)$ stands for the NNN hopping from site $j$ to $i$ along
the clockwise (counterclockwise) path in each square,
and ``K-path,'' ``L-path,'' and ``M-path'' are those represented in Fig.~\ref{fig:Lieb-lattice}
with labels $iK_{\rm eff}, L_{\rm eff}$, and $M_{\rm eff}$, respectively.
The hopping amplitudes in the effective Hamiltonian (\ref{eq:Lieb-Heff}) are given by
\begin{widetext}
\begin{subequations}
\begin{align}
J_\text{eff}&=J \mathcal{J}_0\left(\frac{A}{2}\right)
-\frac{J^3}{\omega^2}\left\{
\sum_{m,n\neq 0}
\frac{\mathcal{J}_m(\frac{A}{2})\mathcal{J}_{m+n}(\frac{A}{2})\mathcal{J}_n(\frac{A}{2})}{mn}
\left[1+2(-1)^n+2\cos\frac{n\pi}{2}\right]
\right.
\notag
\\
&\quad
\left.
+\sum_{n\neq 0} \frac{\mathcal{J}_n^2(\frac{A}{2})\mathcal{J}_0(\frac{A}{2})}{n^2}
\left[3+(-1)^n+\cos\frac{n\pi}{2}\right]
\right\},
\\
iK_\text{eff}&=-i\frac{J^2}{\omega}\sum_{n\neq 0} \frac{\mathcal{J}_n^2(\frac{A}{2})}{n}\sin\frac{n\pi}{2},
\\
L_{\rm eff}
&=
-\frac{J^3}{\omega^2}\left\{
\sum_{m,n\neq 0}
\frac{\mathcal{J}_m(\frac{A}{2})\mathcal{J}_{m+n}(\frac{A}{2})\mathcal{J}_n(\frac{A}{2})}{mn}
(-1)^m\cos\frac{n\pi}{2}
+\frac{1}{2}\sum_{n\neq 0} \frac{\mathcal{J}_n^2(\frac{A}{2})\mathcal{J}_0(\frac{A}{2})}{n^2}
\left[(-1)^n+\cos\frac{n\pi}{2}\right]
\right\},
\\
M_{\rm eff}
&=
-\frac{J^3}{\omega^2}\left\{
\sum_{m,n\neq 0}
\frac{\mathcal{J}_m(\frac{A}{2})\mathcal{J}_{m+n}(\frac{A}{2})\mathcal{J}_n(\frac{A}{2})}{mn}
(-1)^{m+n}
+\sum_{n\neq 0} \frac{\mathcal{J}_n^2(\frac{A}{2})\mathcal{J}_0(\frac{A}{2})}{n^2}(-1)^n
\right\}.
\label{eq:Lieb-Keff}
\end{align}
\end{subequations}
\end{widetext}
The pure imaginary NNN hopping $i K_\text{eff}$ is produced 
for two-stage hopping paths composed of angled two sequential NN hoppings, as depicted
by the arrows in Fig.~\ref{fig:Lieb-lattice}(a). 
The Hamiltonian (\ref{eq:Lieb-Heff}) is similar to the checkerboard lattice~\cite{Sun2011,Neupert2011},
a prototypical model for Chern insulators, in that an electron acquires phases as if alternating positive (negative) fluxes $\pm 2\pi$ are applied in blue (pink) regions in Fig.~\ref{fig:Lieb-lattice}(a). 

Under the irradiation with CPL, the flat band remains flat independent of the value of $K_\text{eff}$,
which is forced by the particle-hole symmetry, with which the band structure is symmetric 
between positive and negative energies.
On the other hand, the energy gap starts to open at the Dirac point in the dispersive bands [Fig.~\ref{fig:Lieb-lattice}(b)].
The numerically exact results for the Chern numbers of the upper and lower bands ($C_\pm$) and that for the flat band in the zero-photon sector
are displayed in the top panel of Fig.~\ref{lieb-chern-number} for $\omega=10$.  
The topological-to-topological phase transitions between $C_\pm=1$ and $C_\pm=-1$
take place at zeros of $K_\text{eff}$ (\ref{eq:Lieb-Keff}) (see the bottom panel of Fig.~\ref{lieb-chern-number}),
which agrees with the analytic BW expansion up to $J\mathcal O(J/\omega)$,
giving $C_\pm=\pm{\rm sign}K_{\rm eff}$. 
In the very vicinity of zeros of $J_\text{eff}$, additional topological-to-topological 
transitions with higher Chern numbers ($C_\pm=\pm 3$) are seen to emerge
(Fig.~\ref{lieb-chern-number}).

In Fig.~\ref{fig:Lieb-phase-diagram}, we show the phase diagram for the Lieb lattice
with a circularly polarized light with the frequency $\omega\ge 1.5$. One can see that 
the region of the phase $C=\pm 3$ starts to expand as the frequency is decreased,
while the phase boundary between $C=+1$ and $C=-1$ does not depend on the frequency.
The result is in good agreement with the prediction from the effective Hamiltonian (\ref{eq:Lieb-Heff})
plotted by dashed lines in Fig.~\ref{fig:Lieb-phase-diagram}. The $n=+1$- and $n=-1$-photon Floquet sidebands 
touch with each other at low amplitude and frequency (in particular, at $\omega=W/2=2\sqrt{2}$ in $A\to 0$), 
where the high-frequency expansion breaks down,
and the Chern number summed over all the bands up to $E=0$ changes to $-1$ (Fig.~\ref{fig:Lieb-phase-diagram}).

The phase boundaries are well understood by the effective Hamiltonian (\ref{eq:Lieb-Heff})
obtained from the BW high-frequency expansion. As in the case of the honeycomb lattice (Sec.~\ref{sec:honeycomb-chern}),
we assume that the gap closing occurs at the symmetric points in the Brillouin zone,
where the eigenvalues of the effective Hamiltonian (\ref{eq:Lieb-Heff}) are analytically calculated as
\begin{align}
&0,\;
\pm 2\sqrt{2}(J_{\rm eff}+2L_{\rm eff}+M_{\rm eff})
\;\;
\mbox{for }\bm k=\bm\Gamma=(0,0),
\\
&0,\; \pm 2K_{\rm eff}
\;\;
\mbox{for }\bm k=\bm K=(\pi,\pi),
\\
&0,\; \pm 2(J_{\rm eff}-2L_{\rm eff}+M_{\rm eff})
\;\;
\mbox{for }\bm k=\bm M=(\pi,0).
\end{align}
The condition for gap closing is either $K_{\rm eff}(A)=0$, or
\begin{align}
\omega
&=
J\sqrt{-\frac{J_{\rm eff}^{(2)}(A)+2L_{\rm eff}^{(2)}(A)+M_{\rm eff}^{(2)}(A)}{J_{\rm eff}^{(0)}(A)}},
\\
\omega
&=
J\sqrt{-\frac{J_{\rm eff}^{(2)}(A)-2L_{\rm eff}^{(2)}(A)+M_{\rm eff}^{(2)}(A)}{J_{\rm eff}^{(0)}(A)}},
\end{align}
where we have defined $J_{\rm eff}(\omega,A)=J_{\rm eff}^{(0)}(A)+(J/\omega)^2 J_{\rm eff}^{(2)}(A)$, $L_\text{eff}(\omega,A)=(J/\omega)^2 L_\text{eff}^{(2)}(A)$, and $M_\text{eff}(\omega,A)=(J/\omega)^2 M_\text{eff}^{(2)}(A)$.
These correspond to the dashed lines shown in Fig.~\ref{fig:Lieb-phase-diagram}.
The boundary between $C=+1$ and $C=-1$ phases is determined by $K_{\rm eff}(A)=0$, which 
explains why the boundary does not depend on $\omega$. In the high-frequency expansion,
the $n=+1$ and $n=-1$ Floquet sidebands touch at
\begin{align}
\omega=2\sqrt{2}(J_{\rm eff}+2L_{\rm eff}+M_{\rm eff}),
\end{align}
which agrees with the boundary marked by $\circ$'s in Fig.~\ref{fig:Lieb-phase-diagram}.

\begin{figure}[t]
\begin{center}
\includegraphics[width=\hsize]{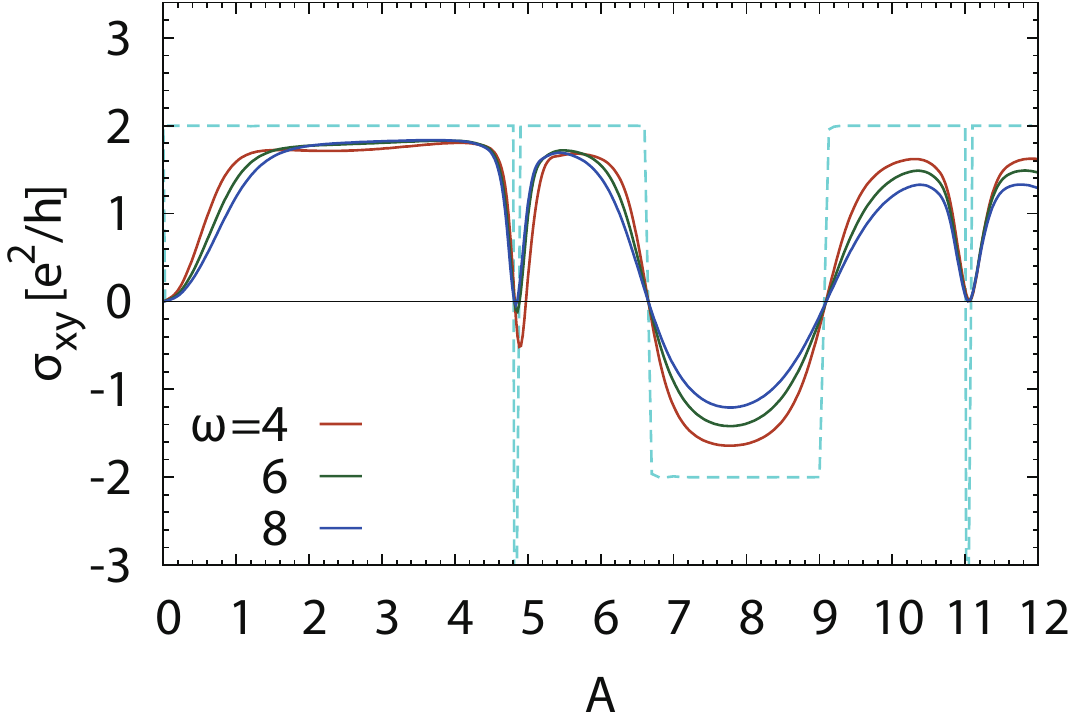}
\caption{
The Hall conductivity $\sigma_{xy}$ for the Lieb tight-binding model (at $U=0$) at half filling
driven by circularly polarized light with the amplitude $A$
and coupled to a heat bath with $\Gamma=T=0.06$
for various values of the frequency $\omega$. 
The dashed line is the Chern number ($\times2$) summed over all the bands below $E=0$
for $\omega=10$ (Fig.~\ref{lieb-chern-number}).}
\label{fig:Lieb-Sxy-Keff}
\end{center}
\end{figure}

\begin{figure}[t]
\begin{center}
\includegraphics[width=\hsize]{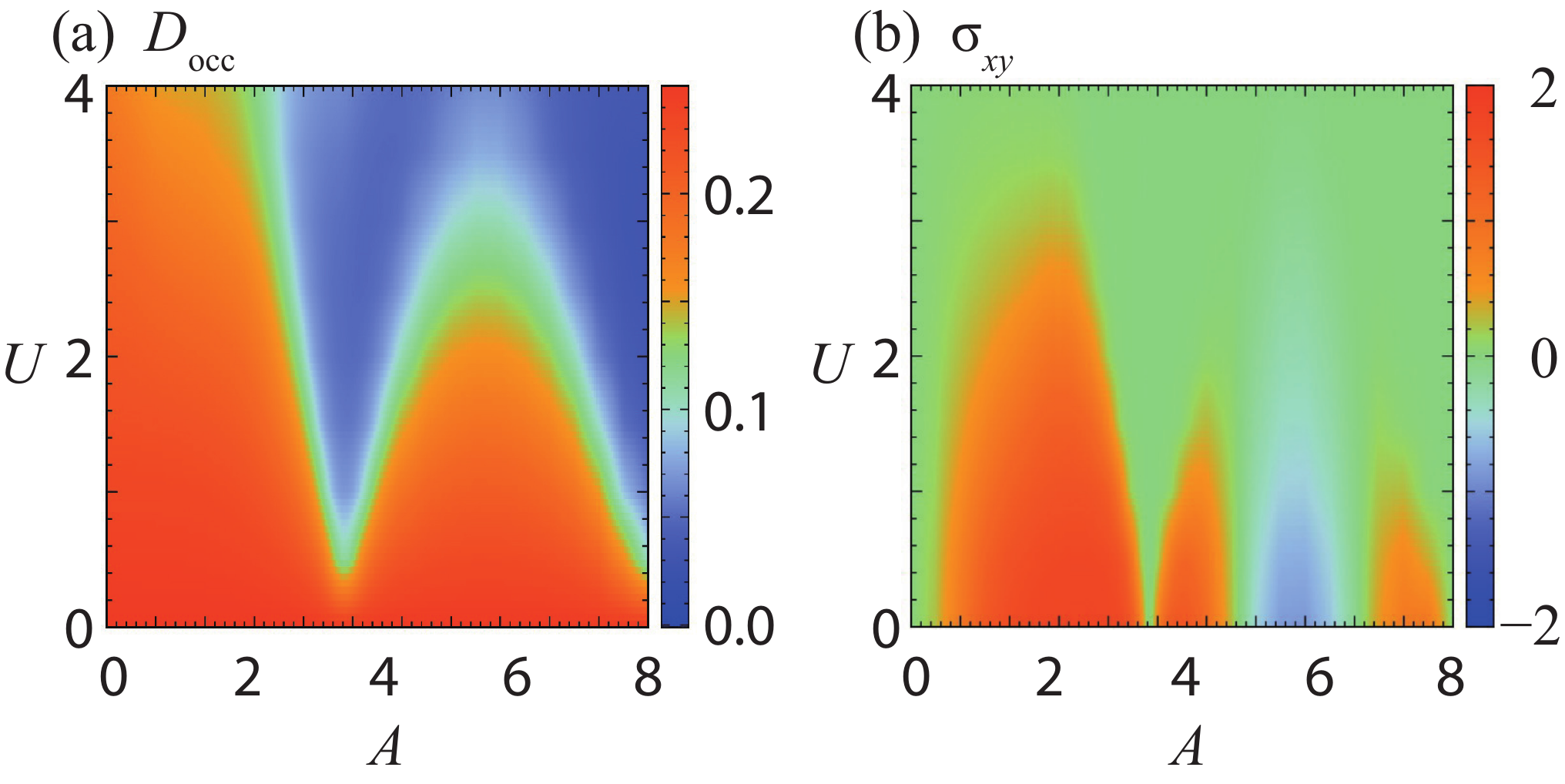}
\caption{Color-coded
(a) double occupancy $D_\text{occ}$ and (b) Hall conductivity $\sigma_{xy}$
against the amplitude $A$ of the circularly polarized light and the Hubbard $U$
for the Lieb-Hubbard model. Here the frequency of CPL is $\omega=10$;
the heat-bath parameters are $\Gamma=T=0.1$.}
\label{fig:Lieb-Docc-Sxy}
\end{center}
\end{figure}

In the presence of dissipation and/or interaction, we can again apply the 
Floquet DMFT to the dissipative Lieb-Hubbard model driven by CPL at half filling. 
In Fig.~\ref{fig:Lieb-Sxy-Keff}, we first look at the Hall conductivity $\sigma_{xy}$ against the amplitude $A$ of CPL at $U=0$ with $\Gamma=T=0.06$ for various values of the frequency $\omega$. 
For comparison, we also display the corresponding Chern number (dashed line in Fig.~\ref{fig:Lieb-Sxy-Keff}).
One can see that $\sigma_{xy}$ approaches its quantized value as $\omega$ is reduced
from 8 to 4. This is because for lower frequencies the band gap $\Delta_G=8|K_{\rm eff}|\propto 1/\omega$
becomes larger, which helps to suppress interband excitations.
The condition for the quantization of $\sigma_{xy}$ is similar to the case of the honeycomb lattice:
$\omega$ should be greater than the effective bandwidth $W|\mathcal{J}_0(A/2)|$ ($W=4\sqrt{2}J$),
and the gap $\Delta_G$ should be greater than $T$ and $\Gamma$.
If we turn on the Hubbard $U$, the phase diagram 
against $A$ and $U$ is as displayed in Fig.~\ref{fig:Lieb-Docc-Sxy}, where the double occupancy is a measure for the Mott 
insulation and the Hall conductivity $\sigma_{xy}$ is a measure for the 
topological properties as before. We have again transitions (crossovers) from
the topological to Mott-insulating phases,
which can be observed as a sharp drop of $D_{\rm occ}$ and $|\sigma_{xy}|$.
This is understood as an analog of the effective bandwidth-controlled Mott transition as in the honeycomb case
(see Sec.~\ref{sec:honeycomb-results}). 
\\\\

\begin{figure}[t]
\begin{center}
\includegraphics[width=.9\columnwidth]{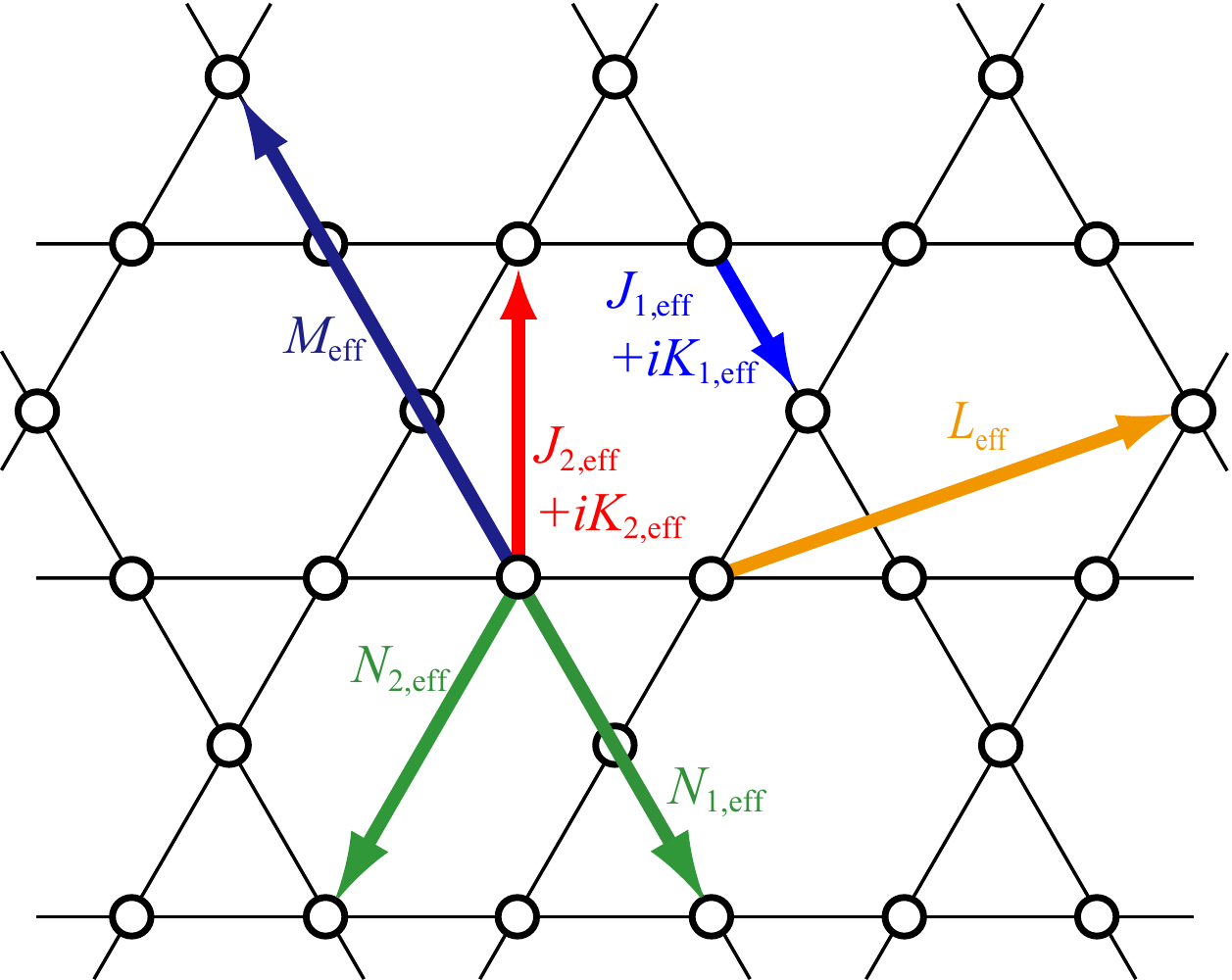}
\caption{
The hoppings in the effective Hamiltonian (\ref{eq:kagome-Heff}) derived from the BW expansion
for the kagom\'e lattice driven by a circularly polarized light. 
}
\label{fig:kagome_NNN}
\end{center}
\end{figure}

\section{Kagom\'e lattice in a circularly polarized light}
\label{sec:kagome}

Let us finally explore another example of Floquet topological insulators, which is the kagom\'e lattice (Fig.~\ref{fig:kagome_NNN}) 
in a circularly polarized light. The Hamiltonian is given by
\begin{align}
\mathcal H^{\mbox{\scriptsize kagom\'e}}(t)
&=
\sum_{i,j}^{\rm NN} J_{i,j}(t)c_i^\dagger c_j,
\end{align}
which has the NN hopping $J_{i,j}(t)$ (\ref{Peierls}) on the kagom\'e lattice.
The length of NN bonds is set to be $1/2$.
The band structure of the kagom\'e lattice, which belongs to the Mielke model among the classes of flat-band models,
consists of one flat band at the lowest energy for $J>0$ and two dispersive bands
in the absence of CPL.
The bottom flat band touches quadratically with the middle band at the $\Gamma$ point,
while the middle and top bands form Dirac cones around $K$ and $K'$ points.

\begin{figure}[t]
\begin{center}
\includegraphics[width=8.3cm]{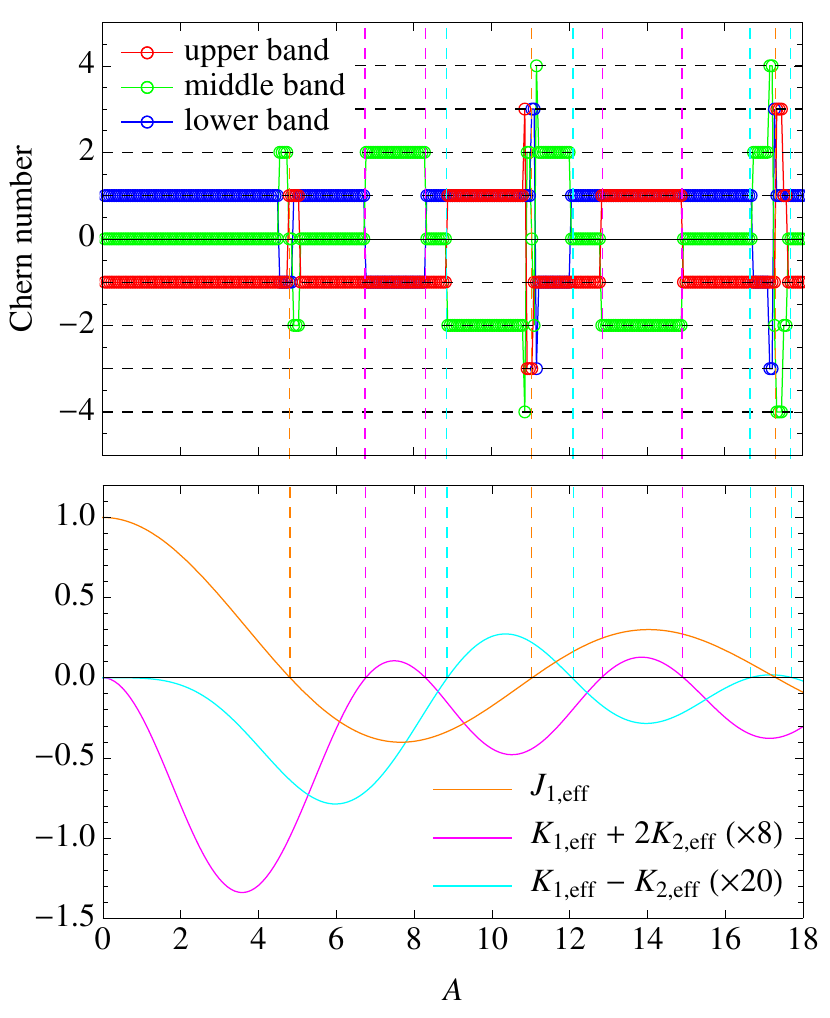}
\caption{
(Top) The Chern number for each band in the zero-photon sector of the kagom\'e tight-binding model driven by circularly polarized light
with $\omega=10$ here. (Bottom) The hopping parameters for the corresponding effective Hamiltonian.
The values of $K_{\rm 1,eff}$ and $K_{\rm 2,eff}$ are rescaled for visibility. 
Dashed lines mark zeros of the curves.
}
\label{fig:kagome-chern-number}
\end{center}
\end{figure}

In the presence of CPL, we apply the BW expansion [Eqs.~(\ref{eq:Heff-formula})--(\ref{eq:Heff-formula2})] to derive
the effective Hamiltonian up to $J\mathcal O(J^2/\omega^2)$,
\begin{align}
\mathcal H_{\rm BW}^{\mbox{\scriptsize kagom\'e}}
&=
\sum_{i,j}^{\rm NN} (J_{\rm 1,eff}+i\tau_{i,j} K_{\rm 1,eff}) c_i^\dagger c_j
\notag
\\
&\quad
+\sum_{i,j}^{\rm NNN} (J_{\rm 2,eff}+i\tau_{i,j} K_{\rm 2,eff})c_i^\dagger c_j
\notag
\\
&\quad
+\sum_{i,j}^{\mbox{\scriptsize L-path}} L_{\rm eff}c_i^\dagger c_j
+\sum_{i,j}^{\mbox{\scriptsize M-path}} M_{\rm eff} c_i^\dagger c_j
\notag
\\
&\quad
+\sum_{i,j}^{\mbox{\scriptsize N$_1$-path}} N_{\rm 1,eff}c_i^\dagger c_j
+\sum_{i,j}^{\mbox{\scriptsize N$_2$-path}} N_{\rm 2,eff}c_i^\dagger c_j
\notag
\\
&\quad
+J\mathcal O\left(\frac{J^3}{\omega^3}\right),
\label{eq:kagome-Heff}
\end{align}
where $\tau_{i,j}=+(-)$ corresponds to the clockwise (counterclockwise) hopping from site $j$ to $i$
in each hexagon (see Fig.~\ref{fig:kagome_NNN}),
and ``L-path,'' etc., are those represented by
the paths shown in Fig.~\ref{fig:kagome_NNN} with labels $M_{\rm eff}$, etc., respectively.
Similar to the case of the Lieb lattice, the BW expansion generates 
a non-Hermitian effective Hamiltonian. In Eq.~(\ref{eq:kagome-Heff}), we drop the anti-Hermitian
part, which does not contribute to the quasienergy up to $J\mathcal O(J^2/\omega^2)$,
as discussed in Sec.~\ref{sec:Lieb}.
The effective hopping parameters then reads
\begin{widetext}
\begin{subequations}
\begin{align}
J_{\rm 1,eff}&=J\mathcal{J}_0\left(\frac{A}{2}\right)
-\frac{J^3}{\omega^2}
\left\{
\sum_{m,n\neq 0} \frac{\mathcal{J}_m(\frac{A}{2})\mathcal{J}_{m+n}(\frac{A}{2})\mathcal{J}_n(\frac{A}{2})}{mn}
\left[1+2(-1)^m+2(1+(-1)^m)\cos\frac{2m\pi}{3}\right]
\right.
\notag
\\
&\quad
\left.
+\sum_{n\neq 0} \frac{\mathcal{J}_{n}^2(\frac{A}{2})\mathcal{J}_{0}(\frac{A}{2})}
{n^2}
\left[4+(-1)^n+(1+(-1)^n)\cos\frac{2n\pi}{3}\right]
\right\},
\\
J_{\rm 2,eff}&=
-\frac{J^3}{\omega^2}
\left\{
2\sum_{m,n\neq 0} \frac{\mathcal{J}_m(\frac{A}{2})\mathcal{J}_{m+n}(\frac{A}{2})\mathcal{J}_n(\frac{A}{2})}{mn}
(-1)^{m+n}\cos\frac{2m\pi}{3}
+\sum_{n\neq 0} \frac{\mathcal{J}_{n}^2(\frac{A}{2})\mathcal{J}_{0}(\frac{A}{2})}
{n^2}
(-1)^n\left(1+\cos\frac{2n\pi}{3}\right)
\right\},
\\
iK_{1, \text{eff}} &= -i \frac{J^2}{\omega} \sum_{n\neq 0} \frac{\mathcal{J}^2_n(\frac{A}{2})}{n} \sin\frac{n\pi}{3}, \\
iK_{2, \text{eff}} &= -i \frac{J^2}{\omega} \sum_{n\neq 0} \frac{\mathcal{J}_n^2(\frac{A}{2})}{n} \sin\frac{2n\pi}{3},
\\
L_{\rm eff}&=
-\frac{J^3}{\omega^2}
\left\{
\sum_{m,n\neq 0} \frac{\mathcal{J}_m(\frac{A}{2})\mathcal{J}_{m+n}(\frac{A}{2})\mathcal{J}_n(\frac{A}{2})}{mn}
(-1)^{m}\cos\frac{2n\pi}{3}
+\frac{1}{2}\sum_{n\neq 0} \frac{\mathcal{J}_{n}^2(\frac{A}{2})\mathcal{J}_{0}(\frac{A}{2})}
{n^2}
\left[(-1)^n+\cos\frac{2n\pi}{3}\right]
\right\},
\\
M_{\rm eff}&=
-\frac{J^3}{\omega^2}
\left\{
\sum_{m,n\neq 0} \frac{\mathcal{J}_m(\frac{A}{2})\mathcal{J}_{m+n}(\frac{A}{2})\mathcal{J}_n(\frac{A}{2})}{mn}
(-1)^{m+n}
+\sum_{n\neq 0} \frac{\mathcal{J}_{n}^2(\frac{A}{2})\mathcal{J}_{0}(\frac{A}{2})}
{n^2}(-1)^n
\right\},
\\
N_{\rm 1,eff}&=
-\frac{J^3}{\omega^2}
\left\{
2\sum_{m,n\neq 0} \frac{\mathcal{J}_m(\frac{A}{2})\mathcal{J}_{m+n}(\frac{A}{2})\mathcal{J}_n(\frac{A}{2})}{mn}
(-1)^{n}\cos\frac{2(m-n)\pi}{3}
+\sum_{n\neq 0} \frac{\mathcal{J}_{n}^2(\frac{A}{2})\mathcal{J}_{0}(\frac{A}{2})}
{n^2}(1+(-1)^n)\cos\frac{2n\pi}{3}
\right\},
\\
N_{\rm 2,eff}&=
-\frac{2J^3}{\omega^2}
\left\{
\sum_{m,n\neq 0} \frac{\mathcal{J}_m(\frac{A}{2})\mathcal{J}_{m+n}(\frac{A}{2})\mathcal{J}_n(\frac{A}{2})}{mn}
\cos\frac{2(m-n)\pi}{3}
+\sum_{n\neq 0} \frac{\mathcal{J}_{n}^2(\frac{A}{2})\mathcal{J}_{0}(\frac{A}{2})}
{n^2}\cos\frac{2n\pi}{3}
\right\}.
\end{align}
\end{subequations}
\end{widetext}
A new ingredient emerges here, where even the NN hopping acquires an imaginary component, $iK_{\rm 1,eff}$, 
in addition to the NNN hopping ($iK_{\rm 2,eff}$). The other hopping amplitudes remain real.

\begin{figure}[t]
\begin{center}
\includegraphics[width=\hsize]{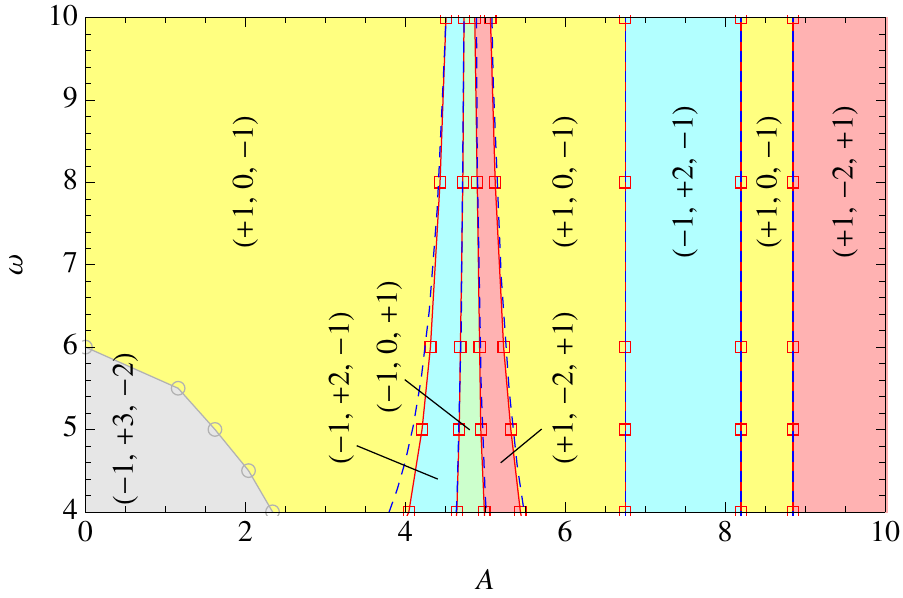}
\caption{
The phase diagram for the tight-binding model on the kagom\'e lattice driven by a circularly polarized light
with the amplitude $A$ and frequency $\omega\ge4$. Each phase is characterized by a set of Chern numbers $(C_1,C_2,C_3)$
for the lower, middle, and upper bands, respectively.
Solid lines marked by {\tiny$\square$}'s represent
the phase boundaries obtained by numerically exact calculations, 
while the dashed lines are obtained from the Brillouin-Wigner expansion up to $J\mathcal O(J^2/\omega^2)$.
Touching between the $n=0$- and $n=\pm1$-photon Floquet bands occurs at the boundary marked by $\circ$'s.}
\label{fig:kagome-phase-diagram}
\end{center}
\end{figure}

In the top panel of Fig.~\ref{fig:kagome-chern-number},
we plot the numerically exact results for the Chern number of the kagom\'e lattice against the amplitude $A$ of CPL ($\omega=10$).
A series of topological-topological transitions are salient, 
which exhibit an even more complex structure than those for the honeycomb (Fig.~\ref{fig:Chern_numbers})
and Lieb (Fig.~\ref{lieb-chern-number}) lattices. 
In the Chern numbers $(C_1,C_2,C_3)$ for the lower, middle, and upper bands, respectively, an essential difference from the Lieb model is 
that even the flat band has nontrivial Chern numbers (with band bending).  
This is allowed since the electron-hole symmetry is 
broken in the kagom\'e lattice.

In Fig.~\ref{fig:kagome-phase-diagram}, we depict the topological phase diagram for the kagom\'e lattice
driven by CPL for the amplitude $A\le 10$ and frequency $\omega\ge 4$, where each phase is labeled
by the set of Chern numbers $(C_1,C_2,C_3)$ in the zero-photon sector.  
At high frequency, the phases that appear around the zeros of $J_{\rm 1,eff}(A)\sim J\mathcal{J}_0(A/2)$ are narrow on the $A$ axis,
whereas they grow wider as one decreases the frequency. The band touching between the zero- and nonzero-photon
sectors occurs at the line marked by $\circ$'s in Fig.~\ref{fig:kagome-phase-diagram}.

The behavior of the Chern numbers can be again captured in terms of the effective Hamiltonian (\ref{eq:kagome-Heff}).
The topological transitions can happen only when the band gap closes, which we assume to take place
at the symmetric points of the Brillouin zone.
The eigenvalues of the effective Hamiltonian (\ref{eq:kagome-Heff}) are analytically calculated as
\begin{align}
&4(J_{\rm 1,eff}+J_{\rm 2,eff}+2L_{\rm eff}+M_{\rm eff}+N_{\rm 1,eff})
+2N_{\rm 2,eff},
\label{eq:kagome-eigenvalue1}
\\
&\!\!-2(J_{\rm 1,eff}+J_{\rm 2,eff})\pm 2\sqrt{3}(K_{\rm 1,eff}-K_{\rm 2,eff})
\notag
\\
&\quad -4L_{\rm eff}-2M_{\rm eff}+4N_{\rm 1,eff}+2N_{\rm 2,eff},
\label{eq:kagome-eigenvalue2}
\end{align}
at $\bm k=\bm\Gamma$, and
\begin{align}
&\!\!-2(J_{\rm 1,eff}-2J_{\rm 2,eff}+2L_{\rm eff}-2M_{\rm eff}+N_{\rm 1,eff})
-N_{\rm 2,eff},
\label{eq:kagome-eigenvalue3}
\\
&J_{\rm 1,eff}-2J_{\rm 2,eff}\pm\sqrt{3}(K_{\rm 1,eff}+2K_{\rm 2,eff})
\notag
\\
&\quad 
+2L_{\rm eff}
-2M_{\rm eff}-2N_{\rm 1,eff}-N_{\rm 2,eff},
\label{eq:kagome-eigenvalue4}
\end{align}
at $\bm k=\bm K,\bm K'$. At $\bm k=\bm M$, the band gaps do not close in the high-frequency regime.
Hence, the topological transitions can take place when two of the three quasienergies meet at $\bm k=\bm\Gamma$ or
$\bm k=\bm K,\bm K'$. In particular, when $A$ is away from the zeros of $\mathcal{J}_0(A/2)$, $J_{\rm 1,eff}$ is much larger than
the other parameters. In this case, the topological transitions take place when
either $K_{\rm 1,eff}-K_{\rm 2,eff}$ or $K_{\rm 1,eff}+2K_{\rm 2,eff}$ vanishes.
In the bottom panel of Fig.~\ref{fig:kagome-chern-number}, we plot
these quantities as well as $J_{\rm 1,eff}$.
We can see that some of the topological transitions observed in the top panel are indeed located
at the zeros of $K_{\rm 1,eff}-K_{\rm 2,eff}$ or $K_{\rm 1,eff}+2K_{\rm 2,eff}$,
which justifies the BW expansion in this parameter regime.
The other transitions occur in the vicinity of the zeros of $J_{1,\rm eff}$,
the place of which can be analytically computed from the eigenvalues (\ref{eq:kagome-eigenvalue1})--(\ref{eq:kagome-eigenvalue4}).
They correspond to the dashed lines in Fig.~\ref{fig:kagome-phase-diagram}.

\section{Conclusion}
\label{sec:concl}
In this paper, we have constructed a general theory for a systematic high-frequency expansion
based on the Brillouin-Wigner theory for quantum systems driven by time-periodic drives.  
We derive the explicit formula for the expansions of the wave operator $\mathit{\Omega}(\epsilon)$ 
(\ref{eq:Omega-eps-series}) and effective Hamiltonian $H_\text{eff}(\epsilon)$ (\ref{eq:heff-eps-series}) up to infinite order in $1/\omega$.
They correctly reproduce the quasienergies and eigenstates as a power series of $1/\omega$.
We have clarified the relation between the BW theory and other existing methods of high-frequency expansions
based on the Floquet-Magnus expansion and van Vleck degenerate perturbation theory.
An advantage of the BW theory is that one can readily calculate higher-order terms
systematically, which greatly facilitates understanding of Floquet topological transitions.

We have then applied the BW theory to the honeycomb, Lieb, and kagom\'e lattice models in a circularly polarized light,
where field-induced Floquet topological-to-topological phase transitions are  observed, 
in a versatile manner that depends on the lattice structure. 
In the high-frequency regime, these transitions can be well understood by the high-frequency expansions.
As the frequency is decreased, the behavior of the system deviates from what is expected from the high-frequency expansions,
and relatively high Chern numbers can be produced with the very complicated phase diagram.

In order to consider realistic situations where the effects of energy dissipation and many-body interaction are relevant,
we investigate the dissipative Hubbard model on the honeycomb and Lieb lattices driven by circularly polarized light
within the Floquet dynamical mean-field theory. We first identify the condition for quantization of the Hall conductivity
in the Floquet topological phases such that $J^2A^2/T, J^2A^2/\Gamma\lesssim\omega\lesssim W$.
We have then found a photo-induced Floquet topological-to-Mott insulator transition (crossover),
which is signaled by a sharp reduction of the double occupancy and the Hall conductivity.
The crossover line can be understood from the high-frequency expansions in BW theory 
as a nonequilibrium analog of bandwidth-controlled Mott transitions.

There are various future problems. For the honeycomb lattice,
we have seen an increasingly intricate phase diagram for the Chern numbers 
in the low-frequency regime. An obvious interest is how the lower-frequency
regime should look for other lattices. Consideration of the Hubbard interaction
is also intriguing for the kagom\'e lattice, where the particle-hole symmetry is broken.
As a quantum state arising from the electron correlation, we have only 
considered the Mott insulator at half filling here, 
but there are a host of other states such as the fractional Chern 
insulator\cite{Tang11,Sun2011,Neupert2011,Grushin14} at fractional filling,
and a possibility of inducing such states in nonequilibrium should be one direction.
An extension of the present line of study to those will be left as a future work.

\begin{acknowledgements}
This work is
supported by Grant-in-Aid for Scientific Research (Grants
No. 26247064 and No. 26400350) from MEXT and
ImPACT from JST (Grant No. 2015-PM12-05-01) (HA).
S.K. is supported by the Advanced
Leading Graduate Course for Photon Science (ALPS).
N.T. is supported by JSPS KAKENHI 
Grants. No. 25104709 and No. 25800192.
\end{acknowledgements}

\appendix

\section{Brillouin-Wigner perturbation theory in quantum mechanics}\label{app:BW}
In this Appendix we recapitulate the Brillouin-Wigner perturbation theory for quantum mechanics with a simple example.
Let us illustrate the theory for a $2\times2$ Hamiltonian,
\begin{equation}
H=\begin{pmatrix}E+m & ke^{-i\theta}\\
ke^{i\theta} & E-m
\end{pmatrix}.
\end{equation}
We want to derive the effective Hamiltonian constructed from a given projection $P$: The effective Hamiltonian is defined as the operator on the space of $P$, whose eigenenergy coincides with original one $\epsilon_\alpha$ and whose eigenvector is given by $P|\Psi_\alpha\rangle$, where $|\Psi_\alpha\rangle$ is that of the original problem. Here we take $P$ as the projection onto the upper component, $P=\text{diag}(1,0)$.

For the construction of the effective Hamiltonian we introduce the {\it wave operator} $\Omega$, which satisfies $|\Psi_\alpha\rangle=\Omega P|\Psi_\alpha\rangle$. With this the effective Hamiltonian is given by $PH\Omega P$, where we can easily check that $PH\Omega PP|\psi_\alpha\rangle=\epsilon_\alpha P|\psi_\alpha\rangle$.
The wave operator is explicitly obtained from the Schr\"odinger equation projected on $(1-P)$, $\epsilon_\alpha d_\alpha=ke^{i\theta}u_\alpha+(E-m)d_\alpha$ for $|\Psi\rangle={}^t(u_\alpha,d_\alpha)$, 
where we have defined the residual part $d_\alpha=(1-P)|\Psi_\alpha\rangle$ as expressed by the projected part $u_\alpha=P|\Psi_\alpha\rangle$ and $\epsilon_\alpha$, namely,
\begin{equation}
|\psi_\alpha\rangle=\begin{pmatrix}1 & 0\\
\dfrac{ke^{i\theta}}{\epsilon_\alpha-E+m} & 0
\end{pmatrix}|\psi_\alpha\rangle=\Omega P|\psi_\alpha\rangle.
\end{equation}
The effective Hamiltonian, for the present choice of the original one, is given as
\begin{equation}
H_{\text{eff}}=PH\Omega P=\begin{pmatrix}E+m+\dfrac{k^{2}}{\epsilon_\alpha-E+m} & 0\\
0 & 0
\end{pmatrix}.
\end{equation}

Here the eigenenergy $\epsilon_\alpha$ must satisfy
\begin{equation}
\epsilon_\alpha=E+m+\frac{k^{2}}{\epsilon_\alpha-E+m},\label{appendix:scf}
\end{equation}
which results in $\epsilon_\alpha=E\pm\sqrt{m^2+k^2}$, reproducing exactly two eigenenergies.

As we have seen here, while the effective Hamiltonian is a $1\times1$ matrix, it can reproduce all the eigenenergies in the original $2\times2$ space, thanks to the $\epsilon$ dependence of the effective Hamiltonian. 
We note that one can eliminate $\epsilon_\alpha$ from the effective Hamiltonian or from the wave operator, by, e.g., expanding both sides of Eq.~(\ref{appendix:scf}) in $k$, but with such a method only one of the solutions, 
\begin{equation}
\epsilon_\alpha=E-m-\frac{k^{2}}{2m}+\frac{k^{4}}{8m^{3}}+\dots,
\end{equation}
is obtained. This is a trivial consequence of the fact that the operator in the lower dimension cannot 
accommodate all the original eigenvalues: This implies that the wave operators and the effective Hamiltonians without $\epsilon$-dependence are not unique, and which one we obtain depends on how we eliminate the $\epsilon$ dependence. In the present case, $P|\Psi_\alpha\rangle=0$ holds for the other solution at $k=0$, which makes $\Omega$ for the other solution singular and forbidden to be expanded around $k=0$.

\section{Exact quasienergy spectrum for the Dirac field around $\Gamma$ point} \label{app:Dirac}
Here, we consider the Dirac model in CPL,
\begin{equation}
H(\bm{k}+\bm{A}(t) ) = \begin{pmatrix} 0 & k_- + A e^{-i\omega t} \\ k_+ + A e^{i\omega t} & 0 \end{pmatrix},
\end{equation}
with $k_\pm \equiv k_x \pm i k_y$ and $\bm{A}(t) = {}^t(A\cos \omega t, A\sin\omega t)$, to obtain an exact expression for quasienergy spectrum. 
Floquet matrices $H_n = (1/\mathscr{T}) \int_0^{\mathscr{T}} H(\bm{k}+\bm{A}(t) ) e^{in\omega t}$ are
\begin{equation}
H_0 = \begin{pmatrix} 0 & k_- \\ k_+ & 0 \end{pmatrix}, \>
H_1 = \begin{pmatrix} 0 & A \\ 0 & 0 \end{pmatrix}, \>
H_{-1} = \begin{pmatrix} 0 & 0 \\ A & 0 \end{pmatrix},
\end{equation}
while the Floquet equation reads
\begin{subequations}
\begin{align}
{\hat H}_F & |\bm{u}_\alpha\rangle = \epsilon |\bm{u}_\alpha \rangle, \\
{\hat H}_F &=\begin{pmatrix}
\ddots & & \vdots & & \\
 & H_{0} - \omega {\bf 1} & H_{-1} & H_{-2} & \\
 \hdots & H_1 & H_0 & H_{-1} & \hdots \\
 & H_2 & H_1 & H_0 + \omega {\bf 1} & \\
 & & \vdots & & \ddots
\end{pmatrix}, \\ 
|\bm{u}_\alpha \rangle &= {}^t \begin{pmatrix} \hdots & \bm{u}_{\alpha,-1} & \bm{u}_{\alpha,0} & \bm{u}_{\alpha,1} & \hdots \end{pmatrix},
\end{align} \label{eq:Dirac_FloquetEq}
\end{subequations}
with ${\bf 1}$ the $2\times2$ identity matrix and $\bm{u}_{\alpha,n} = {}^t (u_{\alpha,n}, v_{\alpha,n})$ a $1\times2$ complex vector. 

At the Dirac point ($\bm{k}=0$), the Floquet matrix ${\hat H}_F$ turns to a $2\times2$ block-diagonal form. As a result, Eq.~(\ref{eq:Dirac_FloquetEq}) is equivalent to a reduced set of eigenvalue problems,
\begin{equation}
\begin{pmatrix}
(m-1)\omega & A \\ A & m\omega
\end{pmatrix}
\begin{pmatrix}
v_{\alpha,m-1} \\ u_{\alpha,m}
\end{pmatrix}
=
\epsilon_\alpha
\begin{pmatrix}
v_{\alpha,m-1} \\ u_{\alpha,m}
\end{pmatrix}, \quad
m \in {\mathbb Z}.
\end{equation}
Thus, for each $m\in{\mathbb Z}$, one can find exact quasienergies, 
\begin{align}
\epsilon^{(0)}_{\pm,m} &= \left(m-\frac 12 \right) \omega \pm \frac{\sqrt{\omega^2 + 4 A^2}}{2},
\label{eq:Dirac_E0}
\end{align}
and eigenvectors,
\begin{subequations}
\begin{align}
|\pm,m&\rangle = {}^t (\hdots,\bm{u}_{-1}, \bm{u}_{0}, \bm{u}_{1}, \hdots), \\
 \bm{u}_{n} &= \frac{1}{(\mp \epsilon_{\mp,0}^{(0)})^{1/2} (\omega^2+4A^2)^{1/4}}
\begin{pmatrix}
-\delta_{n,m} \epsilon_{\mp,0}^{(0)} \\
\delta_{n,m-1} A
\end{pmatrix},
\end{align}
\end{subequations}
for the Floquet equation (\ref{eq:Dirac_FloquetEq}). The quasienergies (\ref{eq:Dirac_E0}) at the Dirac point have also been derived by Oka and Aoki \cite{Oka2009} by solving the time-dependent Schr\"{o}dinger equation.

Next we calculate the quasienergy spectrum around the Dirac point. For that purpose we decompose the Floquet matrix ${\hat H}_F$ into the $\bm{k}=0$ part 
${\hat H}^{(0)} = {\hat H}_F \big|_{\bm{k}=0}$ and the rest 
${\hat H}^{(1)} = {\hat H}_F - {\hat H}_F \big|_{\bm{k}=0} = A\mathcal{O}(k/A)$. Near the Dirac point, ${\hat H}^{(1)}$ is small and one can utilize the well-established perturbation technique. Assuming the unperturbed initial state
as $|\alpha\rangle = |s, m\rangle$ with $s=\pm, m \in {\mathbb Z}$, one can find the quasienergy spectrum up to $A\mathcal{O}(k^2/A^2)$ as
\begin{align}
\epsilon_\alpha 
= \epsilon_\alpha^{(0)}
+ \underbrace{ \langle \alpha | {\hat H}^{(1)} |\alpha\rangle }_{\epsilon_\alpha^{(1)}}
- \underbrace{ \sum_{\beta\neq \alpha} \frac{|\langle \beta | {\hat H}^{(1)}| \alpha\rangle |^2}{\epsilon^{(0)}_{\beta} - \epsilon^{(0)}_{\alpha}} }_{\epsilon_\alpha^{(2)}} + A\mathcal{O}\left(\frac{k^3}{A^3}\right).
\end{align}
We find $\epsilon^{(1)}_\alpha = 0$ since $H^{(1)}|\alpha\rangle$  is always orthogonal to $|\alpha\rangle$.
The lowest-order contribution then comes from 
\begin{equation}
\epsilon_\alpha^{(2)}
=  \frac{\omega^2+A^2}{2A^2\sqrt{\omega^2+4A^2}} \, s k^2.
\end{equation}
This leads us to exact expressions for the topological gap $\Delta\equiv 2\epsilon_{+,0}^{(0)} \big|_{\bm{k}=0}$ at the Dirac point
and the curvature $\kappa \equiv (\partial_{k_x}^2 \epsilon_{+,0}^{(2)})\big|_{\bm{k}=0}$,
\begin{align}
\Delta^\text{(exact)}
&= \sqrt{\omega^2+4A^2} - \omega,
\label{eq:Dirac_LargeW_gap} \\
\kappa^\text{(exact)}
&= \frac{\omega^2+A^2}{A^2\sqrt{\omega^2+4A^2}}.
\label{eq:Dirac_LargeW_kappa}
\end{align}

\section{Appearance of non-Hermitian and many-particle terms in the BW expansion}\label{app:drawback}
In this appendix, we elaborate how the non-Hermitian terms and many-particle terms appear 
in the BW expansion by taking the example given in Sec.~\ref{subsec:dirac}.
Specifically, we show that the non-Hermitian part does not affect
the quasienergies up to the truncation order, and that the many-particle
terms are indeed not necessary in constructing the many-particle eigenstates
for noninteracting problems.
First, to grasp the appearance of the non-Hermitian terms, let us add a scalar term,
\begin{equation}
H^{\prime}(t)=\mu \cos\omega t,
\label{eq:dirac-additional}
\end{equation} 
to the Dirac Hamiltonian Eq.~(\ref{eq:Heff-Dirac}). Eigenvectors for the
new Hamiltonian can be simply obtained by multiplying a factor,
\begin{align}
\eta(t) & =\exp\left(-i\frac{\mu}{\omega}\sin\omega t\right) =\sum_{n=-\infty}^{\infty}\mathcal{J}_{n}\left(\frac{\mu}{\omega}\right)e^{-in\omega t},
\end{align}
to the original eigenvectors for Eq.~(\ref{eq:Heff-Dirac}), since
$i\partial_{t}\eta(t)=H^{\prime}(t)\eta(t)$ so that Eq.~(\ref{eq:dirac-additional}) does not change the quasienergy. For the Floquet-Magnus expansion, all
the terms are given by commutators of
the Floquet matrices, hence no additional terms appear in the effective Hamiltonian, which is consistent
with the above observation. On the other hand, for the Brillouin-Wigner
method additional terms,
\begin{align}
H_{\text{BW}}^{\prime}&=-\frac{2A\mu k_{y}}{\omega^{2}}\begin{pmatrix}i & 0\\
0 & -i
\end{pmatrix}+\frac{A\mu(k^2-A^2)}{\omega^{3}}\begin{pmatrix}0 & 1\\
-1 & 0
\end{pmatrix}\notag\\&+\frac{A\mu}{\omega^{3}}\begin{pmatrix}0 & k_{-}^2\\
-k_{+}^2 & 0
\end{pmatrix}
+\frac{A^2\mu^2k_y}{2\omega^{4}}\begin{pmatrix}0 & -i\\
i & 0
\end{pmatrix}\notag\\
&+\frac{A\mu k_y (16A^2-\mu^2-32k^2)}{8\omega^{4}}\begin{pmatrix}i & 0\\
0 & -i
\end{pmatrix}
+\mathcal{O}\left(\frac{1}{\omega^5}\right),
\end{align}
arise in the effective Hamiltonian, where some of them
are not Hermitian. However, the quasienergies coincide with each other 
up to the truncation order: 
\begin{multline}
\left(\epsilon^{(\le4)}_{\text{BW}}\right)^{2}-\left(\epsilon^{(\le4)}_{\text{FM}}\right)^{2}\\
=-\frac{iA^3 \mu  k_y}{4\omega^5} (16 A^2 - \mu^2 - 48 k^2)
+\mathcal{O}\left(\frac{1}{\omega^6}\right).
\end{multline}

Next we consider Eq.~(\ref{eq:Heff-Dirac}) in the many-particle context. For simplicity, here we use the first-quantization formalism in a two-particle basis.
The two-particle Hamiltonian with momenta $\bm{p}$ and $\bm{q}$ is given as
\begin{equation}
H_{\bm{p},\bm{q}}(t)=H_{\bm{p}}^{\text{Dirac}}(t)\otimes\bm{1}+\bm{1}\otimes H_{\bm{q}}^{\text{Dirac}}(t),\label{eq:two-particle-t}
\end{equation}
whose BW Hamiltonian reads
\begin{align}
(H_{\bm{p},\bm{q}})_{\text{BW}}&=(H_{\bm{p}}^{\text{Dirac}})_{\text{BW}}\otimes\bm{1}+\bm{1}\otimes(H_{\bm{q}}^{\text{Dirac}})_{\text{BW}}\notag\\&+\frac{A^{2}}{\omega^{2}}\begin{pmatrix}0 & q_{-}\\
-q_{+} & 0
\end{pmatrix}\otimes\begin{pmatrix}1 & 0\\
0 & -1
\end{pmatrix}\notag\\&+\frac{A^{2}}{\omega^{2}}\begin{pmatrix}1 & 0\\
0 & -1
\end{pmatrix}\otimes\begin{pmatrix}0 & p_{-}\\
-p_{+} & 0
\end{pmatrix}+\mathcal{O}\left(\frac{1}{\omega^{3}}\right)\notag\\&\equiv H_{\bm{p},\bm{q}}^{(1)}+H_{\bm{p},\bm{q}}^{(2)}+\mathcal{O}\left(\frac{1}{\omega^{3}}\right),
\end{align}
where $H_{\bm{p},\bm{q}}^{(n)}$ denotes the $n$-particle term. Here the BW Hamiltonian is apparently interacting, while Eq.~(\ref{eq:two-particle-t}) is not. In what follows we show that one can construct eigenvectors for this interacting problem from those of a one-particle problem $(H_{\bm{p}}^{\text{Dirac}})_{\text{BW}}$ [Eq.~(\ref{eq:Heff-Dirac-expanded})].

The eigenstates and eigenenergies of the one-particle problem are given as 
\begin{align}
|u_{\bm{p},\pm}^0\rangle & =\begin{pmatrix}\pm p\\
p_{+}
\end{pmatrix}-\frac{A^{2}}{\omega}\begin{pmatrix}1\\
0
\end{pmatrix}-\frac{A^{2}}{p\omega^{2}}\begin{pmatrix}\pm (p^2-A^{2}/2)\\
pp_{+}
\end{pmatrix},\\
\epsilon_{\bm{p},\pm} & =\pm p\left(1-\frac{A^{2}}{\omega^{2}}+\frac{A^{4}}{2\omega^{2}p^{2}}\right),
\end{align}
and the wave operator as
\begin{widetext}
\begin{align}
\Xi_{\bm{p}}(t)&=\begin{pmatrix}1 & 0\\
0 & 1
\end{pmatrix}+
\left[
\frac{A}{\omega}
\begin{pmatrix}0 & 1\\
0 & 0
\end{pmatrix}
-\frac{Ap_{+}}{\omega^2}
\begin{pmatrix}1 & 0\\
0 & -1
\end{pmatrix}
\right]
e^{-i\omega t}+
\left[
-\frac{A}{\omega}\begin{pmatrix}0 & 0\\
1 & 0
\end{pmatrix}
+\frac{Ap_{-}}{\omega^2}\begin{pmatrix}1 & 0\\
0 & -1
\end{pmatrix}
\right]
e^{i\omega t},
\end{align}
up to the second order. Thus the one-particle eigenstates of quasienergy (without a factor $e^{-i\epsilon t}$) for the original time-dependent Hamiltonian [Eq.~(\ref{eq:Heff-Dirac})] are
\begin{align}
\Xi_{\bm{p}}(t)|u_{\bm{p},\pm}^{0}\rangle&=\begin{pmatrix}\pm p\\
p_{+}
\end{pmatrix}-\frac{A^{2}}{\omega}\begin{pmatrix}1\\
0
\end{pmatrix}-\frac{A^{2}}{p\omega^{2}}\begin{pmatrix}\pm (p^2-A^{2}/2)\\
pp_{+}
\end{pmatrix}\notag\\
&+\left[ \frac{A}{\omega}\begin{pmatrix}p_{+}\\
0
\end{pmatrix}+\frac{A}{\omega^{2}}\begin{pmatrix}\mp pp_{+}\\
p_{+}^{2}
\end{pmatrix}\right] e^{-i\omega t}+\left[ \frac{A}{\omega}\begin{pmatrix}0\\
\mp p
\end{pmatrix}+\frac{A}{\omega^{2}}\begin{pmatrix}\pm pp_{-}\\
A^{2}-p^{2}
\end{pmatrix}\right] e^{i\omega t}+\mathcal{O}\left(\frac{1}{\omega^{3}}\right)\\&=|u_{\bm{p},\pm}^{0}\rangle+|u_{\bm{p},\pm}^{+1}\rangle e^{-i\omega t}+|u_{\bm{p},\pm}^{-1}\rangle e^{i\omega t}+\mathcal{O}\left(\frac{1}{\omega^{3}}\right).
\end{align}

The two-particle eigenstates of quasienergy for Eq.~(\ref{eq:two-particle-t}) are obtained as a direct product of one-particle ones, namely  $\Xi_{\bm{p}}(t)|u_{\bm{p},\tau}^{0}\rangle\otimes\Xi_{\bm{q}}(t)|u_{\bm{q},\tau^{\prime}}^{0}\rangle$. This implies that, while $(H_{\bm{p},\bm{q}})_{\text{BW}}$ has an interacting form, 
its eigenstates should simply be obtained as a zero-photon projection of the direct product, $\sum_{m}|u_{\bm{p},\tau}^{m}\rangle\otimes|u_{\bm{q},\tau^{\prime}}^{-m}\rangle$. 
This is readily confirmed from $(H_{\bm{p},\bm{q}})_{\text{BW}}\sum_{m}|u_{\bm{p},\tau}^{m}\rangle\otimes|u_{\bm{q},\tau^{\prime}}^{-m}\rangle=(\epsilon_{\bm{p}\tau}+\epsilon_{\bm{q}\tau^{\prime}})\sum_{m}|u_{\bm{p},\tau}^{m}\rangle\otimes|u_{\bm{q},\tau^{\prime}}^{-m}\rangle+\mathcal{O}(1/\omega^{3})$, because
\begin{align}
(H_{\bm{p},\bm{q}})_{\text{BW}}\sum_{m}|u_{\bm{p},\tau}^{m}\rangle\otimes|u_{\bm{q},\tau^{\prime}}^{-m}\rangle&=(\epsilon_{\bm{p}\tau}+\epsilon_{\bm{q}\tau^{\prime}}+H_{\bm{p},\bm{q}}^{(2)})|u_{\bm{p},\tau}^{0}\rangle\otimes|u_{\bm{q},\tau^{\prime}}^{0}\rangle+H_{\bm{p},\bm{q}}^{(1)}\sum_{m=\pm1}|u_{\bm{p},\tau}^{m}\rangle\otimes|u_{\bm{q},\tau^{\prime}}^{-m}\rangle+\mathcal{O}\left(\frac{1}{\omega^{3}}\right),
\end{align}
\begin{align}
H_{\bm{p},\bm{q}}^{(1)}\sum_{m=\pm1}|u_{\bm{p},\tau}^{m}\rangle\otimes|u_{\bm{q},\tau^{\prime}}^{-m}\rangle&=-\frac{A^{2}}{\omega^{2}}(\tau pp_{-}q_{+}+\tau^{\prime}p_{+}qq_{-})\begin{pmatrix}1\\
0
\end{pmatrix}\otimes\begin{pmatrix}1\\
0
\end{pmatrix}-\frac{A^{2}}{\omega^{2}}(\tau^{\prime}p_{+}^{2}q+\tau pq_{+}^{2})\begin{pmatrix}0\\
1
\end{pmatrix}\otimes\begin{pmatrix}0\\
1
\end{pmatrix}+\mathcal{O}\left(\frac{1}{\omega^{3}}\right),
\end{align}
\begin{align}
H_{\bm{p},\bm{q}}^{(2)}|u_{\bm{p},\tau}^{0}\rangle\otimes|u_{\bm{q},\tau^{\prime}}^{0}\rangle&=\frac{A^{2}}{\omega^{2}}\begin{pmatrix}p_{+}q_{-}\\
-\tau pq_{+}
\end{pmatrix}\otimes\begin{pmatrix}\tau^{\prime}q\\
-q_{+}
\end{pmatrix}+\frac{A^{2}}{\omega^{2}}\begin{pmatrix}\tau p\\
-p_{+}
\end{pmatrix}\otimes\begin{pmatrix}p_{-}q_{+}\\
-\tau^{\prime}p_{+}q
\end{pmatrix}+\mathcal{O}\left(\frac{1}{\omega^{3}}\right)\notag\\&=-H_{\bm{p},\bm{q}}^{(1)}\sum_{m=\pm1}|u_{\bm{p},\tau}^{m}\rangle\otimes|u_{\bm{q},\tau^{\prime}}^{-m}\rangle+(\tau p+\tau^{\prime}q)\frac{A^{2}}{\omega^{2}}\left[ \begin{pmatrix}p_{+}\\
0
\end{pmatrix}\otimes\begin{pmatrix}0\\
-\tau^{\prime}q
\end{pmatrix}+\begin{pmatrix}0\\
-\tau p
\end{pmatrix}\otimes\begin{pmatrix}q_{+}\\
0
\end{pmatrix}\right] +\mathcal{O}\left(\frac{1}{\omega^{3}}\right)\notag\\&=(-H_{\bm{p},\bm{q}}^{(1)}+\epsilon_{\bm{p}\tau}+\epsilon_{\bm{q}\tau^{\prime}})\sum_{m=\pm1}|u_{\bm{p},\tau}^{m}\rangle\otimes|u_{\bm{q},\tau^{\prime}}^{-m}\rangle+\mathcal{O}\left(\frac{1}{\omega^{3}}\right).
\end{align}
\end{widetext}

\bibliography{refs}

\end{document}